\documentstyle[fleqn,11pt,epsf]{article}
\newcommand{\bd}[1]{ \mbox{\boldmath $#1$}  }
\newcommand{\jvec}{\mbox{\boldmath $j$}}

\newcommand{\fbar}{\not{\!f}}

\newcommand{\Qbar}{\not{\!Q}}
\newcommand{\Kbar}{\not{\!K}}

\newcommand{\Hslash}{\not{\!H}}
\newcommand{\Pbar}{\not{\!P}}

\newcommand{\be}{\begin{equation}}
\newcommand{\ee}{\end{equation}}
\newcommand{\ba}{\begin{eqnarray}}
\newcommand{\ea}{\end{eqnarray}}
\newcommand{\uu}{u({\bf p},s)}
\newcommand{\ubar}{\overline{u}({\bf p'},s')}

\newcommand{\sigvec}{\mbox{\boldmath $\sigma$}}
\newcommand{\etavec}{\mbox{\boldmath $\eta$}}

\newcommand{\kappavec}{\mbox{\boldmath $\kappa$}}

\newcommand{\nsigma}{\mbox{\boldmath $\sigma$}}
\newcommand{\ngamma}{\mbox{\boldmath $\gamma$}}
\newcommand{\neta}{\mbox{\boldmath $\eta$}}
\newcommand{\nh}{{\bf      h}}

\newcommand{\nk}{{\bf      k}}
\newcommand{\np}{{\bf      p}}       
\newcommand{\nq}{{\bf      q}}
         
\newcommand{\nJ}{{\bf      J}}
       
\newcommand{\nkappa}{\mbox{\boldmath$\kappa$}}
\newcommand{\columnmatrix}[2]{\left[
                               \begin{array}{cc}
                                  \displaystyle #1 \\[1ex]
                                  \displaystyle #2
                               \end{array}
                            \right]}
\begin{document}
\begin{titlepage}

\mbox{} 
\vspace*{2.5\fill} 
{\Large\bf 
\begin{center}
%
Gauge and Lorentz invariant one-pion 
exchange currents in electron
scattering from a relativistic Fermi gas
%
\end{center}
} 

\renewcommand{\thefootnote}{\fnsymbol{footnote}}

\vspace{1\fill} 
\begin{center}
{\large 
J.E. Amaro$    ^{1,}$%
\footnote[2]{Corresponding author. 
E-mail address: amaro@ugr.es, phone: 34-958-246172, FAX: 34-958-249487}, 
M.B. Barbaro$  ^{2}$, 
J.A. Caballero$^{3}$, 
T.W. Donnelly$ ^{4}$ and 
A. Molinari$   ^{2}$
}
\end{center}
\begin{small}
\begin{center}
$^{1}${\sl 
Departamento de F\'\i sica Moderna,
Universidad de Granada, 
E-18071 Granada, SPAIN 
}\\[2mm]
$^{2}${\sl 
Dipartimento di Fisica Teorica,
Universit\`a di Torino and
INFN, Sezione di Torino \\
Via P. Giuria 1, 10125 Torino, ITALY 
}\\[2mm]
$^{3}${\sl 
Departamento de F\'\i sica At\'omica, Molecular y Nuclear \\ 
Universidad de Sevilla, Apdo. 1065, E-41080 Sevilla, SPAIN 
}\\[2mm]
$^{4}${\sl 
Center for Theoretical Physics, Laboratory for Nuclear Science 
and Department of Physics\\
Massachusetts Institute of Technology,
Cambridge, MA 02139, USA 
} 
\end{center}
\end{small}

\kern 1cm \hrule \kern 3mm 

\begin{small}
\noindent
{\bf Abstract} 
\vspace{3mm} 

A consistent analysis of relativistic pionic correlations and
meson-exchange currents for electroweak quasielastic electron
scattering from nuclei is carried out.  Fully-relativistic
one-pion-exchange electromagnetic operators are developed for use in
one-particle emission electronuclear reactions within the context of
the relativistic Fermi gas model. Then the exchange and pionic
correlation currents are set up fully respecting the gauge invariance
of the theory. Emphasis is placed on the self-energy current which,
being infinite, needs to be renormalized. This is achieved starting in
the Hartree-Fock framework and then expanding the Hartree-Fock current
to first order in the square of the pion coupling constant to obtain a
truly, gauge invariant, one-pion-exchange current.  The model is
applied to the calculation of the parity-conserving (PC) and
parity-violating (PV) inclusive responses of nuclei.  Interestingly,
in the pionic correlations terms exist which arise uniquely from
relativity, although their impact on the responses is found to be
modest.

\kern 2mm 



\end{small}
\kern 2mm \hrule \kern 1cm

\vfill

\noindent MIT/CTP\#3167
\end{titlepage}

\tableofcontents

\newpage


\section{Introduction}
\label{sec:Intro}


In modern experimental studies of electron scattering from nuclei~\cite{Wil97,Vol99,Arr99} 
the typical values of energy and momentum
transfer are comparable to or even larger than the mass scale set by
the nucleon mass and accordingly one must expect relativistic effects
to be important.  Unfortunately the wave functions and operators used
to describe this high-energy regime have been treated only
approximately.  Indeed it is still the case that many calculations
continue to be done at the non-relativistic level with leading-order
expansions of the electroweak currents involved~\cite{Gil97a,Gil97b,Car02,Car98,Fab97}.  
However, a number of studies
in recent years show that such an approach is highly constrained to
work only at relatively low energies and momenta.

In order to gain insight into which ingredients can or cannot be
non-relativistically approximated we have employed a simple model in
which Lorentz covariance and gauge invariance can be maintained,
namely, the relativistic Fermi gas model (RFG). Since our focus is
placed on the quasielastic region where high-energy knockout of
nucleons is kinematically favored, we believe that this model, while
undoubtedly too simple to encompass the aspects of nuclear dynamics is
nevertheless a convenient place to start in such explorations. Indeed,
the problem of relativity in electroweak studies of nuclei is so
difficult~\cite{Kim01,Hor81,Yan92,Kur02,Amo96,Ben99,Gur02,Sim00,Par01,Isg01}
that only in special frameworks such as the RFG can we hope to carry
out all but rather severely approximated modeling.

In the quasielastic regime we also expect pions to play a role that
differs from the dynamics typically occurring near the Fermi surface
where one expects other mesons ($\sigma$ and $\omega$ in particular)
to dominate. For quasielastic scattering the residual interaction of
relevance is principally that between a low-energy hole and a very
high-energy particle, and for this the pion is expected to play an
important role. Accordingly, as the next step after the basic
relativistic Fermi gas of non-interacting nucleons we have
concentrated on one-pion exchange (OPE) effects in our description of
the nuclear responses. These occur as correlation effects and also as
two-body meson-exchange current effects. After developing approximate
methods for modeling to this order, in recent work we have reached the
stage where large classes of effects can be incorporated
fully-relativistically. The present paper is a comprehensive
discussion of what we have learned to this point based on this type of
approach, together with comments on what directions future studies
could follow.

In particular, in a recent paper~\cite{Ama02} we investigated the role
played by pions in inclusive electron scattering from nuclei within
the context of one-particle one-hole (1p-1h) excitations, {\em  i.e.},
for the dominant modes in the quasielastic regime.  There we extended
our previous work~\cite{Ama98a,Ama99a} where a systematic investigation
of relativistic effects in the nuclear electromagnetic responses
spanning a wide range of kinematical conditions and accounting for
both meson-exchange and isobar currents was carried out.  In these
studies a consistent first-order operator, embodying all Feynman
diagrams built out of nucleons and pions with one exchanged pion and
one photon attached to all the possible lines was set up to represent
the two-body current.  Importantly, the latter has been explicitly
proven to be gauge invariant in~\cite{Ama02}.

In addition to the usual contact and pion-in-flight meson-exchange
currents (MEC), this fully-relativistic operator includes as well the
so-called correlation currents.  The latter are often not included in
model calculations because they give rise to contributions assumed
already to be accounted for (at least in part) in the initial and
final nuclear wave functions~\cite{Che71,Ama98c}.  However, our model
is based on an uncorrelated relativistic Fermi gas whose states are
Slater determinants built out of (Dirac) plane waves.  Within a
perturbative approach we are free to consider the one-pion correlation
contributions to the responses as arising either explicitly in the
wave functions or from an appropriate current operator acting on
unperturbed states: our choice has been the latter.  Clearly, should
it be possible to sum up the whole perturbative expansion, then the
results obtained starting with the true ``correlated'' wave function
would be exactly recovered.

In this paper we provide a deeper analysis of the impact of pions on
the nuclear electromagnetic response in the 1p-1h
channel~\cite{Alb90,Alb81,Ama92a,Ama92b,Ama94a}.  Just as for the MEC, the
two-body correlation current also contributes in this sector and, of
course, it should do so consistently, namely fulfilling gauge
invariance at the level of {\em one}-pion-exchange.

When the operator associated with a two-body current acts on the RFG
ground state in general it changes the quantum numbers of two
nucleons: the 1p-1h matrix element is then obtained via the
integration of a one particle state over the Fermi sea.  
In the case of the correlation current two
contributions are thus obtained.  The first one, sometimes referred to
as a vertex correction~\cite{Blu89}, arises from the exchange of a
pion between the particle and hole; the second relates to the Fock
self-energy (SE)~\cite{Blu89,Hor90} and dresses the particle and hole
propagation lines.  This one diverges, since it corresponds to a SE
insertion on an external line, which field theory~\cite{Bjo65,Pes95}
tells us not to include in a perturbative expansion.  Instead one
should apply a renormalization procedure to dress the external lines
by summing up the entire perturbative series of self-energy
insertions.  In the nuclear context this procedure leads to the
relativistic Hartree-Fock (HF) approach.

In some relativistic calculations~\cite{Blu89,Hor90} this contribution
has been treated by introducing from the outset a Hartree-Fock
propagator in the medium, which accounts for the SE diagrams.
In-medium form factors for the 1p-1h current were also introduced
neglecting however any momentum dependence in the self-energy and
effective mass.  Thus in~\cite{Hor90} the self-consistent Hartree mean
field was inserted into the single-particle propagator, automatically
including the Pauli blocking of $N\overline{N}$ pairs, whose
contribution was thus included in the random-phase-approximation (RPA)
responses computed there.  A similar semi-phenomenological treatment
of the nucleon self-energy in the medium at the non-relativistic level
was implemented in~\cite{Gil97a}.  More recently, the Dirac structure
of nucleon self-energy in nuclear matter has been studied in~\cite{Sch00}, 
while a finite nuclei calculation based on the
$\sigma$-$\omega$ model can be found in~\cite{Kim01}, where the
relativistic Hartree model of~\cite{Hor81} is used for the single
particle bound states,

In~\cite{Ama02} the difficulty of the SE insertion in first order was
avoided by computing the associated self-energy response as the
imaginary part of the corresponding polarization propagator with
one-pion-exchange  SE insertions on the particle and hole lines.  A
finite result was thus obtained in first order (one pionic line)
without resorting to the HF approach.  The question then arises
whether it is possible to obtain the same result for the self-energy
response function starting with {\em finite well-defined} matrix
elements of the current operator.

In this paper we answer this question by constructing a renormalized
self-energy current corresponding to one-pion-exchange.  This current
acts over free Dirac spinors and leads to the same response functions
as those obtained by taking the imaginary part of the polarization
propagator computed to first order.  It should be clear that in this
work the concept of renormalization has a many-body significance,
namely it amounts to a relativistic HF approximation and ignores
(see~\cite{Ama02}) the additional vacuum renormalization due to the
change of the negative-energy sea induced by the nuclear
medium~\cite{Hor90}.

The new current is obtained by renormalizing spinors and energies and
by expanding the resulting in-medium one-body current to first order
in the square of the pion-nucleon coupling constant, to be consistent
with the requirement of dealing with diagrams having {\em only one}
pionic line.  The renormalized quantities should then be obtained in
the general case by solving a set of self-consistent relativistic HF
equations numerically.  However, one of the goals of this paper is to
show that to first order the solutions and the corresponding
corrections to the bare single-nucleon current operator can be
expressed analytically in terms of a simple electromagnetic operator.
This operator accounts for two main effects induced by the interaction
of the nucleon with the medium: the first is the enhancement of the
lower-components of the Dirac spinors; the second is a global
renormalization of the spinors in the nuclear medium.  These effects
are genuine relativistic corrections that are absent in a
non-relativistic framework~\cite{Alb90}.  Actually a third
renormalization effect also arises, related to the in-medium
modification of the energy-momentum relation for a nucleon, which is
here treated in first order of the square of the pion-nucleon coupling
constant (in other approaches this effect is embedded in a constant
effective mass~\cite{Blu89}).

Using the renormalized SE current operator together with the MEC and
the vertex exchange operator we prove the full gauge invariance of the
current if account is taken of the change in energy arising from the
HF renormalization to first order.  The results for the inclusive
response functions we obtain with this current agree completely with
the ones of~\cite{Ama02}, where the polarization propagator technique
was used.

The present review is organized as follows: in Section~2 we focus on
parity-conserving electron scattering from nuclei. We begin in
Section~2.1 with some general formalism and then in Section~2.2
discuss the pion exchange and correlation currents. There we revisit
the full set of 1p-1h current operators with one pion-exchange line
which contribute to the electro-excitation process, paying special
attention to the SE contribution.  We show the necessity of
re-defining the otherwise infinite self-energy diagrams. In
Section~2.3 we develop the Hartree-Fock renormalization scheme as a
vehicle to addressing this problem, going on in Section~2.4 to expand
the renormalized spinors and energies to first order in the pion
coupling constant squared obtaining a new self-energy current.  Then
in Section~2.5 we prove the gauge invariance of the theory. To
conclude this section we go on to discuss the hadronic tensor and
electromagnetic response functions (Section~2.6) and present some
typical results (Section~2.7). In Section~3 we briefly discuss
parity-violating electron scattering to place it in context with the
above studies. In Section~4 we make contact with non-relativistic
expansions schemes, both for the pion exchange currents (Section~4.1)
and for the pionic correlations (Section~4.2).  In Section~5 we
summarize our results and draw our conclusions and end with a series
of Appendices where more technical aspects of the formalism are
compiled.


\section{Parity-conserving electron scattering}
\label{sec:PC}


\subsection{General formalism}
\label{sec:PCform}

The general formalism involved in the description of (e,e$'$)
processes for quasielastic kinematics has been derived and discussed
at length in several papers (see for
instance~\cite{Don86,Don92,Bof93,Sar93,Bof96,Bar01,Ama01,Alv01}).
Here we summarize only those aspects that are of special relevance to
the analysis that follows. We limit our attention to the Plane Wave
Born Approximation (PWBA), {\em i.e.}, the electron is described as a
plane wave and interacts with the nuclear target via the exchange of a
virtual photon. The laboratory system variables involved in the
process are $K^\mu=(\varepsilon,\nk)$ and
$K'^\mu=(\varepsilon',\nk')$, the initial and scattered electron
four-momenta, and $P_i^\mu=(E_i,\np_i)=(M_i,{\bf 0})$ and
$P_f^\mu=(E_f,\np_f)$, the initial and final hadronic four-momenta,
respectively. The four-momentum transferred by the virtual photon is
$Q^\mu=(K-K')^\mu=(P_f-P_i)^\mu=(\omega ,\nq)$; for electron
scattering the momentum transfer is spacelike, $Q^2=\omega^2-q^2<0$,
with $q=|\nq|$.  The $S$-matrix element in PWBA can then be written as
\begin{equation}
S_{fi} = -2\pi i \delta(E_f-E_i-\omega)\frac{e^2}{Q^2}
\langle \nk', s'|j_{e\mu}(0)|\nk,s\rangle
\langle f|\hat{J}^{\mu}(Q)|i \rangle \, ,
\end{equation}
where 
\begin{equation}
\langle \nk', s'|j_{e\mu}(0)|\nk,s\rangle
=\left(\frac{m_e}{V\varepsilon'}\frac{m_e}{V\varepsilon}\right)^{1/2}
\overline{u}_{s'}(\nk')\gamma_{\mu}u_s(\nk)
\end{equation}
is the electron current matrix element
and $\hat{J}^{\mu}(Q)$ is the Fourier transform of the nuclear
electromagnetic current operator.

We assume Lorentz invariance, parity conservation and work in the
extreme relativistic limit (ERL), in which the electron energy
$\varepsilon \gg m_e$. Under these conditions the unpolarized,
inclusive (e,e$'$) cross section reads
\begin{equation}
\frac{d\sigma}{d\Omega'_e d\omega}=
\frac{2\alpha^2}{Q^4}\left(\frac{\varepsilon'}{\varepsilon}\right)
\eta_{\mu\nu}W^{\mu\nu}=
\sigma_M\left[v_L R^L(q,\omega) + v_T R^T(q,\omega)\right] \ .
\label{eq_1}
\end{equation}
Here $\alpha$ is the fine structure constant and $\Omega'_e$ the
scattered electron solid angle. The term $\sigma_M$ represents the
Mott cross section which in the ERL reduces to
\begin{equation}
\sigma_{M}=\left(\frac{\alpha\cos\theta_e/2}{2\varepsilon\sin^2\theta_e/2}
\right)^2 \, ,
\label{eq:Mott}
\end{equation}
where $\theta_e$ is the electron scattering angle, and $\eta_{\mu\nu}$
and $W^{\mu\nu}$ are the leptonic and hadronic tensor,
respectively. Within PWBA the leptonic tensor simply reads
\begin{equation}
\eta_{\mu\nu} = K_\mu K'_\nu + K'_\mu K_\nu +\frac{Q^2}{2}g_{\mu\nu} \, .
\end{equation}
The kinematic factors $v_L$ and $v_T$ are evaluated from the leptonic
tensor using standard techniques (see, for example,~\cite{Don86})
\begin{eqnarray} \label{eq:vl}
v_L &=& \left(\frac{Q^2}{q^2}\right)^2 \\
v_T &=& -\frac{1}{2}\left(\frac{Q^2}{q^2}\right)+\tan^2\frac{\theta_e}{2}
\label{eq:vt}
\, ,
\end{eqnarray}
whereas the longitudinal and transverse (with respect to the momentum
transfer $\nq$) response functions $R^L$ and $R^T$ are constructed
directly as components of the hadronic tensor $W^{\mu\nu}$ according
to
\begin{eqnarray}
R^L(q,\omega)&=&\left(\frac{q^2}{Q^2}\right)^2\left[
W^{00}-\frac{\omega}{q}(W^{03}+W^{30})+\frac{\omega^2}{q^2}W^{33}
\right]\label{eq2a} \\
R^T(q,\omega)&=&W^{11}+W^{22} \ ,
\label{eq2b}
\end{eqnarray}
where we use a coordinate system with the $z$-axis in the direction of
the vector $\nq$.  Note that if gauge invariance is fulfilled,
implying that $W^{03}=W^{30}=(\omega/q)W^{00}$ and $W^{33}=
(\omega/q)^2 W^{00}$, then $R^L$ is simply the time component of the
hadronic tensor, namely $W^{00}$.  Hence $R^L$ is determined by the
charge distribution, whereas $R^T$ reflects the current distribution
of the nuclear target.

The hadronic tensor and consequently the response functions derived
from it embody the entire dependence on the nuclear structure,
specifically on the charge and current distributions in nuclei, and
accordingly these provide the prime focus in analyses of electron
scattering.  There are various options on how to proceed in performing
such analyses (see, for example,~\cite{Chi89}), depending on the
specific problem under consideration and on the approximations to be
made. In what follows we recall two common expressions for the
hadronic tensor $W^{\mu\nu}$ and comment briefly on their
applications.

First, the hadronic tensor can be defined according to
\begin{equation}
W^{\mu\nu}
=\overline{\sum_i}\sum_f\langle f|\hat{J}^\mu(Q) |i\rangle^\ast
\langle f|\hat{J}^\nu(Q) |i\rangle \delta(E_i+\omega-E_f) \ ,
\label{eq3}
\end{equation}
where $\hat{J}^\mu(Q)$ represents the nuclear many-body current
operator, the nuclear states $|i\rangle$ and $|f\rangle$ are exact
eigenstates of the nuclear Hamiltonian with definite four-momenta, and
the sum with a bar means average over initial states.  This form is
very general and includes all possible final states that can be
reached through the action of the current operator $\hat{J}^\mu(Q)$ on
the exact ground state.  In our perturbative approach we shall use
eigenstates of the free Hamiltonian $H_0$ (which describes the free
relativistic Fermi gas) and include correlations among nucleons in the
current mediated by the exchange of pions.  This current of course
allows one to reach both the p-h and the 2p-2h sectors in the Hilbert
space of $H_0$. In the present work, however, we shall restrict our
attention to the former.

A different option for evaluating the nuclear responses exploits the
polarization propagator $\Pi^{\mu\nu}$ (also referred to as the
current-current correlation function). The latter can be expressed in
terms of the full propagator, $\hat{G}$, of the nuclear many-body
system, since closure can be used to carry out the sum over the final
states in eq.~(\ref{eq3}). Then one has for the hadronic
tensor~\cite{Fet71}
\begin{equation}
W^{\mu\nu}=
-\frac{1}{\pi} {\rm Im}\, \Pi^{\mu\nu}(q,q;\omega)
=-\frac{1}{\pi} {\rm Im}\,
\overline{\sum_i}\langle i|\hat{J}^{\dagger\mu}(Q)
\hat{G}(\omega +E_i)\hat{J}^\nu(Q) |i\rangle \ .
\label{eq4}
\end{equation}
A possible advantage of this approach relates to the existence of a
well-defined set of rules (the relativistic Feynman diagrams) which
allows one to compute $\Pi^{\mu\nu}$ perturbatively~\cite{Fet71}.

Obviously the two procedures are equivalent and hence the observables
calculated using the expressions for the hadronic tensor given by
eqs.~(\ref{eq3}) or (\ref{eq4}) should be the same. However, notice
that eq.~(\ref{eq3}) is less suitable for dealing with situations
where the nuclear current matrix element $\langle f|\hat{J}^\mu
|i\rangle$ is divergent. In this case one proceeds either by computing
directly the responses via the polarization propagator or by first
renormalizing the current matrix element and then by using
eq.~(\ref{eq3}).

Finally, we remark that gauge invariance must be fulfilled both at the
level of the nuclear current matrix elements and at the level of the
hadronic tensor and/or the polarization propagator. A consequence is
that the electromagnetic continuity equation should be satisfied.  In
other words in momentum space all of the expressions $Q_\mu\langle
f|\hat{J}^\mu(Q) |i\rangle$, $Q_\mu W^{\mu\nu}$ and $Q_\mu\Pi^{\mu\nu}$
should vanish.

\subsection{Pion exchange and correlation currents}
\label{sec:PCpion}

Working within the framework of the relativistic Fermi gas (RFG)
model, {\em i.e.,} for nucleons moving freely inside the nucleus with
relativistic kinematics, in this section we present a detailed study
of the electromagnetic currents accounting for the effects introduced
by pions in first-order perturbation theory (one-pion exchange).

\subsubsection{Feynman diagrams and two-body currents}
\label{sec:PCFey}

The linked, two-body Feynman diagrams that contribute to electron
scattering with one pion-exchange are shown in Fig.~1. The first three
correspond to the usual meson-exchange currents (MEC): diagrams (a),
(b) refer to the contact or seagull current, diagram (c) to the
pion-in-flight current. The four diagrams (d)--(g) represent the
so-called correlation current and are usually not treated as genuine
MEC, but as correlation corrections to the nuclear wave function.
However, again we note that our approach puts all correlation effects
in the current operator and uses an uncorrelated wave function for the
initial and final nuclear states.

\begin{figure}[tb]
\begin{center}
\leavevmode
\def\epsfsize#1#2{0.9#1}
\epsfbox[200 450 400 720]{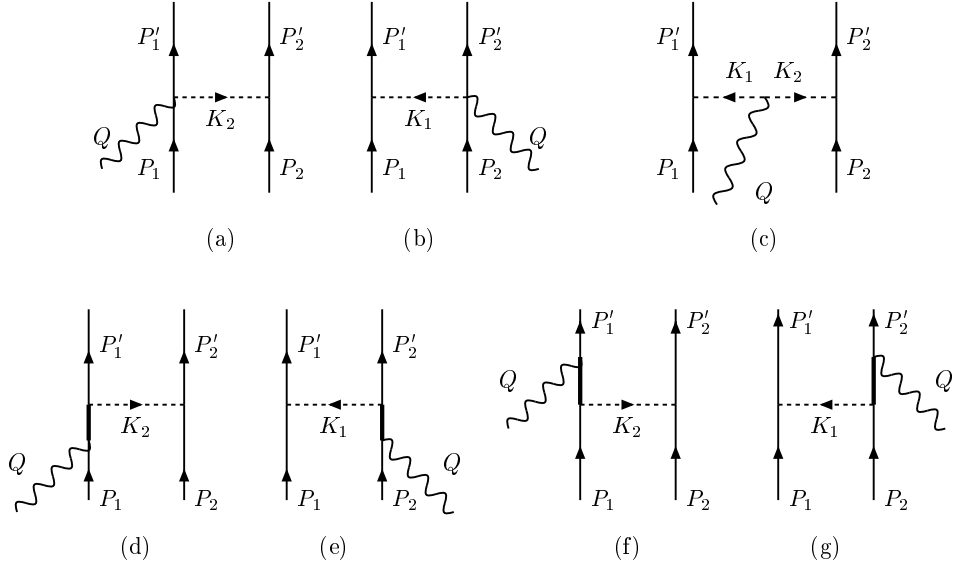}
\end{center}
\caption{Feynman diagrams contributing to the two-body current 
with one pion-exchange. The wide line in the correlation diagrams
(d)--(g) means a fully-relativistic Dirac propagator for the nucleon.}
\end{figure}

In this work we use Bjorken and Drell conventions~\cite{Bjo65} and 
pseudovector $\pi N N$ coupling (the effect of a pseudoscalar
coupling will be commented on later), namely
\begin{equation} 
{\cal H}_{\pi NN} = 
\frac{f}{m_\pi}\overline{\psi}\gamma_5\gamma^{\mu}
(\partial_{\mu}\phi_a)\tau_a \psi \ ,
\label{HamPV}
\end{equation}
where $\psi$ is the nucleon field, $\phi_a$ is the isovector pion
field, $f$ represents the $\pi NN$ coupling constant and $m_\pi$ is 
the pion mass. 
The electromagnetic currents
corresponding to diagrams (a)--(g) are obtained by computing the
$S$-matrix element 
\begin{equation}
S_{fi} = S_{fi}(P'_1,P'_2,P_1,P_2)-S_{fi}(P'_1,P'_2,P_2,P_1)
\end{equation}
for the absorption of a virtual photon by a system of two nucleons,
namely for the process \be \gamma + N_1+ N_2 \rightarrow N_1' + N_2'\
, \ee with $P_1$, $P_2$ ($P'_1$, $P'_2$) being the initial (final)
four-momenta of the two nucleons involved (see Fig.~1). The
electromagnetic current is then defined according to
\begin{equation}\label{S-matrix}
S_{fi}(P'_1,P'_2,P_1,P_2)=
-ie A_{\mu}(Q) 2\pi \delta(E'_1+E'_2-E_1-E_2-\omega)
\langle P'_1 P'_2 | \hat{j}^{\mu}(Q)| P_1 P_2 \rangle ,
\end{equation}
where $A_{\mu}(Q)$ is related to the matrix element of the
electromagnetic field between the incident photon with momentum $Q$
and the vacuum state, namely
\begin{equation}
\langle 0| A_{\mu}(X)|\gamma(Q)\rangle=
A_{\mu}(Q) e^{-iQ\cdot X}\ .
\end{equation}
Finally, the on-shell matrix element of the two-body current
can be  written in terms of a function  $j^{\mu}(\np'_1,
\np'_2,\np_1,\np_2)$
as follows
\begin{eqnarray}
\lefteqn{\langle P'_1 P'_2 | \hat{j}^{\mu}(Q)| P_1 P_2 \rangle =}
\nonumber\\
&=& 
(2\pi)^3 \delta^3(\np'_1+\np'_2-\nq-\np_1-\np_2)
\frac{m^2}{V^2(E_{\np_1} E_{\np_2}
E_{\np'_1} E_{\np'_2})^{1/2}}
j^{\mu}(\np'_1, \np'_2,\np_1,\np_2), 
\end{eqnarray}
where $m$ is the nucleon mass, $V$ is the volume enclosing the system
and $E_{\np}=\sqrt{\np^2+m^2}$ the on-shell energy of a nucleon with
momentum $\np$.  The four-momenta --- indicated by capital letters ---
are defined in Fig.~1.

The general relativistic expressions for the seagull (diagrams a,b),
pion-in-flight (c) and correlation (d-g) current matrix elements
are (isospin summations are understood)
\begin{itemize}
\item
Seagull or contact:
\end{itemize}
\begin{equation}
j^{\mu}_{s}(\np'_1, \np'_2,\np_1,\np_2)
= \frac{f^2}{m_\pi^2}
             i\epsilon_{3ab}
             \overline{u}(\np'_1)\tau_a\gamma_5\Kbar_1 u(\np_1)
             \frac{F_1^V}{K_1^2-m_\pi^2}
             \overline{u}(\np'_2)\tau_b\gamma_5\gamma^{\mu}u(\np_2)
             + (1 \leftrightarrow 2) \label{eq6}
\end{equation}
\begin{itemize}
\item
Pion-in-flight:
\end{itemize}
\begin{eqnarray}
j^{\mu}_{p}(\np'_1, \np'_2,\np_1,\np_2)
&=&
\frac{f^2}{m_\pi^2}
             i\epsilon_{3ab}
             \frac{F_\pi(K_1-K_2)^\mu}{(K_1^2-m_\pi^2)(K_2^2-m_\pi^2)}
     \overline{u}(\np'_1)\tau_a\gamma_5\Kbar_1 u(\np_1)
             \overline{u}(\np'_2)\tau_b\gamma_5\Kbar_2 u(\np_2) \nonumber \\
& & \label{eq7}
\end{eqnarray}
\begin{itemize}
\item
Correlation:
\end{itemize}
\begin{eqnarray}
j^{\mu}_{cor}(\np'_1, \np'_2,\np_1,\np_2)
&=&           \frac{f^2}{m_\pi^2}
              \overline{u}(\np'_1)\tau_a\gamma_5\Kbar_1 u(\np_1)
              \frac{1}{K_1^2-m_\pi^2} \nonumber\\
& & \kern -2cm 
    \mbox{}\times \overline{u}(\np'_2)
    \left[    \tau_a\gamma_5\Kbar_1 
              S_F(P_2+Q)\Gamma^\mu(Q)
            + \Gamma^\mu(Q)S_F(P'_2-Q)
              \tau_a\gamma_5\Kbar_1
    \right]u(\np_2) \nonumber\\
& & \kern -2cm \mbox{}+ (1\leftrightarrow2) \ . \label{eq8}
\end{eqnarray}
In the above, 
$K_1$, $K_2$ are the four-momenta given to the nucleons 1,
2 by the exchanged pion, and they  
are defined in Fig.~1, while $F_1^V$ and $F_\pi$ are the electromagnetic
isovector nucleon and pion form factors, respectively.  Furthermore,
$S_F(P)$ is the nucleon propagator and $\Gamma^\mu(Q)$ the
electromagnetic nucleon vertex, {\em  i.e.,}
\begin{eqnarray}
S_F(P) &=& \frac{\Pbar + m}{P^2-m^2}
\label{NucleonVertex} \\
\Gamma^\mu(Q) &=&
F_1\gamma^\mu+\frac{i}{2m}F_2\sigma^{\mu\nu}Q_\nu \ ,\label{eq10}
\end{eqnarray}
$F_1$ and $F_2$ being the Dirac and Pauli form factors: for these we
use the Galster parameterization~\cite{Gal71}.  Finally, the spinors
(for brevity we denote $u(\np,s_p)$ by $u(\np)$) are normalized
according to the Bjorken and Drell convention~\cite{Bjo65} and the
isospinors are not explicitly indicated.

The seagull and pion-in-flight currents shown above coincide with the
expressions given by Van Orden and Donnelly~\cite{Van81} if account
is taken for the different conventions used for the gamma matrix
$\gamma_5$ and for the metric.  Concerning the correlation current
note that, at variance with~\cite{Van81}, it embodies both the
positive and negative energy components of the nucleon propagator.

A crucial point to be stressed is that the sum of the relativistic
seagull, pion-in-flight and correlation currents satisfy current
conservation, {\em  i.e.} $Q_\mu J^\mu=0$, provided some assumptions
are made for the form factors involved in the various currents. This
is proven in Appendix A (see also~\cite{Ama02}) where we show that
when the seagull and pion-in-flight currents are multiplied by the
same electromagnetic form factor $F_1^V$, gauge invariance is
fulfilled, {\em  i.e.}
\begin{equation} 
Q_{\mu}(j^{\mu}_{s}+j^{\mu}_{p}+j^{\mu}_{cor})=0\ ,
\label{eq23}
\end{equation}
where the two body currents in eq.~(\ref{eq23}) are defined in
eqs.~(\ref{eq6}--\ref{eq8}).

It is also possible~\cite{Gro87} to use different phenomenological
electromagnetic form factors for the nucleon and pion --- even
introducing phenomenological form factors at the strong pion-nucleon
vertices --- without violating current conservation, 
by appropriate modification in the currents through the
generalized Ward-Takahashi identity~\cite{War50,Tak57}.

\subsubsection{Particle-hole matrix elements}
\label{sec:PCph}
In this report we deal with the case of one-particle emission induced
by the two-body currents introduced above. The matrix element of a
two-body operator between the Fermi gas ground state and a 1p-1h
excited state reads
\begin{eqnarray}
\langle ph^{-1}|\hat{j}^{\mu}(Q)|F\rangle 
&\equiv&
(2\pi)^3 \delta^3(\nq+\nh-\np)\frac{m}{V\sqrt{E_\np E_\nh}}j^{\mu}(\np,\nh)
\nonumber\\
&=&
\sum_{k<F} 
\left[ \langle pk |\hat{j}^{\mu}(Q)|hk\rangle
      -\langle pk |\hat{j}^{\mu}(Q)|kh\rangle
\right] \ ,
\label{jmuph}
\end{eqnarray}
where the summation runs over all occupied levels in the ground state,
and thus includes a sum over spin ($s_k$) and isospin ($t_k$) and an
integral over the momentum $\nk$.

The first and second terms in eq.~(\ref{jmuph}) represent the direct
and exchange contribution to the matrix element, respectively.  It can
be easily verified (see, {\em  e.g.},~\cite{Ama02,Ama98a}) that in
spin-isospin saturated systems the direct term vanishes for the MEC
and pionic correlation currents upon summation over the occupied
states.  Hence only the exchange term contributes to the p-h matrix
elements. The associated many-body Feynman diagrams are displayed in
Fig.~2. Diagrams (a,b) and (c) correspond to the seagull (or contact)
and pion-in-flight contributions, respectively.  Diagrams (d--g)
represent instead the correlation contributions. Here we distinguish
the exchange of a pion between a particle and a hole line (d,e),
giving rise to the so-called vertex correlation (VC), and the
self-energy insertions on the nucleonic lines (f,g).  After carrying
out explicitly the sums over the internal spin, $s_k$, and isospin,
$t_k$, the fully-relativistic expressions for the MEC (seagull and
pion-in-flight) and correlation (vertex correlations and self-energy)
currents turn out to be

\begin{itemize}
\item Seagull
\end{itemize}
\be
j^{\mu}_{s}(\np,\nh) = -\frac{f^2}{V m_\pi^2} F_1^{V}
i\varepsilon_{3ab}
\sum_{\nk\leq k_F}\frac{m}{E_{\nk}}
              \overline{u}(\np)\tau_a\tau_b
\left\{ \frac{(\Kbar-m)\gamma^\mu}{(P-K)^2-m_\pi^2}
      + \frac{\gamma^\mu (\Kbar-m)}{(K-H)^2-m_\pi^2} 
\right\}u(\nh)
\label{Sph}
\ee

\begin{itemize}
\item
Pion-in-flight
\end{itemize}
\begin{eqnarray}
\lefteqn{j^{\mu}_{p}(\np,\nh)=}
\nonumber\\
&=&
2m \frac{f^2}{V m_\pi^2} F_1^{V}
i\varepsilon_{3ab} 
\sum_{\nk\leq k_F}
        \frac{m}{E_{\nk}}\frac{(Q+2H-2K)^\mu}{[(P-K)^2-m_\pi^2]
         [(K-H)^2-m_\pi^2]}
\overline{u}(\np)\tau_a\tau_b (\Kbar-m)u(\nh)
\nonumber\\
\label{Pph}
\end{eqnarray}

\begin{itemize}
\item
Vertex correlations
\end{itemize}
\begin{eqnarray}
\lefteqn{j^{\mu}_{VC}(\np,\nh)=}
\nonumber\\
&=& 
\frac{f^2}{Vm_\pi^2} 
\sum_{\nk\leq k_F}\frac{1}{2 E_\nk}
\overline{u}(\np)
\left\{ \frac{\Kbar-\Hslash}{(K-H)^2-m_\pi^2}
\gamma_5 S_F(K+Q) \tau_a \Gamma^{\mu}(Q) \tau_a \gamma_5 
(\Kbar-m)(\Kbar-\Hslash)
\right.
\nonumber\\
&+& 
\left. (\Pbar-\Kbar)(\Kbar-m)
\gamma_5 \tau_a \Gamma^{\mu}(Q) \tau_a S_F(K-Q) \gamma_5
\frac{\Pbar-\Kbar}{(P-K)^2-m_\pi^2} \right\} u(\nh)
\nonumber\\
&\equiv& {\cal F}^{\mu}+{\cal B}^{\mu}
\label{VC}
\end{eqnarray}

\begin{itemize}
\item Self-energy
\end{itemize}
\begin{eqnarray}
j_{SE}^{\mu}(\np,\nh)
&=& 
-\frac{3f^2}{Vm_\pi^2}\sum_{\nk\leq k_F}\frac{1}{2 E_\nk}
\overline{u}(\np)\left\{
\frac{\Pbar-\Kbar}{(P-K)^2-m_\pi^2} 
(\Kbar-m)(\Pbar-\Kbar) S_F(P)\Gamma^{\mu}(Q)\right.
\nonumber\\
&+& \left.
\Gamma^{\mu}(Q)S_F(H) (\Kbar-\Hslash)(\Kbar-m)
\frac{\Kbar-\Hslash}{(K-H)^2-m_\pi^2}
\right\} u(\nh)
\nonumber \\
&\equiv& {\cal H}_p^{\mu}+{\cal H}_h^{\mu}
\label{self-energy-ph}
\end{eqnarray}
The effects of the medium are included through the summation in
eqs.~(\ref{Sph},\ref{self-energy-ph}) over the intermediate momentum
$\nk$ up to the Fermi momentum.

In the thermodynamic limit the sum $\frac{1}{V}\sum_{\nk \leq k_F}$
becomes an integral over the momentum $\int \frac{d^3k}{(2\pi)^3}$ in the
range $0\leq k\leq k_F$, $k_F$ being the Fermi momentum, and over the
angular variables $\theta_k,\phi_k$. Note that, although the global
factor $\displaystyle\frac{m}{V\sqrt{E_\np E_\nh}}$ has been extracted
from the current in eq.~(\ref{jmuph}), the factor $\displaystyle
\frac{m}{V E_\nk}$, associated with the internal line, has to be
retained inside the sum.  Note also that, in order to fulfill gauge
invariance, we have assumed $F_\pi=F_1^V$.

\begin{figure}[tb]
\begin{center}
\leavevmode
\def\epsfsize#1#2{0.9#1}
\epsfbox[200 450 400 720]{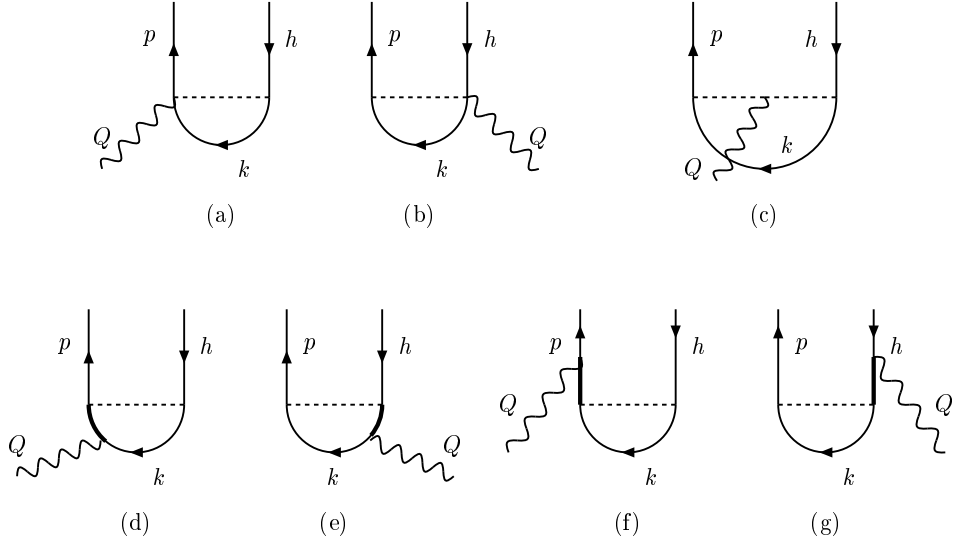}
\end{center}
\caption{Many-body Feynman diagrams contributing to the one-body
current with one pion-exchange. The thick line in the correlation
diagrams (d)--(g)
corresponds to a fully-relativistic Dirac propagator for the
nucleon.  Diagrams (d)-(e) represent the vertex current, while
diagrams (f) and (g) represent the self-energy current of the
hole and of the particle, respectively.
}
\end{figure}

The vertex p-h matrix element splits into two terms ${\cal F}^\mu$ 
and ${\cal B}^\mu$ representing the forward- and
backward-going contributions, respectively (Fig.~2d and 2e). They are 
\be
{\cal F}^\mu
=           -\frac{f^2}{V m_\pi^2}
             \sum_{\nk\leq k_F}\frac{m}{E_{\nk}} \overline{u}(\np)
              \gamma_5(\Kbar-\Hslash)
              S_F(K+Q)\tau_a\Gamma^\mu(Q)\tau_a \gamma_5 
              \frac{(\Kbar - m)}{(K-H)^2-m_\pi^2} u(\nh)
\label{F}
\ee
\be
{\cal B}^\mu =
          -\frac{f^2}{V m_\pi^2}
           \sum_{\nk\leq k_F}\frac{m}{E_{\nk}} \overline{u}(\np)
              \frac{\Kbar-m}{(P-K)^2-m_\pi^2} \gamma_5 
             \tau_a\Gamma^\mu(Q)\tau_a S_F(K-Q)
              \gamma_5
              (\Pbar-\Kbar) u(\nh)\ .
\label{B}
\ee
 
Similarly, the self-energy p-h matrix element splits into two terms,
${\cal H}^\mu_p$ and ${\cal H}^\mu_h$. The former corresponds to the
diagram with the pion inserted in the particle line (Fig.~2g), whereas
the latter describes the diagram with the pion inserted in the hole
line (Fig.~2f). They are given by \be {\cal H}^\mu_p = -\frac{3 f^2}{V
m_\pi^2} \sum_{\nk\leq k_F}\frac{m}{E_{\nk}} \overline{u}(\np) (\Kbar-
m)\frac{\Pbar-\Kbar}{(P-K)^2-m_\pi^2} S_F(P)\Gamma^\mu(Q) u(\nh)
\label{Sp}
\ee
\be
{\cal H}^\mu_h = \frac{3 f^2}{V m_\pi^2}
\sum_{\nk\leq k_F}\frac{m}{E_{\nk}}
\overline{u}(\np)
\Gamma^\mu(Q)S_F(H)\frac{(\Kbar-\Hslash)}{(K-H)^2-m_\pi^2}(\Kbar-m)u(\nh)\ .
\label{Sh}
\ee

Finally, splitting also the electromagnetic nucleon operator
$\Gamma^\mu$ into its isoscalar and isovector parts, one obtains the
isoscalar and isovector contributions to the self-energy and vertex p-h
matrix elements.  The final results can be cast in the form \ba {\cal
F}^{\mu(S)} & =& -\frac{3 f^2}{V m_\pi^2} \sum_{\nk\leq
k_F}\frac{m}{E_{\nk}} \overline{u}(\np) \gamma_5(\Kbar-\Hslash)
S_F(K+Q)\Gamma^{\mu(S)}(Q) \gamma_5 \frac{(\Kbar -
m)}{(K-H)^2-m_\pi^2} u(\nh) \\ {\cal B}^{\mu(S)} &=& -\frac{3 f^2}{V
m_\pi^2} \sum_{\nk\leq k_F}\frac{m}{E_{\nk}} \overline{u}(\np)
\frac{\Kbar-m}{(P-K)^2-m_\pi^2} \gamma_5 \Gamma^{\mu(S)}(Q)S_F(K-Q)
\gamma_5 (\Pbar-\Kbar)u(\nh) \ , \ea for the isoscalar and
\begin{eqnarray}
{\cal F}^{\mu(V)}
&=&            -\frac{f^2}{Vm_\pi^2}
\sum_{\nk\leq k_F}\frac{m}{E_{\nk}}
\overline{u}(\np)
\gamma_5(\Kbar-\Hslash)
\nonumber\\
&\times &              S_F(K+Q)\Gamma^{\mu(V)}(Q)(\tau_3+i\varepsilon_{3ab}
              \tau_a\tau_b)
              \gamma_5\frac{(\Kbar - m)}{(K-H)^2-m_\pi^2} u(\nh)
\label{FV} \\
{\cal B}^{\mu(V)}
&=&            -\frac{f^2}{Vm_\pi^2}
\sum_{\nk\leq k_F}
\frac{m}{E_{\nk}}
\overline{u}(\np)
              \frac{\Kbar-m}{(P-K)^2-m_\pi^2} 
            \gamma_5 \Gamma^{\mu(V)}(Q)(\tau_3+i\varepsilon_{3ab}
            \tau_a\tau_b)
\nonumber\\
&\times & S_F(K-Q) \gamma_5 (\Pbar-\Kbar)u(\nh) \ , 
\label{BV}
\end{eqnarray}
for the isovector vertex p-h matrix elements, and
\begin{eqnarray}
\lefteqn{{\cal H}^{\mu(S,V)}_p =}
\nonumber\\
&=&            -\frac{3f^2}{2 V m_\pi^2}
             \sum_{\nk\leq k_F}\frac{m}{E_{\nk}}
\overline{u}(\np)
\frac{\Pbar-\Kbar}{(P-K)^2-m_\pi^2} (\Kbar - m) (\Pbar-\Kbar)
              S_F(P)\Gamma^{\mu(S,V)}(Q) u(\nh)\nonumber \\
&& \label{se1} 
\end{eqnarray}
and 
\begin{eqnarray} 
\lefteqn{{\cal H}^{\mu(S,V)}_h=}
\nonumber\\
 &=& -\frac{3f^2}{2 V m_\pi^2}
\sum_{\nk\leq k_F}\frac{m}{E_{\nk}} \overline{u}(\np)
\Gamma^{\mu(S,V)}(Q)S_F(H)(\Kbar-\Hslash)(\Kbar-m)
\frac{(\Kbar-\Hslash)}{(K-H)^2-m_\pi^2} u(\nh) \ , \nonumber \\
& & \label{se2} 
\ea 
for the self-energy matrix elements. 
Interestingly,
the isoscalar/isovector ratio is $=-3$ in the vertex matrix element,
whereas in the self-energy case it is the 
unity%
\footnote{ The latter result stems from the relation
$\tau_3+i\varepsilon_{3ab}\tau_a\tau_b = -\tau_3$; however we prefer
to leave the isospin structure of the isovector exchange as in
eqs.~(\ref{FV},\ref{BV}), since it makes more transparent the self-energy
and exchange cancellation in the continuity equation, as shown in
Appendix B.}.  
Note that the MEC (pion-in-flight and seagull) p-h
matrix elements are purely isovector, whereas the vertex and
self-energy correlations get both isoscalar and isovector
contributions.

The VC and SE p-h matrix elements involve the nucleon
propagator $S_F(P)$ which in some situations may imply the occurrence
of singularities. In the case of the vertex diagrams, the four-momenta
appearing in the propagators are $K+Q$ and $K-Q$ for the forward-
(Fig.~2d) and backward-going (Fig.~2e) contributions, respectively,
and an integration over $\nk$ should be done. For $q\geq 2 k_F$ (no
Pauli blocking) it can be proven (see~\cite{Ama02}) that only the
forward diagram contains a pole, {\em  i.e.,} a value of the inner
momentum $\nk$ exists such that the nucleon carrying a four-momentum
$K+Q$ is on-shell.  In this situation the forward vertex p-h matrix
element is evaluated by taking the principal value in the integral
over $\cos\theta_k$. In the case of the backward-going diagram the
nucleon propagator $S_F(K-Q)$ has no singularity for the kinematics in
which we are interested.

The case of the self-energy diagrams is clearly different. Here
the particle ($p$) and hole ($h$) are described in the Fermi gas by
unperturbed plane waves, {\em  i.e.,} they are on-shell, and hence the
propagators $S_F(P)$ and $S_F(H)$ diverge. 
The divergence of the diagrams (f)--(g) is reminiscent of the
well-known infinity occurring in standard perturbative quantum field
theory, when self-energy insertions in the external legs are included
in Feynman diagrams~\cite{Pes95}.  As is well-known, there one should
renormalize the theory by dressing the external legs, propagators and
vertices. In the nuclear matter case we assume that the
particle-physics effects are already accounted for by the physical
masses and electromagnetic form factors. However, an additional
nuclear physics renormalization, arising from the interaction of a
nucleon with the nuclear medium, should be included at the
one-pion-exchange level to account for the self-energy diagram.

The self-energy current in eq.~(\ref{self-energy-ph}) can be
written in the following form:
\begin{equation}\label{SEC}
j^{\mu}_{SE}(\np,\nh) =
\overline{u}(\np)\Sigma(P)S_F(P)\Gamma^{\mu}(Q)u(\nh)
+\overline{u}(\np)\Gamma^{\mu}(Q)S_F(H)\Sigma(H)u(\nh) \ ,
\end{equation}
where $\Sigma(P)$ is the nucleon self-energy matrix that in first order reads
\begin{equation}\label{SE1}
\Sigma(P) = 
-\frac{f^2}{Vm_\pi^2}\sum_{\nk\leq k_F}\sum_{s_k,t_k} \frac{m}{E_\nk}
\tau_a\gamma_5(\Pbar-\Kbar)
\frac{u(\nk)\overline{u}(\nk)}{(P-K)^2-m_\pi^2}
\tau_a\gamma_5(\Pbar-\Kbar)\ .
\end{equation}
This is diagrammatically displayed in Fig.~3. The SE
matrix, shown in Fig.~3(c), corresponds
to the Fock term of the mean-field potential (the Hartree or direct term is
zero for pion exchange, since it involves a pion carrying zero momentum).

\begin{figure}[tb]
\begin{center}
\leavevmode
\def\epsfsize#1#2{0.9#1}
\epsfbox[200 450 400 720]{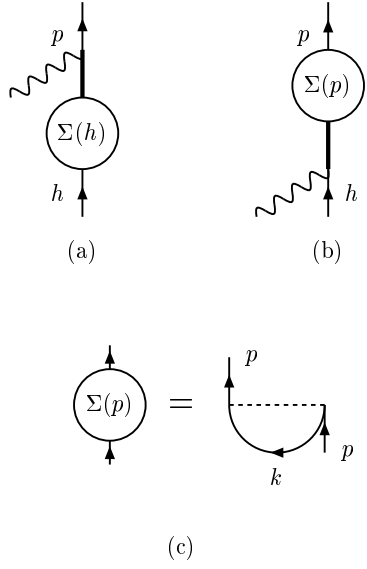}
\end{center}
\caption{ Diagrammatic representation of the self-energy current for
a hole (a) and a particle (b). The self-energy is defined to first
order as the Fock insertion shown in (c) with one pion-exchange.  }
\end{figure}

Performing the sum over the internal spin ($s_k$) and isospin ($t_k$) indices
and using the commutation
properties of the gamma matrices to eliminate $\gamma_5$, the
self-energy can be cast in the form
\begin{equation}\label{SE2}
\Sigma(P) = -\frac{3f^2}{m_\pi^2}
\int\frac{d^3 k}{(2\pi)^3}\theta(k_F-k)
\frac{1}{2E_\nk}
\frac{(\Pbar-\Kbar)(\Kbar-m)(\Pbar-\Kbar)}{(P-K)^2-m_\pi^2}\ ,
\end{equation}
where the sum over $\nk$ has been converted into an integral.  Note
that the self-energies $\Sigma(P)$ and $\Sigma(H)$ appearing in
eq.~(\ref{SEC}) are evaluated for free particles and holes, {\it
i.e.}, for $P^{\mu}$ and $H^{\mu}$ on-shell.  Hence the self-energy
contributions to the current are divergent, since so are the free
propagators $S_F(P)$ and $S_F(H)$ in eq.~(\ref{SEC}). Therefore they
should not be computed using eq.~(\ref{SEC}), but rather one should
first renormalize the wave function and the propagator of the
particles in the medium. This is achieved through the summation of the
full series of diagrams with repeated self-energy insertions displayed
in Fig.~4.

Now the energy of a particle in nuclear matter is modified by the
interaction with the medium and, as well, through its energy-momentum
relation. Thus the associated momentum is no longer on-shell and
therefore in the next section we shall evaluate the self-energy for
{\em off-shell} particles. In the first iteration, corresponding to
one pion-exchange, the particle $P^{\mu}$ is off-shell, but the
intermediate interacting hole $K^{\mu}$ is still
on-shell%
\footnote{Note that in deriving eq.~(\ref{SE2}) we have
assumed free spinors $u(\nk)$; hence eq.~(\ref{SE2}) is only valid for
$K^{\mu}$ on-shell. The off-shell case requires a redefinition of the
spinors $u(\nk)$ according to an interacting Dirac equation, as is
shown later.}.
 In this case, with the help of Dirac spinology, one
writes
\begin{equation}
(\Pbar-\Kbar)(\Kbar-m)(\Pbar-\Kbar)
= 
2(P\cdot K-m^2)(\Pbar+m) 
-(P^2-m^2)(\Kbar+m)\ ,
\end{equation}
which allows one to recast the self-energy in eq.~(\ref{SE2}) for the
off-shell momentum $P$ in the form
\begin{equation}\label{SE3}
\Sigma(P) = -\frac{3f^2}{m_\pi^2}
\int\frac{d^3 k}{(2\pi)^3}\theta(k_F-k)
\frac{1}{2E_\nk}
\frac{2(P\cdot K-m^2)(\Pbar+m)-(P^2-m^2)(\Kbar+m)}{(P-K)^2-m_\pi^2} \ .
\end{equation}
Note that the second term inside the integral vanishes for $P$ on-shell.

In general the self-energy of a nucleon in nuclear matter can be written in
the form~\cite{Cel86}:
\begin{equation}\label{spin}
\Sigma(P) = mA(P)+B(P)\gamma_0 p^0 -C(P)\ngamma\cdot\np \ .
\end{equation}
In contrast to the quantum field-theory decomposition $\Sigma(P)=
mA+B\Pbar$, owing to the non-invariance under a boost of the step
function $\theta(k_F-k)$ appearing in the self-energy, in nuclear
matter $B(P)\ne C(P)$. This in turn reflects the existence of a
privileged system, namely the lab system where the Fermi gas has total
momentum ${\bf p}_{FG}=0$.  Here it is natural to compute the
self-energy.  Under a boost, the Fermi gas ground state is no longer
characterized by $k<k_F$ and also the self-energy takes a different
form.

In the case of the Fock self-energy in eq.~(\ref{SE3}) 
the functions $A,B,C$ can be expressed in terms of the integrals
(for $K^{\mu}$ on-shell)
\begin{eqnarray}
I(P) &\equiv&
\int\frac{d^3 k}{(2\pi)^3}\theta(k_F-k)
\frac{1}{2E_\nk}
\frac{1}{(P-K)^2-m_\pi^2}
\label{I}
\\
L^{\mu}(P) &\equiv&
\int\frac{d^3 k}{(2\pi)^3}\theta(k_F-k)
\frac{1}{2E_\nk}
\frac{K^{\mu}}{(P-K)^2-m_\pi^2} \, .
\label{Lmu}
\end{eqnarray}
Indeed one gets%
\footnote{
${\bf L}$ is parallel to ${\bf p}$ since, choosing ${\bf p}$ along the $z$
axis, the azimuthal integration in eq.~(\ref{Lmu}) yields $L_x=L_y=0$.}
\begin{eqnarray}
A(P) &=&
-3\frac{f^2}{m_\pi^2}
\left\{ 
   2\left[P_\mu L^\mu(P) -m^2 I(P)\right]- (P^2-m^2)I(P)
\right\}
\label{A-definition}\\ 
B(P) &=&
-3\frac{f^2}{m_\pi^2}
\left\{
   2\left[P_\mu L^\mu(P) -m^2 I(P)\right]- (P^2-m^2)\frac{L_0(P)}{p_0}
\right\}
\label{B-definition}\\ 
C(P) &=&
-3\frac{f^2}{m_\pi^2}
\left\{
   2\left[P_\mu L^\mu(P) -m^2 I(P)\right]- (P^2-m^2)\frac{L_3(P)}{p}
\right\} \ .
\label{C-definition}
\end{eqnarray}
Note that $A=B=C$ for $P$ on-shell. In this case one simply has
$\Sigma(P)_{on-shell}= A(P)(m+\Pbar)$.

\subsection{Hartree-Fock renormalization in nuclear matter}
\label{sec:PCHF}
In this section we discuss the renormalization of the nucleon
propagator and spinors associated with the pionic self-energy
in a Hartree-Fock scheme.

\subsubsection{Nucleon propagator}
\label{sec:PCprop}
The Hartree-Fock (HF) nucleon propagator in the nuclear medium is the
solution of Dyson's equation
\begin{equation}
S_{HF}(P) = S_0(P)+S_0(P) \Sigma(P) S_{HF}(P) \ ,
\label{Dyson}
\end{equation}
where $\Sigma(P)$ is the HF proper self-energy and
\be
S_0(P) =
\frac{\theta(p-k_F)}{\Pbar+m+i\epsilon}+
\frac{\theta(k_F-p)}{\Pbar+m-i\epsilon p_0}
\label{S0}
\ee
is the free propagator in the medium.
Equation~(\ref{Dyson}) results from summing 
up a series with an infinite number 
of self-energy insertions (see Fig.~4) for each of the two terms in
(\ref{S0})%
\footnote{No interference term arises, since $\theta(k-k_F)\theta(k_F-k)=0.$}
, namely 
\begin{figure}[tb]
\begin{center}
\leavevmode
\def\epsfsize#1#2{0.9#1}
\epsfbox[170 600 470 730]{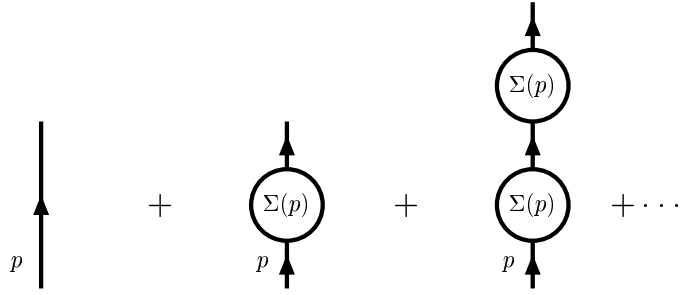}
\end{center}
\caption{
Diagrammatic series for the nucleon propagator in the medium.
}
\end{figure}
\begin{equation}
\frac{1}{\Pbar-m}+
\frac{1}{\Pbar-m}\Sigma(P)\frac{1}{\Pbar-m}+
\frac{1}{\Pbar-m}\Sigma(P)\frac{1}{\Pbar-m}\Sigma(P)\frac{1}{\Pbar-m}+
\cdots
=
\frac{1}{\Pbar-m-\Sigma(P)}\ .
\label{propHF}
\end{equation}
Using the spin decomposition of the self-energy in eq.~(\ref{spin}), we
can write
\begin{equation}
\Pbar-m-\Sigma(P) 
= \left[1-B(P)\right]\gamma_0 p_0-
\left[1-C(P)\right]\ngamma\cdot\np -
\left[1+A(P)\right] m\ .
\end{equation}
Now the new four-momentum
$f^{\mu}=f^{\mu}(P)$, which is related to $P^{\mu}$ as follows
\begin{eqnarray}
f_0(P) &=& \frac{1-B(P)}{1-C(P)}\ p_0 \label{f_0}\\
{\bf f}(P) &=& \np \ ,
\end{eqnarray}
and the functions
\begin{eqnarray}
\widetilde{m}(P) &=& \frac{1+A(P)}{1-C(P)}\ m  \label{m-tilde}\\
z(P) &=& \frac{1}{1-C(P)} \, ,
\label{zp}
\end{eqnarray}
allow one to recast eq.~(\ref{propHF}) in the form 
\begin{equation} \label{propagator}
\frac{1}{\Pbar-m-\Sigma(P)}
= \frac{z(P)}{\gamma_0 f_0(P)-\ngamma\cdot\np -\widetilde{m}(P)}
= \frac{z(P)}{\fbar(P)-\widetilde{m}(P)}\ .
\end{equation}
For a nucleon with a fixed three-momentum $\np$, the pole of the
propagator in eq.~(\ref{propagator}) in the variable $p_0$ defines the
new energy of the nucleon in the medium. To find the latter we
introduce
\begin{equation}\label{Etilde}
\widetilde{E}(P)\equiv 
E(\np,\widetilde{m}(P))=\sqrt{\np^2+\widetilde{m}(P)^2}\ .
\end{equation}
Then the propagator reads
\begin{equation}
\frac{1}{\Pbar-m-\Sigma(P)}
= \frac{z(P)}{f_0(P)-\widetilde{E}(P)}
  \frac{\fbar(P)+\widetilde{m}(P)}{f_0(P)+\widetilde{E}(P)}
\label{propse}
\end{equation}
and its pole $p_0$ is found by solving the implicit equation 
\begin{equation}
f_0(P)=\widetilde{E}(P)\ ,
\end{equation}
which, exploiting eq.~(\ref{f_0}), can be recast as follows
\begin{equation}\label{dispersion}
p_0= \frac{1-C(P)}{1-B(P)}\sqrt{\np^2+\widetilde{m}(P)^2}
\equiv \frac{1-C(p_0,\np)}{1-B(p_0,\np)}
\sqrt{\np^2+\widetilde{m}(p_0,\np)^2}\ .
\label{p0eq}
\end{equation}
The solution of eq.~(\ref{p0eq}) for fixed $\np$ 
defines the new dispersion relation $p_0=\epsilon(\np)$
for interacting nuclear matter. 
Once the above equation has been solved,
the field strength renormalization constant
\begin{equation}
Z_2(\np)= {\rm Res}
\left. \frac{z(P)}{f_0(P)-\widetilde{E}(P)}\right|_{p_0=\epsilon(\np)}\ ,
\end{equation}
defined as the residue of the first factor on the right-hand side of
eq.~(\ref{propse}) at $p_0=\epsilon(\np)$, can be computed.
Indeed using eq.~(\ref{zp}), $Z_2(\np)$ is obtained 
by expanding the denominator around the pole $\epsilon(\np)$,
{\em  i.e.,}
\begin{equation}
\left[1-C(P)\right]\left[f_0(P)-\widetilde{E}(P)\right]
= Z_2(\np)^{-1}\left[p_0-\epsilon(\np)\right]+ \cdots\ ;
\label{nearpole}
\end{equation}
hence
\begin{eqnarray}
Z_2(\np)^{-1}
&=&\left.\frac{\partial}{\partial p_0}\right|_{p_0=\epsilon(\np)}
\left\{\left[1-C(P)\right]\left[f_0(P)-\widetilde{E}(P)\right]\right\}
\nonumber\\
&=& \left.\frac{\partial}{\partial p_0}\right|_{p_0=\epsilon(\np)}
\left\{\left[1-B(P)\right]p_0-\left[1-C(P)\right]\widetilde{E}(P)\right\}\ .
\end{eqnarray}
With the help of eq.~(\ref{Etilde}) 
the derivative can be easily evaluated, the result being
\begin{equation} \label{Z_2}
Z_2(\np)^{-1} = 
\left[ 1-B
-\frac{\partial B}{\partial p_0}p_0 
-m\frac{\widetilde{m}}{\widetilde{E}}\frac{\partial A}{\partial p_0} 
+\frac{\np^2}{\widetilde{E}}\frac{\partial C}{\partial p_0} 
\right]_{p_0=\epsilon(\np)}\ .
\end{equation}

\subsubsection{Nucleon spinors}
\label{sec:PCspin}
The self-energy modifies not only the propagator and the
energy-momentum relation of a nucleon,
but, as well, the free Dirac spinors. In fact
the spinors are now solutions of the Dirac equation in
the nuclear medium~\cite{Cel86}, {\em  i.e.},
\begin{equation}\label{new_spinors}
[\Pbar-m-\Sigma(P)]\phi(\np)=0\ ,
\end{equation}
which, again using the decomposition in eq.~(\ref{spin}),
can be recast as follows
\begin{equation}\label{Dirac}
\left[ \gamma_0 f_0(P)
       -\ngamma\cdot\np
       -\widetilde{m}(P)
\right]\phi(\np)=0\ ,
\end{equation}
the functions $f_0(P)$ and $\widetilde{m}(P)$ being defined in
eqs.~(\ref{f_0}) and (\ref{m-tilde}), respectively.
Equation~(\ref{Dirac}) has the same structure as the free Dirac
equation; hence for the positive-energy eigenvalue one has
\begin{equation}
f^2_0(P)= \np^2+\widetilde{m}^2(P)\ ,
\end{equation}
which implicitly yields the energy $p_0=\epsilon(\np)$ of the nucleon
in the nuclear medium. This result was already obtained as the pole of
the nucleon propagator. Then the corresponding positive-energy spinors
($s$ being the spin index) read
\begin{equation}
\phi_s(\np) = 
\sqrt{Z_2(\np)}
\left(
\frac{\widetilde{E}(\np)+\widetilde{m}(\np)}{2\widetilde{m}(\np)}
\right)^{1/2}
\columnmatrix{\chi_s}{
\frac{\nsigma\cdot\np}{\widetilde{E}(\np)+\widetilde{m}(\np)}\chi_s}
=
\sqrt{Z_2(\np)}u_s(\np,\widetilde{m}(\np)),
\label{phispin}
\end{equation}
where the  two functions  {\em of the three-momentum $\np$}
\be
\widetilde{m}(\np)
\equiv
\widetilde{m}(\epsilon(\np),\np)
\label{Dirac-mass}
\ee
and
\be
\widetilde{E}(\np) 
\equiv
\widetilde{E}(\epsilon(\np),\np) 
= \sqrt{\np^2+\widetilde{m}(\np)^2}\ ,
\label{Dirac-energy}
\ee represent the nucleon effective mass and effective energy
corresponding to $p_0=\epsilon(\np)$. The field strength
renormalization constant, $\sqrt{Z_2(\np)}$, of the new spinors,
defined in eq.~(\ref{Z_2}), is required by renormalization theory,
since the propagator in eq.~(\ref{propagator}) for $p_0$ close to the
pole $\epsilon(\np)$ reads, from eq.~(\ref{nearpole}),
\begin{equation}
\frac{1}{\Pbar-m-\Sigma(P)} 
\sim 
\frac{Z_2(\np)}{p_0-\epsilon(\np)}
\frac{\fbar(\np)+\widetilde{m}(\np)}{2\widetilde{E}(\np)}
=
\frac{1}{p_0-\epsilon(\np)}
\frac{\widetilde{m}(\np)}{\widetilde{E}(\np)}
\sum_s \phi_s(\np)\overline{\phi}_s(\np)\ .
\end{equation}
Once the new spinors have been computed, the self-energy can be
evaluated by inserting $\phi(\nk)$ instead of $u(\nk)$ into
eq.~(\ref{SE1}). Then the Dirac equation should be solved again with
the new self-energy and so on.  This self-consistent procedure leads
to the relativistic Hartree-Fock model which has to be dealt with
numerically.

In this paper we do not attempt to solve the HF equations, since we
are interested only in the OPE first iteration correction to the
single-nucleon current. Although the latter cannot be derived by
directly applying the Feynman rules, it can still be identified with
the self-energy diagrams of Fig.~2 (f)--(g).  Thus in the next section
we shall compute the renormalized one-body current using the new
spinors and energy-momentum relation and then expand it in powers of
the square of the pion-coupling constant $f^2/m_\pi^2$.  As we shall
see, the unperturbed one-body current is thus recovered as the
leading-order term whereas the first-order term is the searched for
self-energy contribution.

It is also important to remark that the use of the new `renormalized'
wave functions $\phi_s$ leads to a slightly modified global momentum
distribution as shown in~\cite{Ama02b}. Note however that the
number of particles is conserved without modifying the value of the
Fermi momentum selected.

\subsection{Self-energy current to first order}
\label{sec:PCSE}

The particle-hole (p-h) current matrix element in the HF approximation reads
\begin{equation}\label{HF-current}
j_{HF}^{\mu}(\np,\nh)= \overline{\phi}(\np)\Gamma^{\mu}(Q)\phi(\nh) \ ,
\end{equation}
where the spinors  $\phi(\np)$, the first
iteration solution of the Hartree-Fock
equation, are given by eq.~(\ref{phispin}). 
Hence eq.~(\ref{HF-current}) represents the
electromagnetic excitation of the p-h pair with dressed external lines
corresponding to the sum of the diagrams shown in Fig.~5.
\begin{figure}[tb]
\begin{center}
\leavevmode
\def\epsfsize#1#2{0.9#1}
\epsfbox[160 580 570 750]{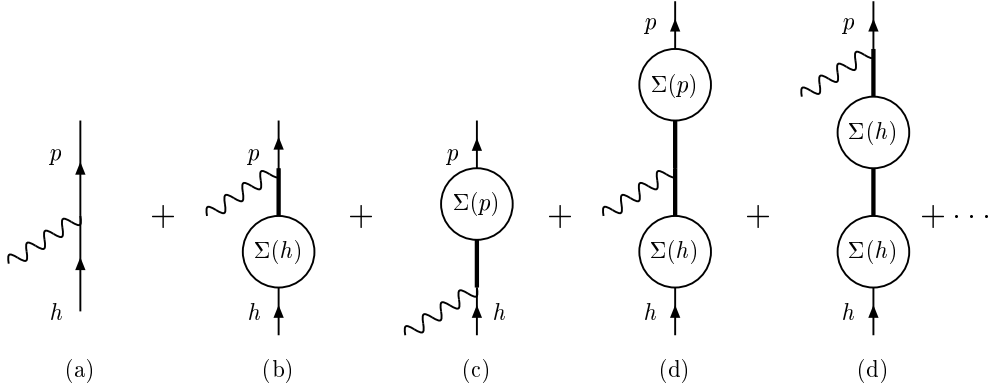}
\end{center}
\caption{
Diagrammatic series for the one-body electromagnetic current with 
dressed external lines.
}
\end{figure}
In order to obtain a genuine one-pion-exchange expression we expand
eq.~(\ref{HF-current}) in powers of the square of the pion coupling
constant $f^2/m_\pi^2$ and single out the first-order term, {\em i.e.}
the one linear in $f^2/m_\pi^2$. We shall still refer to the current
thus obtained, representing the OPE contribution, as the
``self-energy'' current and, importantly, we shall show that it yields
a finite contribution, free from the divergence problem of the current
in eq.~(\ref{SEC}).

To proceed we start by deriving the HF energy $\epsilon(\np)$ to first
order in $f^2/m_\pi^2$.  For this purpose we note that the functions
$A(P)$, $B(P)$ and $C(P)$ defined in
eqs.~(\ref{A-definition}--\ref{C-definition}) are of order
$O(f^2/m_\pi^2)$.  Hence the following expansion of the Dirac mass in
eq.~(\ref{m-tilde}) holds:
\begin{equation}\label{m-expansion-1}
\widetilde{m}(P)=m\frac{1+A(P)}{1-C(P)}
= m\left[1+A(P)+C(P)\right]+O\left(\frac{f^4}{m_\pi^4}\right)\ .
\end{equation}
Inserting this into eq.~(\ref{dispersion}) for the energy
and expanding again to first order in $f^2/m_\pi^2$, we get
\begin{eqnarray}
p_0=\epsilon(\np)
&\simeq& 
[1-C(P)+B(P)]\sqrt{\np^2+m^2+2m^2\left[A(P)+C(P)\right]}
\nonumber\\
&=& 
E_\np+\Delta E(p_0,\np)\ ,
\end{eqnarray}
where $E_\np= \sqrt{\np^2+m^2}$ is the unperturbed free energy and
\begin{equation}\label{energycorrection}
\Delta E(p_0,\np)\equiv 
\frac{1}{E_\np}
\left[m^2 A(P)+E_\np^2 B(P)-\np^2 C(P)\right] 
+O\left(\frac{f^4}{m_\pi^4}\right)
\end{equation}
is the first-order correction to the energy.
Next we can insert the above value of $p_0$ 
{\em inside} the argument of the functions $A$, $B$, $C$.
Expanding the latter around the on-shell value $p_0=E_\np$ we get
\begin{equation}
A(P)=A(p_0,\np)=A(E_\np+\Delta E,\np)= 
A(E_\np,\np)+O\left(\frac{f^4}{m_\pi^4}\right)
\simeq A_0(\np)\ ,
\end{equation}
where $A_0(\np) \equiv A(E_p,\np)$.
Likewise to first order we obtain 
\begin{eqnarray}
B(P) &\simeq & B(E_\np,\np)\equiv B_0(\np) \\
C(P) &\simeq & C(E_\np,\np)\equiv C_0(\np) \ .
\end{eqnarray}
Recalling that for $P$ on-shell
the functions $A$, $B$, $C$ coincide, {\em  i.e.,}
$A_0(\np)= B_0(\np)= C_0(\np)$,
we can insert these on-shell values 
into eq.~(\ref{energycorrection}) and, neglecting terms of second order,
{\em  i.e.,} $O\left(\frac{f^4}{m_\pi^4}\right)$, 
we finally arrive at the result
\begin{equation}
p_0=\epsilon(\np)= E_\np+\frac{1}{E_\np}B_0(\np)(m^2+E_\np^2-\np^2)
+O\left(\frac{f^4}{m_\pi^4}\right)=
E_\np+\frac{1}{E_\np}2m^2B_0(\np)
+O\left(\frac{f^4}{m_\pi^4}\right)\ .
\label{HFE}
\end{equation}
The above expression can be recast in terms of the
on-shell value of the self-energy 
\begin{equation}\label{Sigma_0}
\Sigma_0(\np)\equiv 2mB_0(\np)\ ,
\end{equation}
which satisfies the relation
\begin{equation}
\Sigma(E_\np,\np)u(\np) = \Sigma_0(\np)u(\np)\ ,
\end{equation}
thus showing that the free spinors are eigenvectors of the {\em
on-shell} self-energy matrix $\Sigma(E_\np,\np)$ corresponding to the
eigenvalue $\Sigma_0(\np)$.  Hence to first order in $f^2/m_\pi^2$,
the HF energy in eq.~(\ref{HFE}) is found to read
\begin{equation}\label{energy}
\epsilon(\np)
\simeq 
E_\np+ \frac{m}{E_\np}\Sigma_0(\np)
\end{equation}
in terms of the on-shell self-energy eigenvalue $\Sigma_0(\np)$.  When
compared with the non-relativistic HF energy (see eq.~(\ref{nr-energy})
in Section~\ref{sec:NRcorr}) it appears that, beyond the different
expressions of the self-energy functions that hold in the relativistic
and non-relativistic frameworks, an extra multiplicative factor
$m/E_\np$ occurs in the relativistic case.

Once the HF energy $\epsilon(\np)$ is known to first order
in $f^2/m_\pi^2$, we expand as well the renormalized
spinors, namely
\begin{equation}\label{ren-spinor}
\sqrt{\frac{\widetilde{m}(\np)}{\widetilde{E}(\np)}}
u(\np,\widetilde{m}(\np))
= 
\sqrt{\frac{\widetilde{E}(\np)+\widetilde{m}(\np)}{2\widetilde{E}(\np)}}
\columnmatrix{\chi}%
{\frac{\nsigma\cdot\np}{\widetilde{E}(\np)+\widetilde{m}(\np)}\chi}\ .
\end{equation}
Actually, for later use in the calculation of the hadronic tensor, it is
convenient to expand the spinor multiplied by the factor
$\sqrt{\frac{\widetilde{m}(\np)}{\widetilde{E}(\np)}}$. 

Thus we start by expanding the
Dirac mass in eq.~(\ref{m-expansion-1}) around the on-shell energy,
obtaining
\begin{equation}
\widetilde{m}(\np)
=
m\left[1+A_0(\np)+C_0(\np)\right]+O(f^2/m_\pi^2)
\simeq  m+\Sigma_0(\np)\ ,
\label{m-expansion}
\end{equation}
where use has been made of the on-shell self-energy in eq.~(\ref{Sigma_0}).
Likewise, using the HF equation (eq.~(\ref{dispersion})),
the Dirac energy $\widetilde{E}(\np)$ defined in eq.~(\ref{Dirac-energy})
is given by
\begin{eqnarray}
\widetilde{E}(\np)
&=& \frac{1-B}{1-C}\epsilon(\np)
\simeq
\left[1-B_0(\np)+C_0(\np)\right]\left[E_\np+\frac{m}{E_\np}\Sigma_0(\np)
\right]
\nonumber\\
&\simeq& E_\np+ \frac{m}{E_\np}\Sigma_0(\np)
\simeq \epsilon(\np)\ .
\label{E-expansion}
\end{eqnarray}
After some algebra the following first-order expressions are obtained
\begin{eqnarray}
\sqrt{\frac{\widetilde{E}+\widetilde{m}}{2\widetilde{E}}}
&\simeq&
\sqrt{\frac{m+E_\np}{2E_\np}}
\left( 1+\frac{E_\np-m}{2E_\np}\frac{\Sigma_0}{E_\np}\right)
\label{norm}\\
\frac{1}{\widetilde{E}+\widetilde{m}}
&\simeq&
\frac{1}{m+E_\np}\left(1-\frac{\Sigma_0}{E_\np}\right)\ .
\label{1Em}
\end{eqnarray}
Inserting eqs.~(\ref{norm}) and~(\ref{1Em}) into the renormalized spinor
in eq.~(\ref{ren-spinor}) we get
\begin{eqnarray}
\sqrt{\frac{\widetilde{m}(\np)}{\widetilde{E}(\np)}}
u(\np,\widetilde{m}(\np))
&\simeq&
\sqrt{\frac{m+E_\np}{2E_\np}}
\left[ 1+\frac{E_\np-m}{2E_\np}\frac{\Sigma_0}{E_\np}\right]
\columnmatrix{\chi}%
{\frac{\nsigma\cdot\np}{m+E_\np}\left(1-\frac{\Sigma_0}{E_\np}\right)\chi}
\nonumber\\
&\simeq&
\sqrt{\frac{m}{E_\np}}u(\np)
+
\frac{\Sigma_0}{E_\np}
\sqrt{\frac{m}{E_\np}}
\sqrt{\frac{m+E_\np}{2m}}
\columnmatrix{\frac{E_\np-m}{2E_\np}\chi}%
{-\frac{E_\np+m}{2E_\np}\frac{\nsigma\cdot\np}{m+E_\np}\chi}\ .
\end{eqnarray}
Since
\begin{equation}
\columnmatrix{(E_\np-m)\chi}%
{-(E_\np+m)\frac{\nsigma\cdot\np}{m+E_\np}\chi}
=
(E_\np\gamma_0-m)
\columnmatrix{\chi}{\frac{\nsigma\cdot\np}{m+E_\np}\chi}\ .
\end{equation}
the first-order (in $f^2/m_\pi^2$) renormalized spinor 
can be cast in the form
\begin{equation} \label{A1}
\sqrt{\frac{\widetilde{m}(\np)}{\widetilde{E}(\np)}}
u(\np,\widetilde{m}(\np))
\simeq
\sqrt{\frac{m}{E_\np}}
\left[ 
      u(\np)
     +\frac{\Sigma_0(\np)}{E_\np}\frac{E_\np\gamma_0-m}{2E_\np}u(\np)
\right]\ .
\end{equation}
The above expansion transparently displays the effect of the
self-energy on the free spinor $u(\np)$.  Indeed the second term in
the square brackets of eq.~(\ref{A1}) corresponds to a negative-energy
component with momentum $\np$. In fact, the Dirac equation for a
positive-energy spinor is given by
\begin{equation}
(\np\cdot\ngamma+m)u(\np)=E_\np \gamma_0 u(\np),
\kern 1cm
\mbox{with $E_\np >0$} \ .
\end{equation}
Now if we apply the operator $(\np\cdot\ngamma+m)$ to the spinor
$(E_\np\gamma_0-m)u(\np)$ we obtain
\begin{eqnarray}
(\np\cdot\ngamma+m)(E_\np\gamma_0-m)u(\np)
&=&
\np\cdot\ngamma(E_\np\gamma_0-m)u(\np)
+m(E_\np\gamma_0-m)u(\np)
\nonumber\\
&=&
(-E_\np\gamma_0-m)\np\cdot\ngamma u(\np)
+m(E_\np\gamma_0-m)u(\np)
\nonumber\\
&=&
(-E_\np\gamma_0-m)(E_\np\gamma_0-m) u(\np)
+m(E_\np\gamma_0-m)u(\np)
\nonumber\\
&=&
-E_\np\gamma_0(E_\np\gamma_0-m) u(\np).
\label{A129}
\end{eqnarray}
Hence $(E_\np\gamma_0-m)u$ is an eigenvector of the free Dirac
Hamiltonian with eigenvalue $-E_\np$. Therefore the operator
$E_\np\gamma_0-m$ transforms a positive-energy spinor $u(\np)$ into a
negative-energy one.

Moreover, it is useful to write down 
the correction to the free spinor (see eq.~(\ref{A1}))
in an alternative
form. Using the identity in eq.~(\ref{A129}) we can write
\begin{equation}
(\Pbar-m)(E_\np\gamma_0-m)u(\np)=2E_\np\gamma_0(E_\np\gamma_0-m) u(\np) \ .
\end{equation}
Multiplying by $[2E_\np(\Pbar-m)]^{-1}$ we then obtain
\begin{equation}
\frac{E_\np\gamma_0-m}{2E_\np}u(\np)
= \frac{1}{\Pbar-m}\gamma_0(E_\np\gamma_0-m) u(\np) \ .
\end{equation}
Hence the second term in the square brackets of the right-hand 
side of eq.~(\ref{A1})
can be recast in the form
\begin{eqnarray}
\frac{\Sigma_0}{E_\np}
\frac{E_\np\gamma_0-m}{2E_\np}
u(\np)
&=&
\frac{\Sigma_0}{E_\np}\frac{1}{\Pbar-m}\gamma_0(E_\np\gamma_0-m)u(\np)
=
\frac{1}{\Pbar-m}
\left(1-\frac{m}{E_\np}\gamma_0\right)
\Sigma(\np)u(\np)
\nonumber\\
&=&
S_F(\np)\left(1-\frac{m}{E_\np}\gamma_0\right)\Sigma(\np)u(\np) \ .
\label{A132}
\end{eqnarray}
The first term in eq.~(\ref{A132}), $S_F(\np)\Sigma(\np)u(\np)$,
corresponds to the one that enters in the original (divergent)
self-energy current for a nucleon on-shell (eq.~(\ref{SEC})).  The
subtracted term, with the factor $\frac{m}{E_\np}\gamma_0$ inserted
between the propagator and the self-energy, cancels the divergence and
yields a finite result.  Thus it can be viewed as a ``recipe'' to
renormalize the self-energy current.

We turn now to an expansion of the field-strength renormalization
function defined in eq.~(\ref{Z_2}). For this purpose we use
eqs.~(\ref{m-expansion},\ref{E-expansion}), obtaining
\begin{equation}
Z_2(\np) \simeq 
\left[ 1+B_0(\np)
       +\frac{m^2}{E_\np}\frac{\partial A}{\partial p_0} 
       +E_\np\frac{\partial B}{\partial p_0}
       -\frac{\np^2}{E_\np}\frac{\partial C}{\partial p_0} 
\right]_{p_0=E_\np}\ ,
\end{equation}
which implies that
\begin{equation}
\sqrt{Z_2(\np)} \simeq 1+\frac12\alpha(\np)
\end{equation}
with
\begin{equation} \label{alpha}
\alpha(\np) \equiv
B_0(\np)
+\left[\frac{m^2}{E_\np}\frac{\partial A}{\partial p_0} 
+E_\np\frac{\partial B}{\partial p_0}
-\frac{\np^2}{E_\np}\frac{\partial C}{\partial p_0} 
\right|_{p_0=E_\np}\ .
\end{equation}

Hence, collecting the above results and inserting them into 
eq.~(\ref{phispin}), we get to first order 
\begin{equation}\label{expanded-spinor}
\sqrt{\frac{\widetilde{m}(\np)}{\widetilde{p_0}(\np)}}
\phi(\np)
\simeq 
\sqrt{\frac{m}{E_\np}}
\left[ 
      u(\np)
     +\frac{\Sigma_0}{E_\np}\frac{E_\np\gamma_0-m}{2E_\np}u(\np)
     +\frac12\alpha(\np) u(\np)
\right]\ .
\end{equation}
Thus, within the OPE approach
the renormalized HF spinors in the nuclear medium are characterized by
two new elements with respect to the bare $u(\np)$: the term
$\frac{\Sigma_0}{E_\np}\frac{E_\np\gamma_0-m}{2E_\np}u(\np)$
introduces negative-energy components in the wave function, while the
term $\frac12\alpha(\np) u(\np)$ comes from the field-strenght
renormalization which modifies the occupation number of the
single-particle states.

Using the above expressions for the renormalized spinors, 
we now expand the renormalized one-body current matrix element
to first order in $f^2/m_\pi^2$, getting
\begin{eqnarray}
  \sqrt{\frac{\widetilde{m}(\np)}{\widetilde{E}(\np)}}
  \sqrt{\frac{\widetilde{m}(\nh)}{\widetilde{E}(\nh)}}
  j^{\mu}_{HF}(\np,\nh)
&\simeq&
\sqrt{\frac{m}{E_\np}\frac{m}{E_\nh}}
\overline{u}(\np)
\left[ 
      \Gamma^{\mu}
     +\Gamma^{\mu}
       \frac{\Sigma_0(\nh)}{E_\nh}\frac{E_\nh\gamma_0-m}{2E_\nh}
     +\frac{\alpha(\nh)}{2}
      \Gamma^{\mu}
\right.
\nonumber\\
&&
\kern 6em
\left.\mbox{}
     +\frac{\Sigma_0(\np)}{E_\np}\frac{E_\np\gamma_0-m}{2E_\np}
      \Gamma^{\mu}
     +\frac{\alpha(\np)}{2}
      \Gamma^{\mu}
\right]
u(\nh)
\nonumber\\
&\equiv&
\frac{m}{\sqrt{E_\np E_\nh}}
\left[ j^{\mu}_{OB}(\np,\nh)+j^{\mu}_{RSE}(\np,\nh) \right]\ .
\label{HF-current-expansion}
\end{eqnarray}
In eq.~(\ref{HF-current-expansion}) the term $j^{\mu}_{OB}$
represents the usual one-body current matrix element evaluated 
with free spinors, {\em  i.e.,}
\begin{equation}
j^{\mu}_{OB}(\np,\nh)= \overline{u}(\np)\Gamma^{\mu}(Q)u(\nh) \ ,
\end{equation}
whereas $j^{\mu}_{RSE}$ is a new renormalized self-energy current matrix 
element
that includes the effects of the renormalization of the spinors. 
It can be decomposed according to
\begin{equation}\label{RSE}
j^{\mu}_{RSE}(\np,\nh)
= j^{\mu}_{RSE1}(\np,\nh)+j^{\mu}_{RSE2}(\np,\nh)\ ,
\end{equation}
where  $j^{\mu}_{RSE1}$ embodies the correction arising from
the new spinor solution of the Dirac equation in the medium
and $j^{\mu}_{RSE2}$ the one stemming from the field-strength 
renormalization function $\sqrt{Z_2}$ in the medium. 
Their expressions are the following:
\begin{eqnarray} 
j^{\mu}_{RSE1}(\np,\nh)
&=&
\overline{u}(\np)
\left[ 
     \Gamma^{\mu}
       \frac{\Sigma_0(\nh)}{E_\nh}\frac{E_\nh\gamma_0-m}{2E_\nh}
     +\frac{\Sigma_0(\np)}{E_\np}\frac{E_\np\gamma_0-m}{2E_\np}
      \Gamma^{\mu}
\right]
u(\nh)
\label{RSE1}
\\
j^{\mu}_{RSE2}(\np,\nh)
&=&
\left[\frac{\alpha(\nh)+\alpha(\np)}{2}\right]j^{\mu}_{OB}(\np,\nh)\ .
\label{RSE2}
\end{eqnarray}

\subsection{Gauge invariance of the theory}
\label{sec:PCgauge}

A crucial feature of the present theory is that the hadronic tensor,
computed either through the p-h matrix elements or through the
polarization propagator, is gauge invariant.  This may be somewhat
surprising because, as shown in Appendix B (see also~\cite{Ama02}),
current conservation is already obtained at the level of the MEC and
correlation p-h matrix elements: hence the one-body current p-h matrix
element also has to be independently conserved. This however occurs
only in zeroth order of perturbation theory. To be dealt with
properly, the situation clearly requires the renormalization of the
p-h energies and of the Dirac spinors (see previous sections).  Only
then does it become possible to set up a renormalized SE current which
leads to a hadronic tensor coinciding with the one obtained through
the polarization propagator~\cite{Ama02}.

As shown in the previous section, the renormalized HF current matrix
element, expanded to first order in $f^2/m_\pi^2$, has been split into
the usual one-body current and into a new renormalized self-energy
current. In order to be consistent with the one-pion-exchange model,
we should add the contribution of the seagull, pion-in-flight and
vertex correlation currents corresponding to the diagrams shown in
Fig.~2(a--e). We point out once more that the self-energy diagrams (f)
and (g), of Fig.~2, corresponding to insertions in external legs,
should be disregarded in computing amplitudes (or currents) in
perturbation theory.  Rather, their contributions should be taken into
account via renormalized energies and spinors as solutions of the
relativistic HF equations. We have expressed the latter, to first
order in $f^2/m_\pi^2$, in the form of a new current operator (denoted
as RSE current).

Then the total current in our model reads
\begin{equation}\label{jfull1}
j^{\mu}(\np,\nh)= j^{\mu}_{OB}(\np,\nh)+ j^{\mu}_{OPE}(\np,\nh)\ ,
\label{jtotal}
\end{equation}
where $j^{\mu}_{OPE}$ embodies the seagull, pion-in-flight, vertex
correlation and renormalized self-energy currents, namely
\begin{equation}
j^{\mu}_{OPE}= j^{\mu}_{s}+ j^{\mu}_{p}+j^{\mu}_{VC}+j^{\mu}_{RSE}\ .
\label{jfull}
\end{equation}
In what follows we shall prove the gauge invariance of this current to
first order in $f^2/m_\pi^2$. In so-doing it is crucial to take into
account not only the full current in eqs.~(\ref{jfull1},\ref{jfull}),
but also the first-order correction to the energy of the particles and
holes due to the self-energy interaction in eq.~(\ref{energy}).  In
other words, for a given momentum transfer $\nq=\np-\nh$, the energy
transfer should be computed as the difference between the particle and
hole HF energies and not using the free values $E_\np$ and $E_\nh$.
Thus the energy transfer is
\begin{equation}
\omega_{HF} = E_\np-E_\nh +
\frac{m}{E_\np}\Sigma_0(\np)-\frac{m}{E_\nh}\Sigma_0(\nh)
\end{equation}
and the associated four-momentum transfer is
$Q_{HF}^{\mu}=(\omega_{HF},\nq)$. To make the following discussion clearer 
we denote with
$Q_{HF}$ the HF four-momentum and with  $\omega_{HF}$ the HF  energy 
transfer, to 
distinguish them from the on-shell values $Q$ and $\omega$.

\paragraph{\bf{Divergence of the one-body current}}
\mbox{}

The divergence of the zeroth-order one-body current computed using the
HF four-momentum transfer $Q_{HF}$ is given by
\begin{equation}
Q_{HF,\mu}j^{\mu}_{OB}(\np,\nh)
= \overline{u}(\np)Q_{HF,\mu}
\Gamma^{\mu}\left( Q_{HF} \right) u(\nh)
= \overline{u}(\np)F_1\left(Q_{HF}\right)\Qbar_{HF}u(\nh)\ ,
\end{equation}
where the nucleon vertex $\Gamma^{\mu}(Q_{HF})$ is also evaluated at the
momentum transfer $Q_{HF}$. Because of 
$\overline{u}(\np)\!\!\!\Qbar u(\nh)=0$, only the first-order contribution
arising from the
self-energy correction survives, namely
\begin{equation}  \label{OB-div}
Q_{HF,\mu}j^{\mu}_{OB}(\np,\nh)
= \overline{u}(\np)F_1(Q)
\left[\frac{m}{E_\np}\Sigma_0(\np)-\frac{m}{E_\nh}\Sigma_0(\nh)\right]
\gamma_0 u(\nh)\ .
\end{equation}
In the above the Dirac form factor $F_1$ is computed at the
unperturbed value $Q^{\mu}$, since we disregard second-order
contributions.  Note that the one-body current itself is not gauge
invariant --- its divergence yields a first-order term which turns out
to be essential for the gauge invariance of the full current, as we
shall see below.

\paragraph{\bf{Divergence of the MEC}}
\mbox{}

The seagull and pionic $1p-1h$ currents given in
eqs.~(\ref{Sph},\ref{Pph}) are already of first order in
$f^2/m_\pi^2$; thus in computing their divergence we use the
unperturbed value of the energy transfer, neglecting a term of order
$O\left(\frac{f^4}{m_\pi^4}\right)$.  Using the free Dirac
equation and exploiting the kinematics we obtain 
\begin{eqnarray}
\lefteqn{Q_{\mu} j_{s}^{\mu}(\np,\nh) }
\nonumber\\
&=&
-\frac{f^2}{Vm_\pi^2} 
F_1^V i \epsilon_{3ab}
\overline{u}(\np)\tau_a\tau_b
\sum_{\nk\leq k_F}\frac{m}{E_\nk}
\left\{
\frac{2(K\cdot P-m\Kbar)}{(P-K)^2-m_\pi^2}
-\frac{2(K\cdot H-m\Kbar)}{(K-H)^2-m_\pi^2}
\right\}
u(\nh)
\nonumber\\
\label{seagull-div}\\
\lefteqn{Q_{\mu} j_{p}^{\mu}(\np,\nh) }
\nonumber\\
&=&
-\frac{f^2}{Vm_\pi^2} 
F_1^V i \epsilon_{3ab}
\overline{u}(\np)\tau_a\tau_b
\sum_{\nk\leq k_F}\frac{m}{E_\nk}
\left\{
\frac{2m(\Kbar-m)}{(P-K)^2-m_\pi^2}
-\frac{2m(\Kbar-m)}{(K-H)^2-m_\pi^2}
\right\}
u(\nh)\ .
\nonumber\\
\label{pion-in-flight-div}
\end{eqnarray}
In deriving these equations we have used the relations
$Q_{\mu}(Q+2H-2K)^{\mu}=-2K\cdot Q$ and
\begin{equation}
\frac{1}{(K-H)^2-m_\pi^2}-
\frac{1}{(P-K)^2-m_\pi^2}
=\frac{-2P\cdot Q}{[(K-H)^2-m_\pi^2][(P-K)^2-m_\pi^2]}\ .
\end{equation}
Upon addition of Eqs.~(\ref{seagull-div},\ref{pion-in-flight-div})
the terms containing $\Kbar$ cancel,
leaving for the total divergence of the seagull and pion-in-flight
the expression
\begin{eqnarray}
\lefteqn{Q_{\mu}(j_{s}^{\mu}+j_p^{\mu}) }
\nonumber\\
&=&
-\frac{f^2}{Vm_\pi^2} 
F_1^V i \epsilon_{3ab}
\overline{u}(\np)\tau_a\tau_b
\sum_{\nk\leq k_F}\frac{m}{E_\nk}
\left\{
\frac{2(K\cdot H-m^2)}{(K-H)^2-m_\pi^2}
-\frac{2(K\cdot P-m^2)}{(P-K)^2-m_\pi^2}
\right\}
u(\nh)\ , 
\end{eqnarray}
which can be further simplified by exploiting the self-energy of
eq.~(\ref{SE3}) for on-shell momenta. One finally obtains
\begin{equation}
Q_{\mu}(j_{s}^{\mu}+j_p^{\mu})= 
\frac{i}{3} F_1^V  \epsilon_{3ab}
\overline{u}(\np)\tau_a\tau_b
[\Sigma(\np)-\Sigma(\nh)]u(\nh)\ .
\label{MEC-div}
\end{equation}

\paragraph{\bf{Divergence of the vertex correlation current}}
\mbox{}

Starting from the 1p-1h matrix element of the VC current in
eq.~(\ref{VC}) and applying the Dirac equation, we get
\begin{eqnarray}
Q_{\mu}j_{VC}^{\mu}(\np,\nh)
&=& 
\frac{f^2}{Vm_\pi^2} 
\overline{u}(\np)\tau_a F_1\tau_a 
\sum_{\nk\leq k_F}\frac{1}{2E_\nk}
\gamma_5(\Pbar-\Kbar)
\frac{\Kbar+m}{(P-K)^2-m_\pi^2}
\gamma_5(\Pbar-\Kbar)
u(\nh)
\nonumber\\
&-& 
\frac{f^2}{Vm_\pi^2} 
\overline{u}(\np)
\tau_a F_1 \tau_a
\sum_{\nk\leq k_F}\frac{1}{2E_\nk}
\gamma_5(\Kbar-\Hslash)
\frac{\Kbar+m}{(K-H)^2-m_\pi^2}
\gamma_5(\Kbar-\Hslash)u(\nh)\ ,
\nonumber\\
\end{eqnarray}
where we recognize again the expression of the self-energy 
matrix in eq.~(\ref{SE2}). 
Since the Dirac form factor can be split into an isoscalar and an 
isovector component according to
\begin{equation}
F_1 = \frac12(F_1^S+F_1^V\tau_3)\ ,
\end{equation}
which yields
\begin{equation}
\tau_a F_1 \tau_a = 3 F_1 +iF_1^V\epsilon_{3ab}\tau_a\tau_b\ ,
\end{equation}
the divergence of the VC current written
in terms of the self-energy function reads
\begin{equation}
Q_{\mu}j_{VC}^{\mu}(\np,\nh) = 
\overline{u}(\np)
\left(
       F_1 + \frac{i}{3}F_1^V\epsilon_{3ab}\tau_a\tau_b
\right)
\left[\Sigma(\nh)-\Sigma(\np)\right] u(\nh)\ .
\end{equation}
Comparing this result with
eq.~(\ref{MEC-div}) we note that the term above containing
$\epsilon_{3ab}\tau_a\tau_b$ cancels with the MEC contribution. Hence
\begin{equation}\label{Ward}
Q_{\mu}\left[j_{MEC}^{\mu}(\np,\nh)+j_{VC}^{\mu}(\np,\nh)\right] =
\overline{u}(\np)F_1\left[\Sigma(\nh)-\Sigma(\np)\right]u(\nh)\ .
\end{equation}
The above relation just expresses the Ward-Takahashi identity~\cite{Gro87} relating the
full vertex correction, namely MEC plus VC
(diagrams 2 (a)--(e)), to the self-energy matrix element.  

\paragraph{\bf{Divergence of the RSE current}}
\mbox{}

Finally we compute the divergence of the renormalized self-energy
(RSE) current defined in eqs.~(\ref{RSE}), (\ref{RSE1}) and
(\ref{RSE2}).  For this purpose we first note that the divergence of
$j^\mu_{RSE2}$ vanishes to first order because it is proportional to
the OB current. Hence we write
\begin{equation} 
Q_{\mu}j^{\mu}_{RSE}(\np,\nh)
=
\overline{u}(\np)
\left[ 
     F_1\Qbar
       \frac{\Sigma_0(\nh)}{E_\nh}\frac{E_\nh\gamma_0-m}{2E_\nh}
     +\frac{\Sigma_0(\np)}{E_\np}\frac{E_\np\gamma_0-m}{2E_\np}
     F_1\Qbar
\right]
u(\nh).
\end{equation}
Using the relation 
$\overline{u}(\np)\Qbar u(\nh)=0$ and  
\begin{eqnarray}
\overline{u}(\np)\Qbar\gamma_0 u(\nh)
&=&
\overline{u}(\np)2(m\gamma_0-E_\nh) u(\nh)
\\
\overline{u}(\np)\gamma_0\Qbar u(\nh)
&=&
\overline{u}(\np)2(E_\np-m\gamma_0) u(\nh)
\end{eqnarray}
it is straightforward to obtain
\begin{equation}
Q_{\mu}j^{\mu}_{RSE}(\np,\nh) =
\overline{u}(\np)F_1\left[\Sigma(\np)-\Sigma(\nh)\right] u(\nh)
+\overline{u}(\np)F_1
\left[\frac{m}{E_\nh}\Sigma_0(\nh)-\frac{m}{E_\np}\Sigma_0(\np)
\right] u(\nh)\ .
\end{equation}

Remarkably the first term of this equation cancels with the divergence
of the MEC plus the VC current, given by the Ward-Takahashi identity
in eq.~(\ref{Ward}), whereas the second term cancels with the
divergence of the OB current in eq.~(\ref{OB-div}). We have thus
proven that, within the present model up to first order in
$f^2/m_\pi^2$, the total current in eq.~(\ref{jtotal}) satisfies the
continuity equation, namely
\begin{equation}
Q_{HF,\mu}(j_{OB}^{\mu}+j_{MEC}^{\mu}+j_{VC}^{\mu}+j_{RSE}^{\mu})=0\ .
\end{equation}

\subsection{Nuclear hadronic tensor and electromagnetic response functions}
\label{sec:PCresp}

\begin{figure}[t]
\begin{center}
\leavevmode
\def\epsfsize#1#2{0.9#1}
\epsfbox[100 420 500 760]{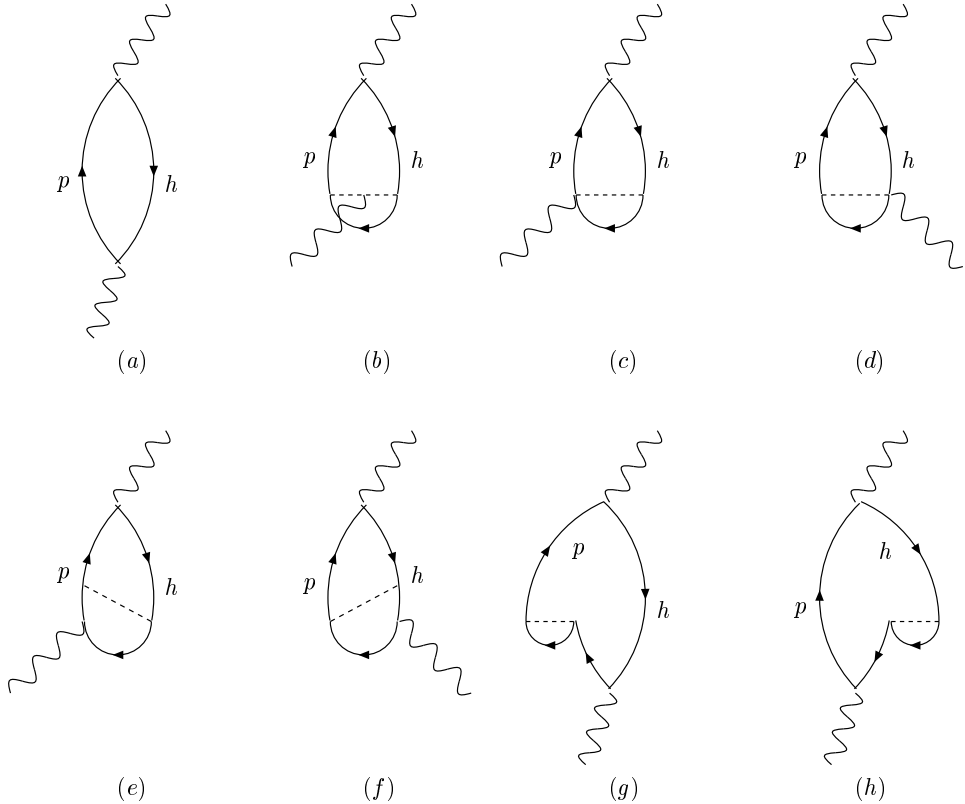}
\end{center}
\caption{
Feynman diagrams of the free (a) and first-order pion-in-flight (b), 
seagull (c and d), vertex correlation (e and f) and self-energy (g and h) 
polarization propagator.}
\end{figure}

In this section we compute the electromagnetic inclusive response
functions for one-particle emission reactions within the RFG model.
As discussed in previous sections, the p-h matrix elements
corresponding to the different pionic diagrams are all well-defined
except for the self-energy term which diverges, and consequently needs
to be renormalized. In what follows, we evaluate the hadronic tensor
starting from the current p-h matrix elements in the case of the
one-body, MEC and vertex correlation diagrams. These are shown
diagrammatically in Fig.~6.  
On the contrary, for the self-energy diagrams we calculate
the hadronic tensor in two at first sight different
ways: on the one hand, from the polarization propagator $\Pi^{\mu\nu}$
(see Appendix C), and on the other, using the
renormalized, well defined, SE p-h matrix elements 
(Appendix D). We prove that the two formalisms are equivalent.

The formalism of the nuclear hadronic tensor set up with the p-h matrix
elements has been presented in detail, within the RFG model, in
previous papers~\cite{Alb90,Bar93}. Hence, here we simply summarize
the results needed for later discussions. 
Before starting the analysis of pionic contributions, we recall the
analytic expressions for the OB, leading order 
electromagnetic responses of the RFG (see, for
example,~\cite{Don92,Alv01} for details):

\be
R^{L,T}(q,\omega)=R_0(q,\omega) \left[U^{L,T}_p(q,\omega)+U^{L,T}_n(q,\omega)
\right]
\ ,
\label{RLT}
\ee
where $p$ and $n$ refer to protons and
neutrons, respectively, and, for $Z=N$,
\begin{equation}
R_0(q,\omega)\equiv \frac{3 Z}{4 m \kappa\eta_F^3} 
(\varepsilon_F-\varepsilon_0) \theta(\varepsilon_F-\varepsilon_0)
\label{R0}
\end{equation}
with
\be
\varepsilon_0=
{\rm Max}\,\left\{ \varepsilon_F-2\lambda,
\kappa\sqrt{1+\frac{1}{\tau}}-\lambda
\right\} \ .
\ee
In the above the usual dimensionless variables
\be
\lambda=\frac{\omega}{2m},\,\,\,\, 
\tau=\frac{|Q^2|}{4m^2},\,\,\,\,
\kappa=\frac{q}{2m},\,\,\,\,
\eta_F=\frac{k_F}{m},\,\,\,\,
\varepsilon_F=\frac{E_F}{m}
\ee
have been introduced and  $E_F=\sqrt{k_F^2+m^2}$ is the Fermi energy.
The functions $U^{L,T}$ in eq.~(\ref{RLT}) are
\begin{eqnarray}
U^L_{p(n)}(q,\omega) &=& \frac{\kappa^2}{\tau} \left\{G_{Ep(n)}^2+
\frac{\Delta}{1+\tau}\left[G_{Ep(n)}^2+\tau G_{Mp(n)}^2\right]\right\} 
\\
U^T_{p(n)}(q,\omega) &=& 2\tau G_{Mp(n)}^2+
\frac{\Delta}{1+\tau}\left[G_{Ep(n)}^2+\tau G_{Mp(n)}^2\right] \ ,
\end{eqnarray}
where
\begin{equation}
\Delta\equiv \frac{\tau}{\kappa^2} \left[\frac{1}{3}\left(\varepsilon_F^3
+\varepsilon_F\varepsilon_0+\varepsilon_0^2\right)+\lambda\left(
\varepsilon_F+\varepsilon_0\right)+\lambda^2\right]-\left(1+\tau\right)\ .
\label{Delta}
\end{equation}

\subsubsection{MEC and vertex pionic contributions}
\label{sec:PCMECVC}

The hadronic tensor that
arises from the interference of the single-nucleon, OB current,
$j_{OB}^{\mu}$, with the one-pion-exchange current $j_a^{\mu}$, with
$a=s$ (seagull), $p$ (pion-in-flight) and $vc$ (vertex correlation), 
is for the RFG model with $Z=N$ (see eq.~(\ref{relativistic-response})
below) 
\begin{equation}
W^{\mu\nu}
=\frac{3Z}{8\pi k_F^3q}
\int_{h_0}^{k_F} h dh (\omega+E_{\nh}) \int_0^{2\pi}d\phi_h
\sum_{s_p,s_h} 
\frac{m^2}{E_\np E_\nh} 2{\rm Re}\, 
\left[j^\mu_{OB}(\np,\nh)^*
      j^\nu_a(\np,\nh)
\right] 
\ , \label{eq38} 
\end{equation}
where 
$j^\mu_{OB}(\np,\nh)=\overline{u}({\bf p}) \Gamma^\mu u({\bf h})$ 
is the single-nucleon p-h
matrix element with $\Gamma^\mu$ the electromagnetic nucleon current
from eq.~(\ref{eq10}) and $j^\nu_a(\np,\nh)$ is the p-h
matrix element for the seagull, pion-in-flight or vertex current as
given in eqs.~(\ref{Sph}), (\ref{Pph}) and (\ref{F}-\ref{B}),
respectively.

Note that in eq.~(\ref{eq38}) the integral over the hole polar angle,
$\cos\theta_h$, has been performed explicitly by exploiting the
energy-conserving $\delta$-function. This fixes the minimum momentum
of the hole according to \be h_0=m\sqrt{\varepsilon_0^2-1}\ .
\label{neededforpsi}
\ee
Moreover, the hole
three-momentum 
\be
{\bf h}=h\left(\sin\theta_0\cos\phi_h,\sin\theta_0\sin\phi_h,
\cos\theta_0\right)\ ,
\ee
involved in the hadronic tensor, must be evaluated for the following
specific value of the polar angle
\be
\cos\theta_0=\frac{\lambda\varepsilon-\tau}{\eta\kappa} \ ,
\ee
with $\eta=h/m$.

The hadronic tensor, as was the case for the current, can be also split into 
isoscalar and isovector parts, since there is no interference between
the two isospin channels.

An important issue relates to the form factor of the $\pi NN$ vertex,
$\Gamma_\pi$, which incorporates some aspects of the short-range
physics affecting the pionic correlations. In all of the above
expressions $\Gamma_\pi$ has not been explicitly indicated for sake of
simplicity. In~\cite{Ama02} the analysis of the gauge invariance at
the level of the particle-hole channel, performed by deriving the
contribution to the continuity equation of the isoscalar and isovector
SE, VC and MEC p-h matrix elements, is presented. There, it is shown
that the SE and VC contributions cancel in the isoscalar channel, in
contrast with the non-relativistic result~\cite{Alb90}, where the SE
is by itself gauge invariant.  Furthermore, the SE and VC contribution
in the isovector channel is exactly canceled by that of the MEC
(seagull and pion-in-flight).  It is crucial to recall that the
inclusion of $\Gamma_\pi$ in the p-h current matrix elements is not
without consequences in connection with gauge invariance.  
In fact, in this case, the model is not gauge invariant
unless new terms are added to the MEC
(see~\cite{Bar93,Dub76,Ris89,Mat89,Bar94} and~\cite{Ama96c,Bar99} for
recent work on the restoration of current conservation in model
calculations).  Lacking a fundamental theory for $\Gamma_\pi$, in the
calculations reported in this work we use the phenomenological
expression
\begin{equation}
\Gamma_\pi(P)=\frac{\Lambda^2-m_\pi^2}{\Lambda^2-P^2}
\label{eq44}
\end{equation}
with $\Lambda=1.3$ GeV. As long as the dependence upon $\Lambda$ is
not too strong the gauge invariance of the theory should not be too
badly affected. Within a non-relativistic approach for the pion
currents, a detailed discussion on the breakdown of the gauge
invariance induced by $\Gamma_\pi$, and on the dependence of the
responses upon the cutoff value can be found in~\cite{Bar93,Bar94}.

In~\cite{Ama93,Ama93b} the effects of the MEC upon the 
transverse response in a non relativistic shell model for finite
nuclei were studied as a function of the cutoff $\Lambda$.

\subsubsection{Relativistic self-energy responses}
\label{sec:PCrSE}

As already discussed in previous sections, a crucial point to be
emphasized is that the self-energy p-h matrix element,
eq.~(\ref{SEC}), is divergent. Hence it cannot be used directly in the
evaluation of the hadronic tensor.  Instead one should use
renormalized spinors with the corresponding renormalized energies.
Above we have taken account of the effect of renormalization to first
order in $f^2/m_\pi^2$ by introducing an extra term in the current:
the RSE current defined in eq.~(\ref{RSE}). In addition there is also
a $O(f^2/m_\pi^2)$ modification of the energy of the particles,
eq.~(\ref{energy}).  These two modifications of the free current and
energy in turn give a contribution to the hadronic tensor of order
$O(f^2/m_\pi^2)$, which we will refer to as {\em renormalized-self
energy contribution} (RSE), which is of the same order as the MEC and
VC currents and should be included in any consistent calculation to
first order in $f^2/m_\pi^2$. In addition this contribution is needed
for the gauge invariance of the results.

In what follows we derive the RSE contribution to the nuclear response
functions. This RSE contribution should replace the SE Feynman
diagrams shown in fig. 6 (g),(h). As a matter of fact, these two
diagrams can be computed using the polarization propagator formalism
(see Appendix C and Ref.~\cite{Ama02}), where one does not need to
appeal to renormalization since the SE diagrams are finite in this
case. Our goal is to show that the results for the response functions
obtained in the two ways coincide, although they stem from different
approaches. This is proved in Appendix D.  The RSE contribution,
therefore, can be identified with the contribution coming from the two
diagrams (g), (h) of fig. 6.

The one-body hadronic tensor in HF approximation reads
\begin{eqnarray}
W^{\mu\nu}_{HF}(\omega,\nq)
&=& V \sum_{s_ps_h}\sum_{t_pt_h}
\int\frac{d^3 h}{(2\pi)^3}
\frac{\widetilde{m}(\np)\widetilde{m}(\nh)}%
{\widetilde{E}(\np)\widetilde{E}(\nh)}
j_{HF}^{\mu}(\np,\nh)^* j_{HF}^{\nu}(\np,\nh)
\nonumber\\
&\times&
\delta(\omega+\epsilon(\nh)-\epsilon(\np))\theta(k_F-h)\ ,
\end{eqnarray}
where $\np=\nh+\nq$ and $j_{HF}^{\mu}(\np,\nh)$ is the one-body HF 
current in eq.~(\ref{HF-current})
computed using the renormalized HF spinors and HF
energies of the particle and the hole.

Next we use the expansions in eqs.~(\ref{HF-current-expansion}) 
for the current $j_{HF}$ and (\ref{energy}) for the HF energies.
In addition we expand the energy delta function to first order  
in $f^2/m_\pi^2$ according to
\begin{equation}
\delta(\omega+\epsilon(\nh)-\epsilon(\np))
\simeq 
\delta(\omega+E_\nh-E_\np)
+\frac{d \delta(\omega+E_\nh-E_\np)}{d\omega}
\left[
       \frac{m}{E_\nh}\Sigma_0(\nh)
      -\frac{m}{E_\np}\Sigma_0(\np)
\right]\ .
\end{equation}
Inserting all of these relations into the hadronic tensor and
neglecting terms of second order we get for the diagonal elements of
the hadronic tensor
\footnote{We only work out the diagonal elements of the hadronic tensor,
since these are the ones that contribute to the unpolarized inclusive 
longitudinal and transverse response functions.}
\begin{equation}
W^{\mu\mu}_{HF}(\omega,\nq)
\simeq 
W^{\mu\mu}_{OB}(\omega,\nq)
+\Delta W^{\mu\mu}_{RSE}(\omega,\nq)
\label{Wmumu}
\end{equation}
(the summation convention is not in force in eq.~(\ref{Wmumu})),
where $W^{\mu\mu}_{OB}(\omega,\nq)$ is the usual OB hadronic tensor of
a RFG, {\em  i.e.},
\begin{equation}
W^{\mu\mu}_{OB}=V\sum_{s_ps_h}\sum_{t_pt_h}
\int\frac{d^3h}{(2\pi)^3}
\frac{m^2}{E_\np E_\nh}
       |j^{\mu}_{OB}(\np,\nh)|^2
\delta(\omega+E_\nh-E_\np)\theta(k_F-h)\ ,
\end{equation}
and $\Delta W^{\mu\mu}_{RSE}(\omega,\nq)$ is the first-order
self-energy correction 
\begin{eqnarray}
\Delta W^{\mu\mu}_{RSE}
&=&
V\sum_{s_ps_h}\sum_{t_pt_h}
\int\frac{d^3h}{(2\pi)^3}
\frac{m^2}{E_\np E_\nh}
\left\{ \phantom{\frac12}
        2{\rm Re}\; j^{\mu}_{OB}(\np,\nh)^*j^{\mu}_{RSE}(\np,\nh)
        \delta(\omega+E_\nh-E_\np)
\right.
\nonumber\\
&& \mbox{}+\left.
       |j^{\mu}_{OB}(\np,\nh)|^2
\left[
       \frac{m}{E_\nh}\Sigma_0(\nh)
      -\frac{m}{E_\np}\Sigma_0(\np)
\right]
       \frac{d}{d\omega} \delta(\omega+E_\nh-E_\np)
\right\}\theta(k_F-h)\ .
\label{RSE-response}
\end{eqnarray}
In eq.~(\ref{RSE-response}) the first term corresponds to the interference 
between the OB and the RSE
currents, while the second one, which shifts the
allowed kinematical region because of the derivative 
of the energy delta function, is due to the modification of the
nucleon energies in the medium.

Carrying out the spin traces for the single-nucleon current 
\begin{equation}
\sum_{s_ps_h}
|j^{\mu}_{OB}(\np,\nh)|^2
=
\frac{1}{4m^2}
{\rm Tr}
\left\{
     \Gamma^{\mu}(Q)
(\Hslash+m)\Gamma^{\mu}(-Q)(\Pbar+m)
\right\} \ ,
\end{equation}
we get for the renormalized self-energy response function 
\begin{eqnarray}
\lefteqn{\Delta W^{\mu\mu}_{RSE} }
\nonumber\\
&=&
V\int\frac{d^3h}{(2\pi)^3}
\frac{1}{4E_\np E_\nh}
{\rm Tr}
\left\{
     \Gamma^{\mu}(Q)
\left[ 
       \frac{\Sigma_0(\nh)}{E_\nh}\frac{E_\nh\gamma_0-m}{2E_\nh}
      +\frac{\alpha(\nh)}{2}
\right]
(\Hslash+m)\Gamma^{\mu}(-Q)(\Pbar+m)
\right.
\nonumber\\
&&
\kern 3cm
+
\Gamma^{\mu}(Q)
(\Hslash+m)\Gamma^{\mu}(-Q)(\Pbar+m)
\left[ 
     \frac{\Sigma_0(\np)}{E_\np}\frac{E_\np\gamma_0-m}{2E_\np}
+
\frac{\alpha(\np)}{2}
\right]
\nonumber\\
&&
\kern 3cm
+
  \Gamma^{\mu}(Q)
  (\Hslash+m)
  \left[ 
        \frac{\Sigma_0(\nh)}{E_\nh}\frac{E_\nh\gamma_0-m}{2E_\nh}
       +\frac{\alpha(\nh)}{2}
  \right]
  \Gamma^{\mu}(-Q)
  (\Pbar+m)
\nonumber\\
&&
\kern 3cm
+
\left.
\Gamma^{\mu}(Q)
(\Hslash+m)
\Gamma^{\mu}(-Q)
\left[ 
     \frac{\Sigma_0(\np)}{E_\np}\frac{E_\np\gamma_0-m}{2E_\np}
     +\frac{\alpha(\np)}{2}
\right]
(\Pbar+m)
\right\}
\nonumber\\
&&        
\kern 3cm
\times
\delta(\omega+E_\nh-E_\np)\theta(k_F-h)
\nonumber\\
&& \mbox{}+
V\int\frac{d^3h}{(2\pi)^3}
\frac{1}{4E_\np E_\nh}
{\rm Tr}
\left\{
     \Gamma^{\mu}(Q)
(\Hslash+m)\Gamma^{\mu}(-Q)(\Pbar+m)
\right\}
\nonumber\\
&&
\kern 3cm
\left(
       \frac{m}{E_\nh}\Sigma_0(\nh)
      -\frac{m}{E_\np}\Sigma_0(\np)
\right)
       \frac{d}{d\omega} \delta(\omega+E_\nh-E_\np)\theta(k_F-h)\ .
\label{W-RSE}
\end{eqnarray}
More precisely, one should add two copies of eq.~(\ref{W-RSE}),
one with the form factors appropriate to the proton and one to the neutron.

In Appendix D we show that this contribution to the response function
is identical to the one obtained in Appendix C by computing the
imaginary part of the polarization propagator corresponding to the two
SE diagrams (g), (h) of Fig.~6.  This identity is not trivial: indeed
in the case of the polarization propagator the response functions,
with the Fock self-energy dressing the particle and the hole lines,
are computed by representing the product of two nucleon propagators as
the derivative of a single one to deal with the presence of a double
pole in the integrand.  In the present paper the problem has been
solved differently. First the entire perturbative series with Fock
self-energy insertions has been summed up and then the result has been
expanded to first order, thus obtaining a finite first-order current
operator. Because of the equivalence of these two procedures we are
confident about the validity of the results we have obtained for the
self-energy contribution to the nuclear responses.

\subsection{Analysis of results}
\label{sec:PCresults}

In this section we report the numerical results obtained for the
pionic MEC (pion-in-flight and seagull) and for the correlation
(vertex and self-energy) contributions to the quasielastic peak (QEP) in the
1p-1h sector.  The calculation is fully-relativistic. We have taken
$Z=N=20$ and set $k_F=237$ MeV/c, which is representative of nuclei in
the vicinity of $^{40}$Ca.

The 5-dimensional integrations of the MEC and correlation responses
implicit in eq.~(\ref{eq38}) have been performed numerically.  The
reliability of the numerical procedure has been proven by checking
that the free RFG responses coincide with their analytic expressions
(see, {\em  e.g.},~\cite{Don92}). 

\subsubsection{MEC}
\label{sec:MEC}

\begin{figure}[hp]
\begin{center}
\leavevmode
\def\epsfsize#1#2{0.9#1}
\epsfbox[100 200 500 730]{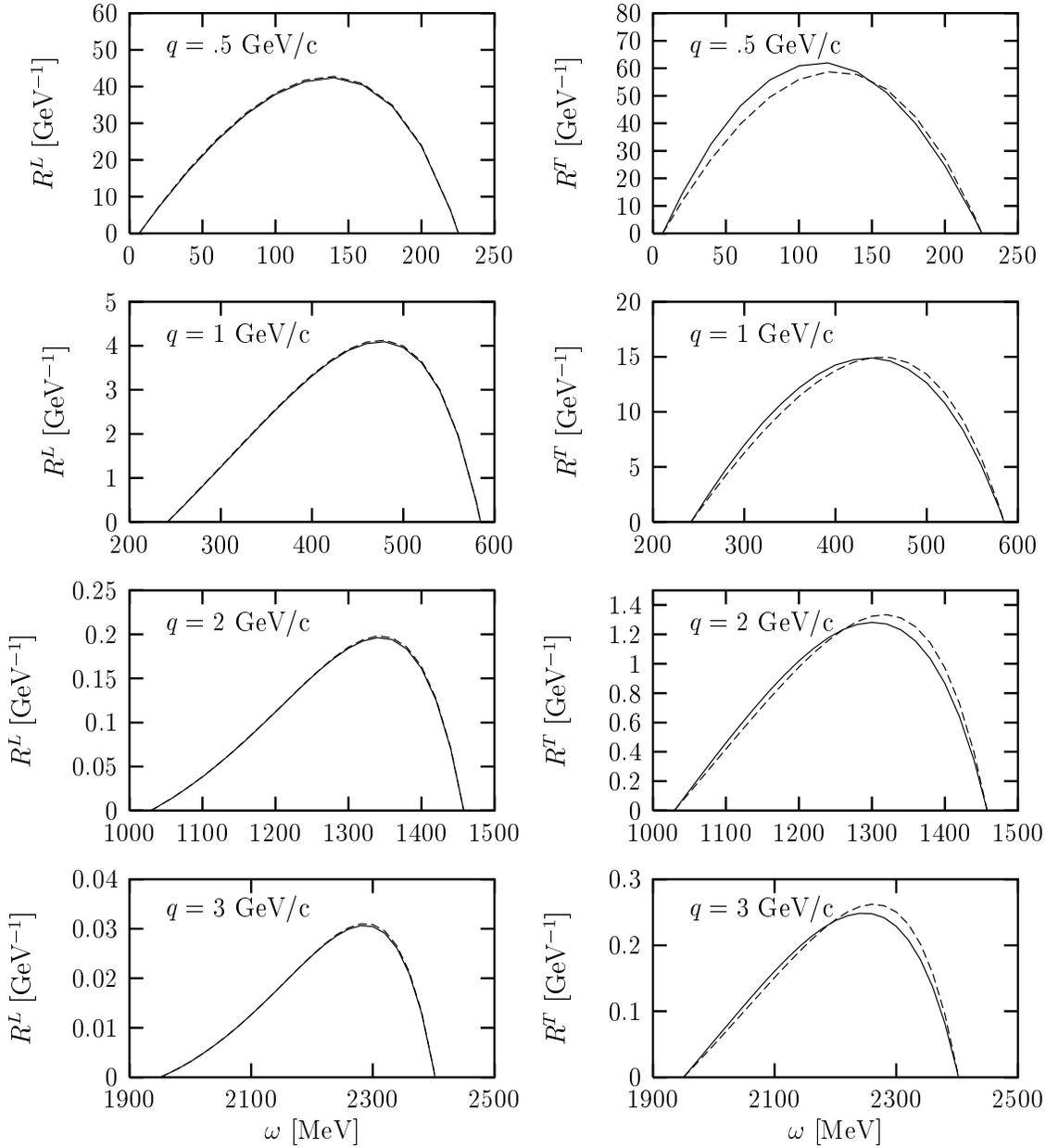}
\end{center}
\caption{
Longitudinal (left panels) and transverse (right panels) electromagnetic
response functions
versus $\omega$. Dashed: free RFG; solid: RFG+MEC contribution.
Here and in all the figures that follow, unless explicitly indicated, the
nucleus is
$^{40}$Ca, corresponding to a Fermi momentum $k_F=237$ MeV/c.}
\end{figure}

\begin{figure}[hp]
\begin{center}
\leavevmode
\def\epsfsize#1#2{0.9#1}
\epsfbox[100 200 500 730]{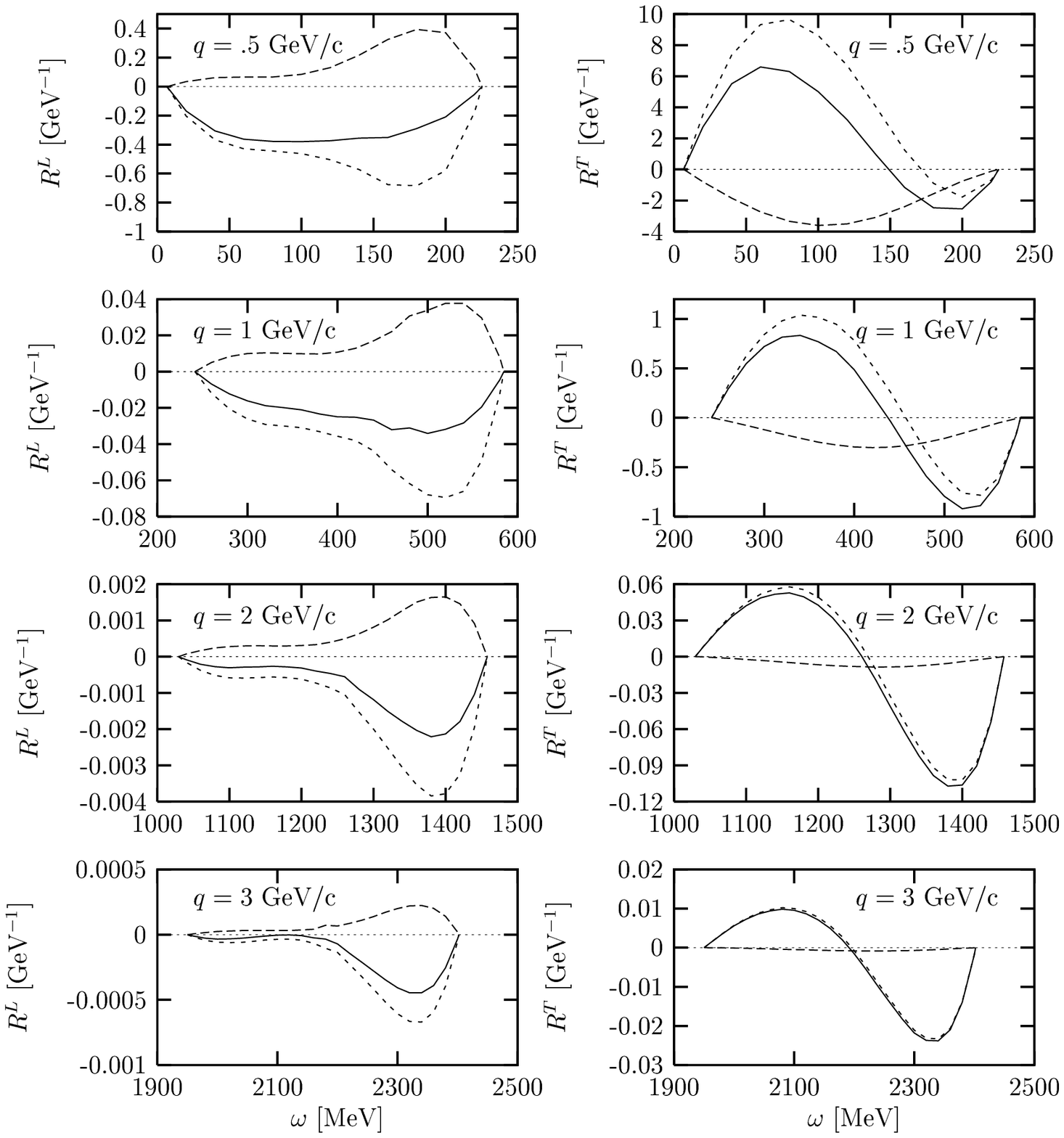}
\end{center}
\caption{
Separate MEC contribution to the longitudinal (left panels) and
transverse (right panels) responses. Dashed: pion-in-flight; short-dashed:
seagull and solid: MEC (pion-in-flight + seagull) contribution. 
}
\end{figure}

We start by analyzing the effects introduced by the MEC. These are
presented in Figs.~7 and 8 where we show the longitudinal (left
panels) and transverse (right panels) response functions versus the
transferred energy $\omega$ for four different values of the
transferred momentum $q$: 0.5, 1, 2 and 3 GeV/c.  First, in Fig.~7 we
compare the free RFG responses (dashed) with the responses obtained
including the global MEC contribution (solid). As shown, while for the
longitudinal responses the MEC are hardly visible, in the T channel
they contribute somewhat more, 
typically by about 5--10\%,
depending upon $q$ and $\omega$ (see discussion later).

  In Fig.~8 we display the separate pion-in-flight (dashed) and
seagull (short-dashed) contribution to $R^L$ and $R^T$ for various
values of $q$. The total MEC (seagull + pion-in-flight) contribution is
also shown (solid line).  In the transverse channel (right panels) it
appears that the seagull term is always larger than the pion-in-flight
term, a dominance that increases with $q$ and reflecting the spin
nature of the photon-MEC interaction.  Moreover, whereas the
pion-in-flight term is always negative, the seagull changes sign with
$\omega$, inducing a (mild) softening of the response, {\em  i.e.}, a
shift to lower energy. Within the longitudinal channel (left panels),
the seagull term, now always negative, also dominates. Note however
that the relative difference between the seagull and pion-in-flight
contributions is not as large as in the previous case.  Moreover, the
behavior of the seagull and pion-in-flight terms in the longitudinal
channel as $q$ increases displays a different pattern from the one
shown in the transverse channel, since for high $q$ the pionic current
is not negligible compared with the seagull one.

To complete this discussion we briefly comment on the MEC dependence
upon the momentum transfer $q$ and the Fermi momentum $k_F$,
associated with scaling of first and second kind, respectively
(see~\cite{Don99a,Don99b,Mai02}).

In~\cite{Ama02} we have explored in detail the evolution with $q$ of
the MEC in the transverse channel (as they are negligible in the
longitudinal channel).  We have proven that their relative
contribution to $R^T$ decreases with $q$, but does not vanish for
large values of $q$. In fact, the relative MEC contribution decreases
in going from 0.5 to 1 GeV/c, but then it rapidly saturates at or
slightly above $q$=1 GeV/c, where its value stabilizes, typically
around 10$\%$. Thus, one can conclude that at momentum transfers above
1 GeV/c scaling of the first kind is satisfied for the MEC
contributions considered in this work.  Moreover, for high $q$ the MEC
almost vanish for $\omega$ in the vicinity of the QEP.

A detailed analysis of the $k_F$ dependence of the MEC contribution in
the transverse response has also been presented in~\cite{Ama02}.  The
MEC contribution is found to {\em grow} with $k_F$, in contrast with
the free response which decreases as $k_F^{-1}$.  It is also shown
that the two-body MEC processes violate the second-kind scaling by
roughly three powers of $k_F$.  This effect is a rapid function of the
Fermi momentum (or equivalently, of the density): for example, if one
considers the cases $^2$H/$^4$He/heavy nuclei with Fermi momenta of
approximately 55/200/260 MeV/c, respectively, then the 1p-1h MEC
contributions amount to 0.1/5/10\% of the total transverse response,
respectively (normalizing to 10\% for the heavy nucleus case).

\subsubsection{Correlations}
\label{sec:PCcorr}

\begin{figure}[hp]
\begin{center}
\leavevmode
\def\epsfsize#1#2{0.9#1}
\epsfbox[100 195 500 710]{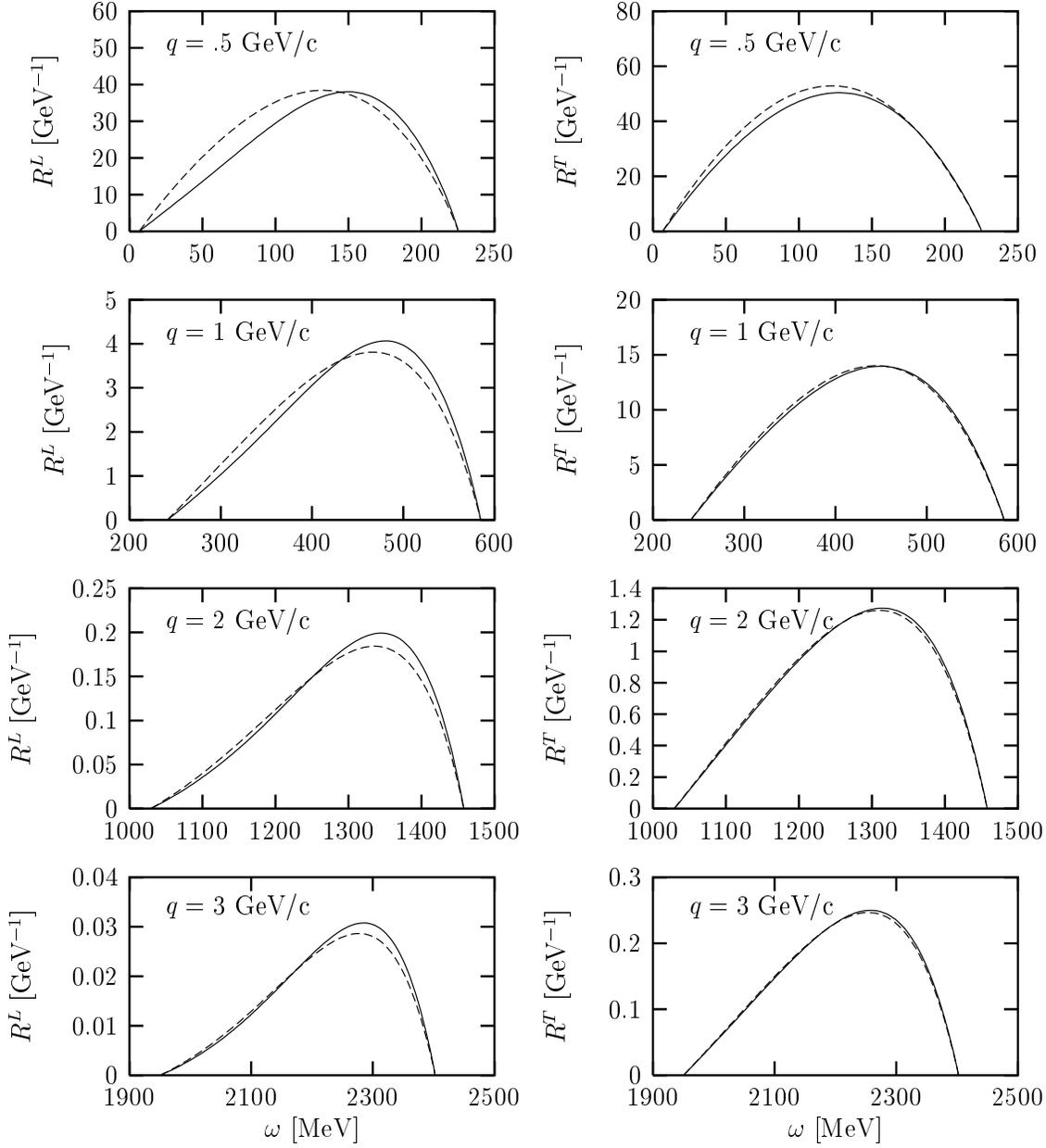}
\end{center}
\caption{Same as Fig.~7 but for the vertex correlation. Dashed: RFG responses;
solid: RFG + VC contribution.}
\end{figure}

In Fig.~9 we display the vertex correlation contribution to the
longitudinal and transverse responses by comparing the free RFG
responses (dashed) with the responses obtained including the VC
contribution (solid). As noted, the VC action, while substantial in
both the longitudinal and transverse channel, is actually dominant in
the former by roughly a factor of three.  This outcome relates to the
minor role played by the isoscalar contribution in the transverse
response, in turn due to the smallness of the isoscalar magnetic
moment.

The evolution with $q$ of the VC in the longitudinal and
transverse channels has been discussed at length 
in~\cite{Ama02}. Let us summarize the basic findings.  First, the
VC do not saturate quite as rapidly as the MEC, although their
behavior is rather similar and saturation again occurs somewhere
above $q=1$--1.5 GeV/c: 
thus, once more,
scaling of the first kind is
achieved at high momentum transfers for these contributions.
Moreover, similarly to the MEC case, for high $q$ the VC almost
vanish around the QEP.

Finally, the vertex correlations are found to {\em grow} with $k_F$,
much as the MEC do. From a semi-relativistic point of view, we find a
behavior that goes as $k_F^2$.  The basic conclusion is similar to
that made above for the seagull contribution and hence for the total
MEC at high $q$, namely, scaling of the second kind is badly broken by
effects that go roughly as $k_F^3$.

\begin{figure}[hp]
\begin{center}
\leavevmode
\def\epsfsize#1#2{0.9#1}
\epsfbox[100 195 500 705]{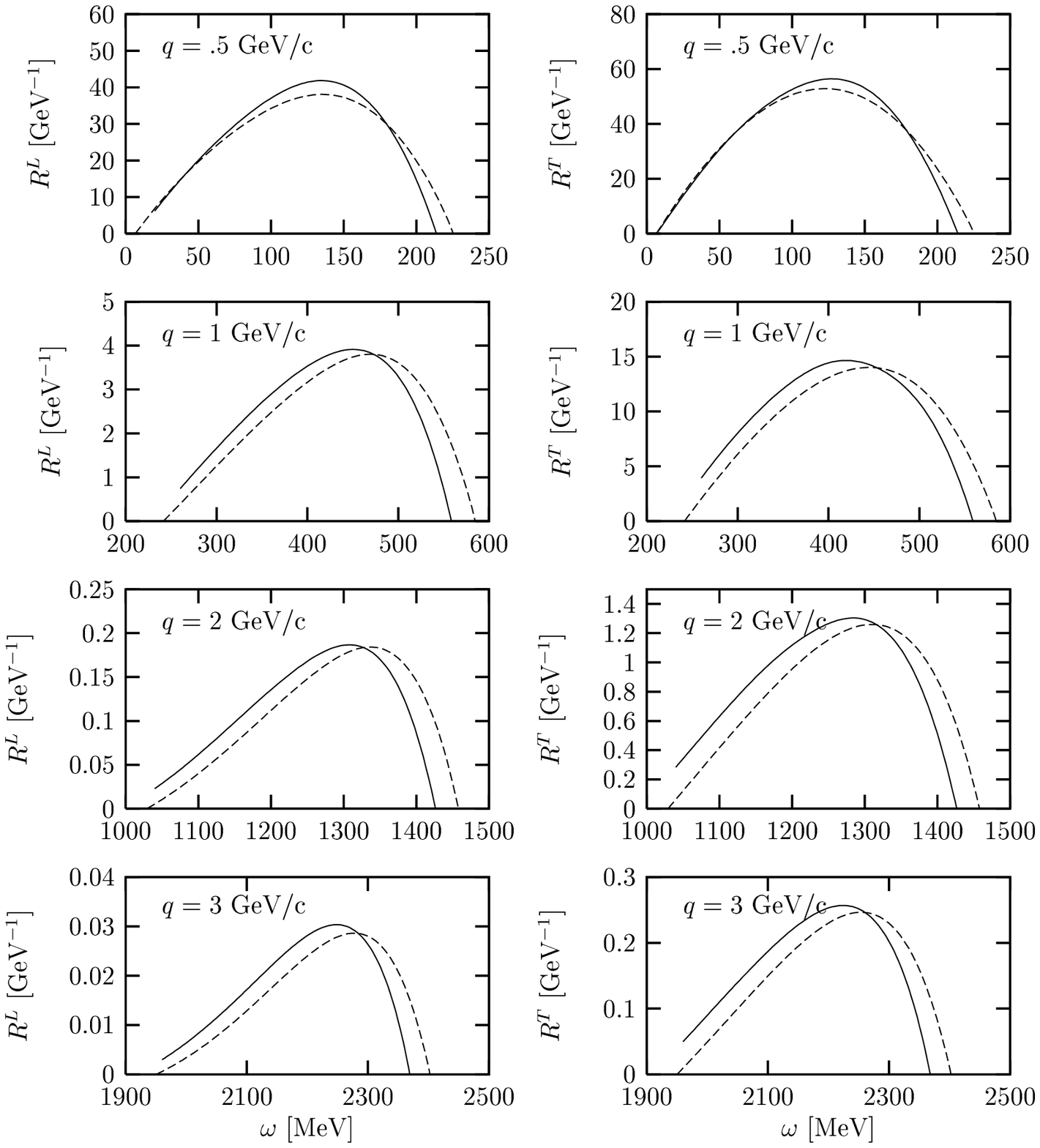}
\end{center}
\caption{Same as Fig.~7 but for the self-energy. Dashed: RFG responses; solid:
RFG + SE contribution.}
\end{figure}

\begin{figure}[hp]
\begin{center}
\leavevmode
\def\epsfsize#1#2{0.9#1}
\epsfbox[100 210 500 710]{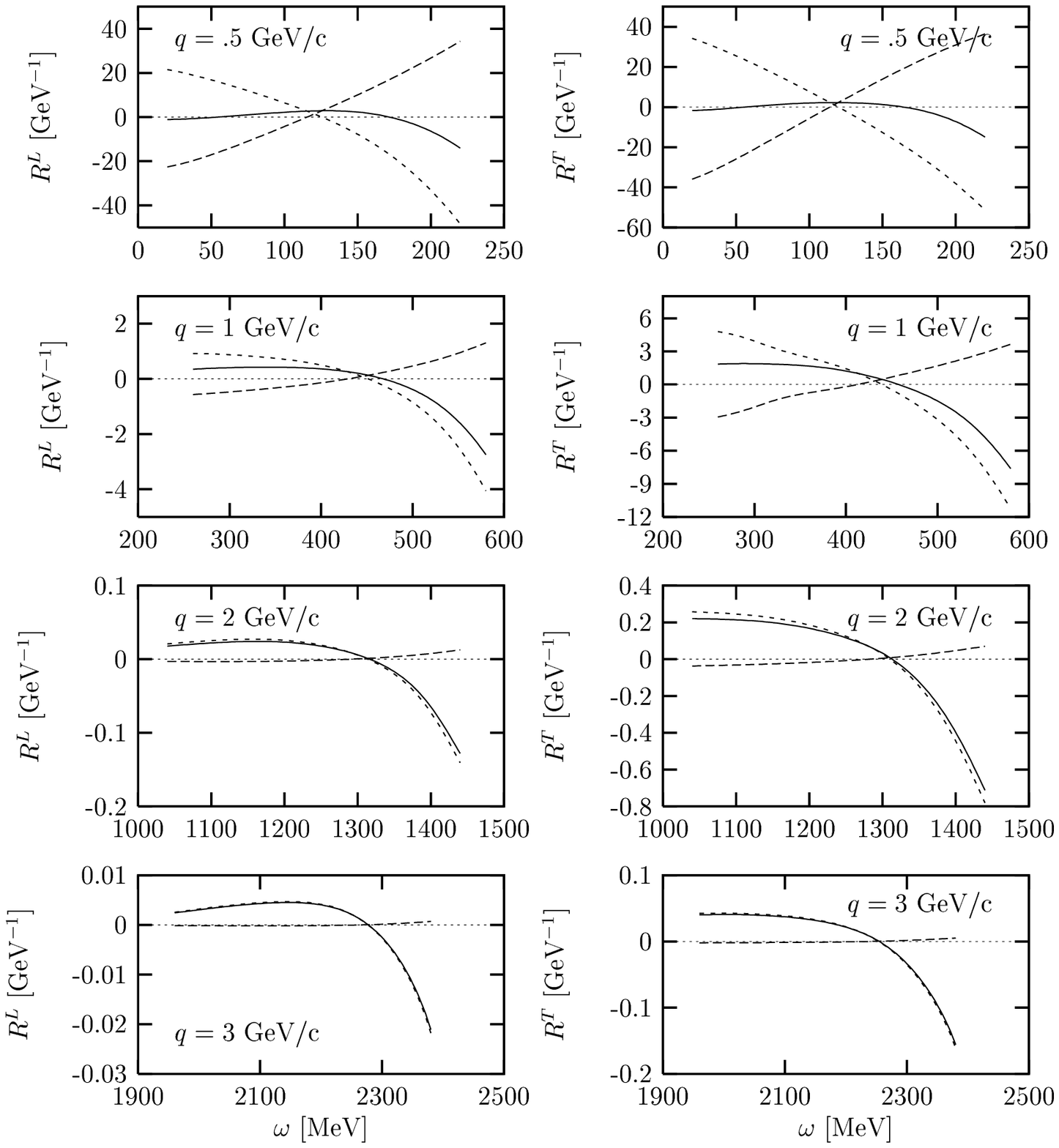}
\end{center}
\caption{Particle (dashed) and hole (short-dashed) contributions to
the longitudinal (left panels) and transverse (right panels)
self-energy. The solid line represents the total SE contribution.}
\end{figure}

The role played by the SE contribution is displayed in
Figs.~10 and 11.  In Fig.~10 we show the total RFG + SE
responses (solid line) compared with the free RFG responses
(dashed). Note that, in contrast with the MEC and vertex
correlations, which mostly contribute to only one channel (transverse
and longitudinal, respectively), the impact of the self-energy on
$R^L$ and $R^T$ is similar, leading in both cases 
to a softening of the responses for high $q$.

The separate particle (dashed) and hole (short-dashed) SE
contributions to the longitudinal and transverse responses are
presented in Fig.~11. Here, also the total SE contribution
(solid) is displayed. 
We observe that the self-energy contribution results from a quite
delicate cancellation between the responses having only the particle
or only the hole dressed (Fig.~11). This was already pointed out
in~\cite{Bar93} within the framework of a treatment in which
relativistic effects were partially incorporated and it is now
confirmed within a fully-relativistic context.

Whereas this cancellation is very substantial at $q$=0.5 GeV/c, as the
momentum transfer increases the imbalance between the two
contributions grows.  Indeed the response associated with the particle
self-energy is suppressed by the form factors and by the pion
propagator, but that coming from the hole self-energy is not. As a
result, for $q\geq$ 2 GeV/c the total self-energy response 
is almost entirely due to the hole dressing and induces a moderate
softening to the free response.  Note that the SE contribution does
not vanish on the borders of the response region. Moreover for high
values of $\omega$ (close to the upper border) it becomes very large
(Fig.~11) and yields a significant lowering of the upper $\omega$
limit in the responses. This clearly points to the insufficiency of a
first-order perturbative treatment in this kinematical region, an
effect already present in the partially relativized analysis
of~\cite{Bar93} and emphasized by our fully-relativistic calculation.
Therefore the summation of the full Fock series becomes necessary near
the upper boundary of the response.

\begin{figure}[htb]
\begin{center}
\leavevmode 
\def\epsfsize#1#2{0.9#1}
\epsfbox[170 495 470 690]{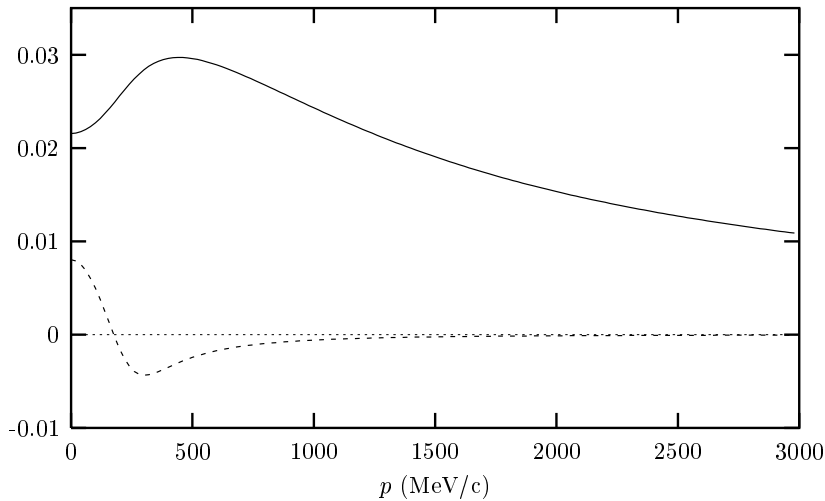}
\end{center} 
\caption{The on-shell self-energy $\Sigma_0(p)/E_\np$ defined in
eq.~(\protect\ref{Sigma_0}) (solid line) and the field-strength
renormalization function $\alpha(p)$ given in
eq.~(\protect\ref{alpha}) (dashed line) plotted versus the momentum
$p$.}
\end{figure}

\begin{figure}[hp]
\begin{center}
\leavevmode
\def\epsfsize#1#2{0.9#1}
\epsfbox[100 210 500 720]{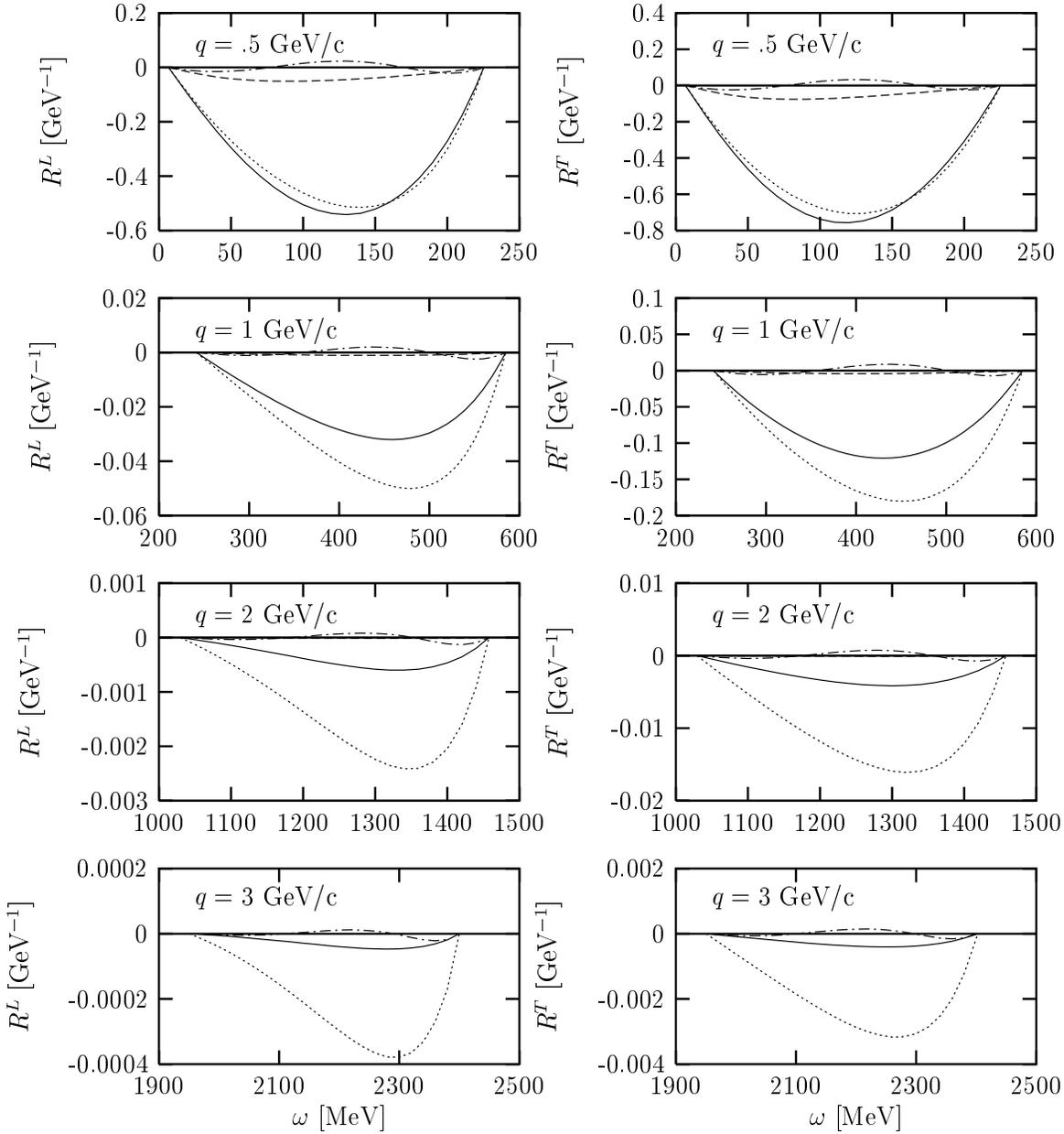}
\end{center}
\caption{The contribution of the renormalized self-energy current to
the longitudinal (left panels) and transverse (right panels) responses
plotted versus $\omega$.  The separate contributions of the current
$j^\mu_{RSE1}$ for the particle (solid) and hole (dotted) and of the
current $j^\mu_{RSE2}$ for the particle (dashed) and hole (dot-dashed)
are displayed.  }
\end{figure}

\begin{figure}[hp]
\begin{center}
\leavevmode
\def\epsfsize#1#2{0.9#1}
\epsfbox[100 210 500 710]{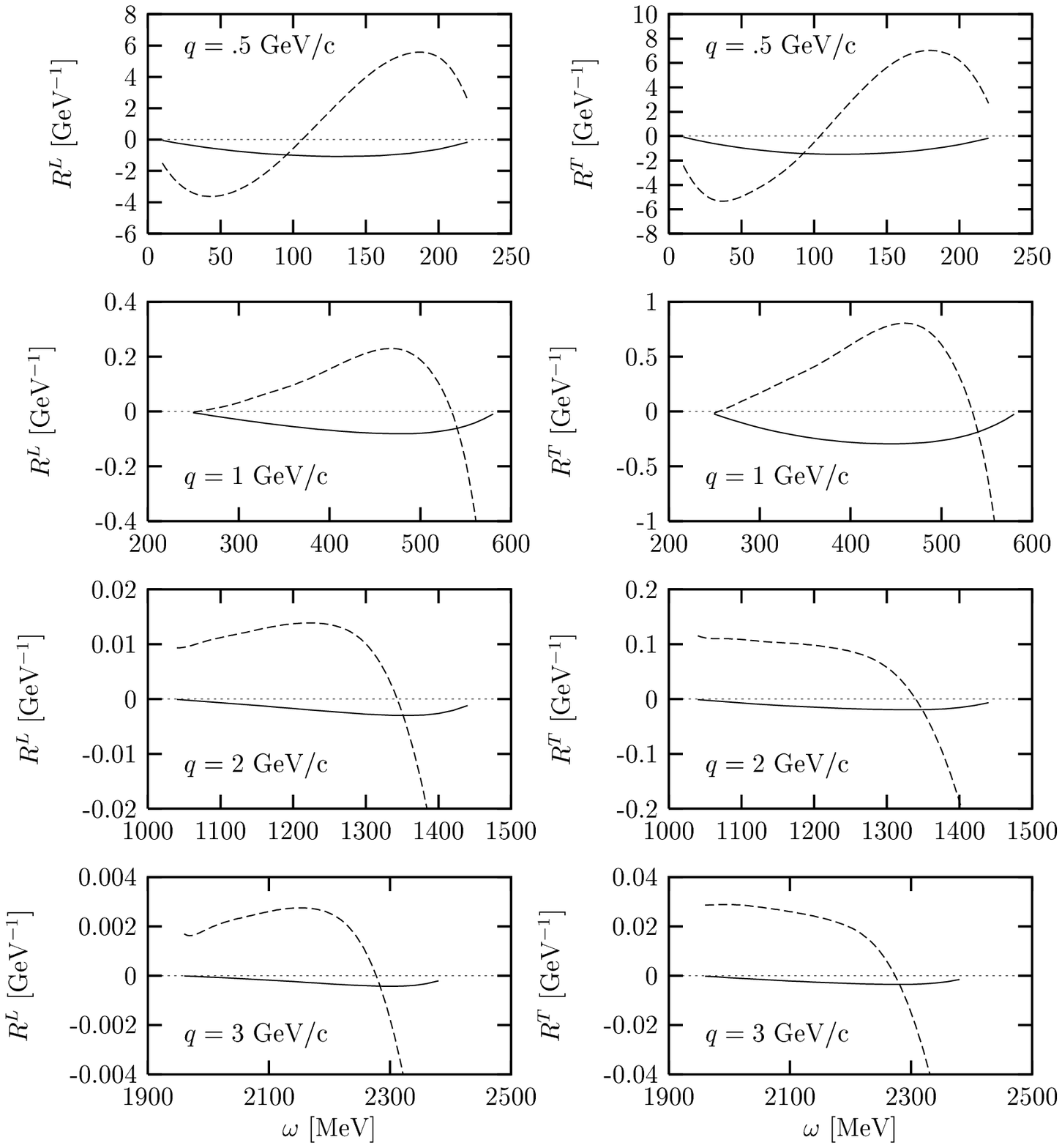}
\end{center}
\caption{The contributions of the first (solid) and second (dashed)
term in eq.~(\protect\ref{RSE-response}) to the longitudinal (left
panels) and transverse (right panels) responses.  }
\end{figure}

The analysis of the scaling and superscaling properties of the
self-energy correlations has been presented in~\cite{Ama02}.  
In accord with the above,
the particle contribution decreases with $q$, going to zero
at $q\simeq 2$ GeV/c, whereas the hole contribution, although also
decreasing with $q$ when not too high, saturates for $q\geq$1 GeV/c 
(see Fig.~11). 
As a result the total self-energy grows with $q$ in the
range $q$=0.5--2 GeV/c, then stabilizes typically at about 30-40\% of
the free response to the left of the QEP, thus inducing an important
softening of the longitudinal and transverse responses.  In summary,
again scaling of the first kind is achieved at momentum transfers
somewhat below 2 GeV/c. Finally, we also prove that the self-energy
relative contribution grows with $k_F$, although not uniformly in the
scaling variable $\psi$ (see~\cite{Don99a,Don99b,Mai02}) --- recall
that in the first-order analysis presented in this paper the edges of
the response region are not treated adequately for the self-energy
contribution and thus should not be taken too seriously. Where the
self-energy contribution is correctly modeled (away from the edges) we
again see breaking of second-kind scaling by roughly $k_F^3$.

In what follows we explore the impact on the responses of the new
currents $j^\mu_{RSE1}$ and $j^\mu_{RSE2}$ that arise from the
enhancement of the lower components of the spinors and from the field
strength renormalization $\sqrt{Z_2(\np)}$, respectively.  In Fig.~12
we show the on-shell self-energy (solid curve) and the field strength
renormalization function (dashed curve) given by eqs.~(\ref{Sigma_0})
and (\ref{alpha}), respectively. The explicit expressions for
$\Sigma_0(p)$ and $\alpha(p)$ are derived in Appendix E.  The
$\Sigma_0(p)$ obtained here is in good agreement with the results
of~\cite{Ana81} and its effect on the single-particle energy in
eq.~(\ref{energy}) and on the effective mass in
eq.~(\ref{m-expansion}) is very small (less than $\sim 3\%$).  Note
that $\alpha$, which is linked to the current $j^\mu_{RSE2}$ of
eq.~(\ref{RSE2}), is much smaller than $\Sigma_0(p)/E_\np$, which
enters in $j^\mu_{RSE1}$ through eq.~(\ref{RSE1}).  Thus the effect of
the enhancement of the lower components of the spinors dominates over
the field-strength renormalization. This is very clearly seen in
Fig.~13, where the various contributions to the longitudinal and
transverse responses stemming from $j^\mu_{RSE1}$ and $j^\mu_{RSE2}$
are displayed versus the transferred energy $\omega$ for momentum
transfer $q=$ 0.5, 1, 2 and 3 GeV/c.  It is evident that the effect of
$j^\mu_{RSE2}$ is negligible with respect to that of $j^\mu_{RSE1}$.
The separate contributions of the particle and hole self-energies are
also shown: as $q$ increases the contribution of the particle is
suppressed, whereas the one of the hole survives.

In Fig.~14 we compare the contribution to the longitudinal and
transverse responses due to renormalization of the wave functions
(solid) with that arising from renormalization of the energies
(dashed).  The effect linked to modification of the energy due to the
medium is the dominant one, the other being very small, especially for
large values of $q$.

\begin{figure}[htp]
\begin{center}
\leavevmode
\def\epsfsize#1#2{0.9#1}
\epsfbox[100 210 500 710]{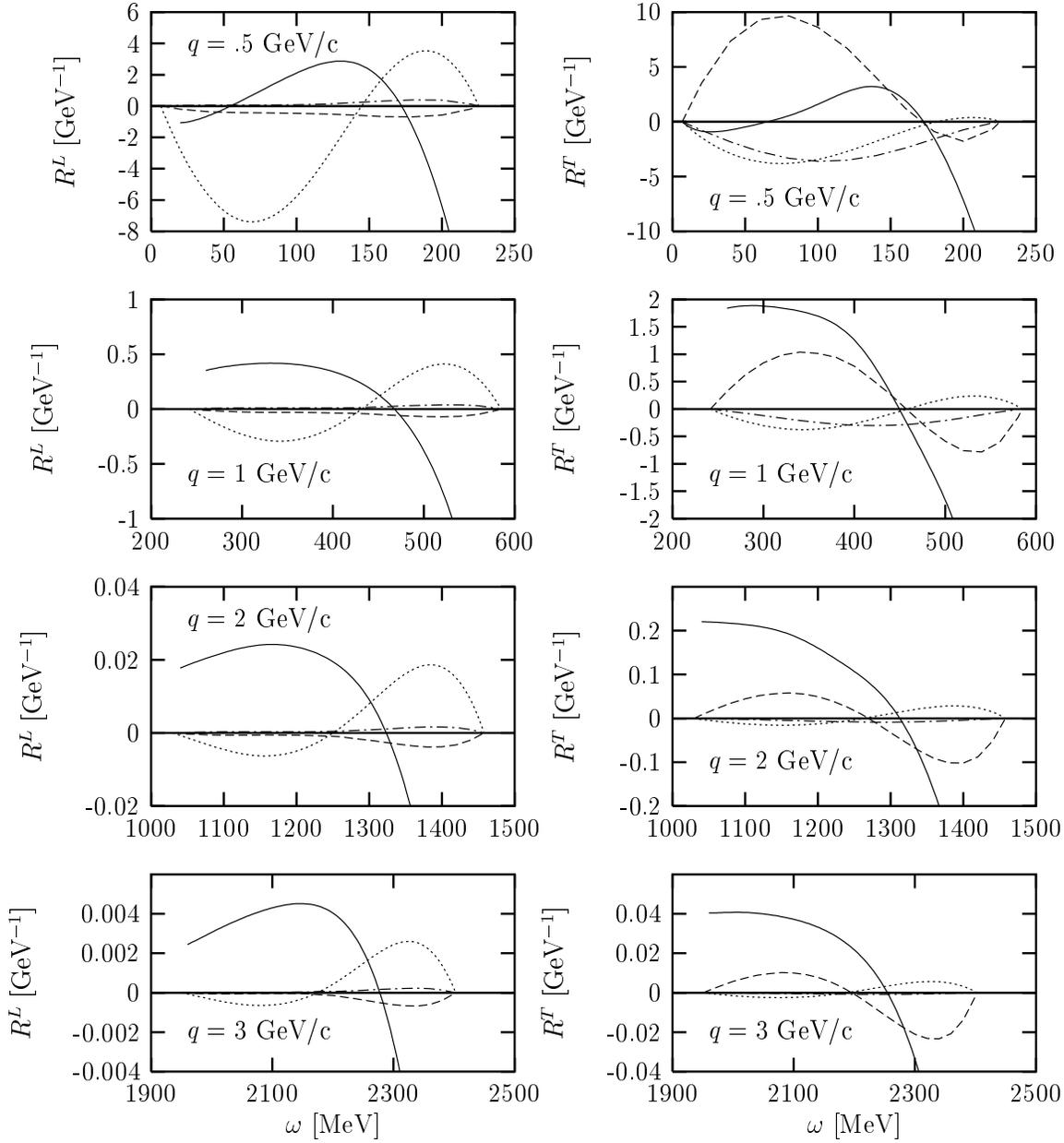}
\end{center}
\caption{Separate pion-in-flight (dot-dashed), seagull (dashed),
vertex correlation (dotted) and self-energy (solid) contributions to
the longitudinal (left panels) and transverse (right panels)
responses.  }
\end{figure}

\begin{figure}[htp]
\begin{center}
\leavevmode
\def\epsfsize#1#2{0.9#1}
\epsfbox[100 270 500 750]{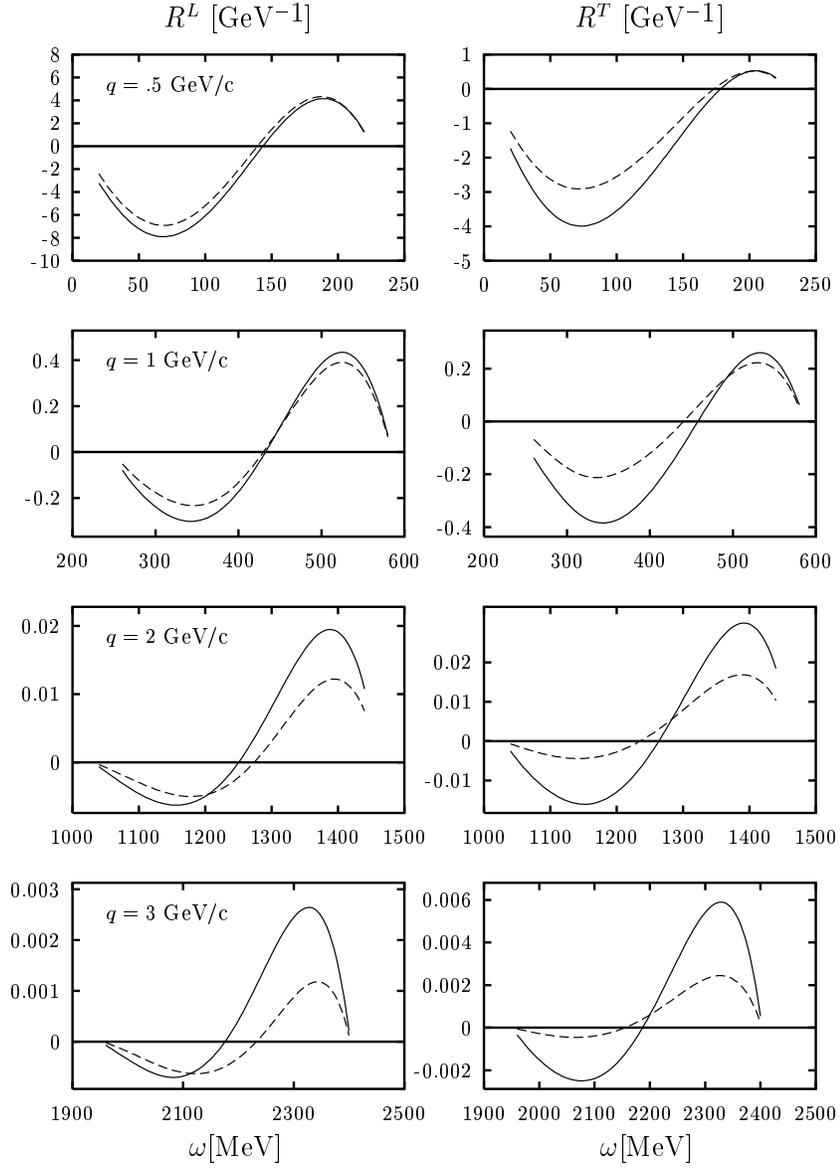}
\end{center}
\caption{Longitudinal and transverse vertex correlation responses 
versus $\omega$ in the pseudovector (solid) and pseudoscalar (dashed)
$\pi$-$N$ coupling.}
\end{figure}

\begin{figure}[htp]
\begin{center}
\leavevmode
\def\epsfsize#1#2{0.9#1}
\epsfbox[100 210 500 710]{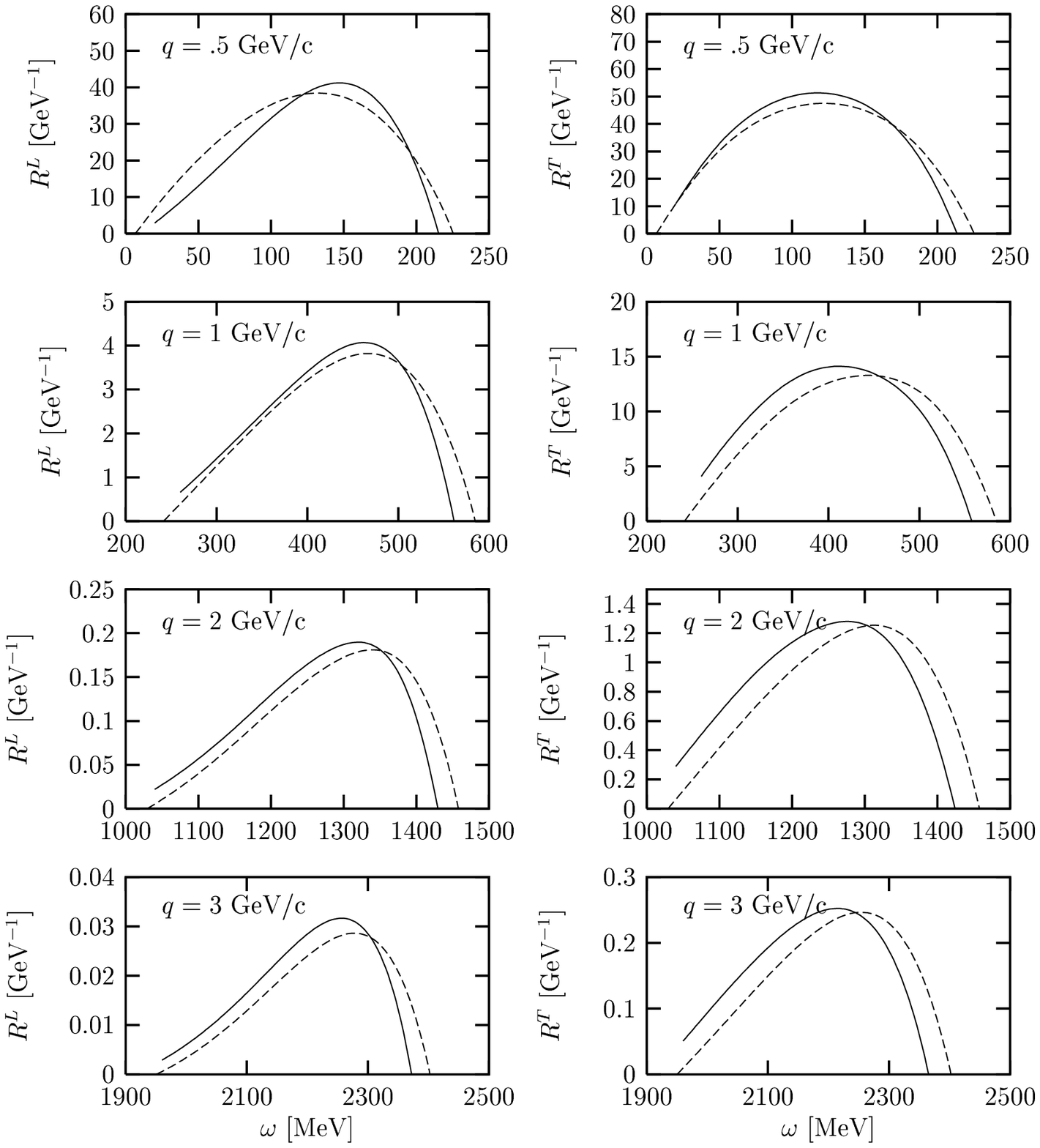}
\end{center}
\caption{Longitudinal and transverse responses versus $\omega$
including all first-order contributions (solid) compared with the free
result (dashed).  }
\end{figure}

To complete this section we display in Fig.~15 the separate
contributions of seagull (dashed), pion-in-flight (dot-dashed), VC
(dotted) and SE (solid) to the longitudinal and transverse responses.
Worth pointing out is the oscillatory behavior versus $\omega$ of the
vertex correlations, which induces a hardening of the responses.  In
addition the seagull and vertex correlations tend to {\em cancel} in
the transverse channel, especially for low values of $q$, whereas for
higher $q$ the MEC dominate.  Note that in the T channel both the
seagull and VC exactly vanish at the same value of $\omega$, the
latter coinciding with the QEP for high momentum transfers, as said
above. It is also important to point out that the net effect
introduced by the SE contribution is in general the largest one for
transfer momentum values $q\geq 1$ GeV/c. Within the L channel, the
pionic correlations (VC and SE) clearly dominate over the MEC. In the
transverse channel, apart from the SE contribution which seems to
dominate for high $q$, the seagull term is clearly more important than
the VC one, whereas the pion-in-flight only enters for $q$ not very
high.

Up to now we have considered a pseudovector coupling for the
pion, eq.~(\ref{HamPV}).
We now shortly investigate the effects 
on the responses of using a pseudoscalar pion-nucleon Hamiltonian
\begin{equation}\label{HamPS}
{\cal H}^{(PS)}_{\pi NN} = ig\overline{\psi}\gamma_5\phi_a\tau_a \psi 
\end{equation}
instead of the pseudovector one.  For on-shell nucleons the
Hamiltonians in eqs.~(\ref{HamPV}) and (\ref{HamPS}) are equivalent
provided $f/m_\pi=g/(2m)$, but for off-shell nucleons this is not
so. Among the diagrams considered in our approach the only one
involving off-shell nucleons is the one associated with the vertex
correlations (Figs.~6e and f). Hence in Fig.~16 we compare the VC
contribution to $R^L$ and $R^T$ obtained with the pseudovector (solid)
and pseudoscalar (dashed) couplings. The difference between the two is
especially sizable in the transverse channel (where the impact of VC
is smaller) and increases with the momentum transfer.

In conclusion, in Fig.~17 we display the total responses in first
order of perturbation theory and compare them with the zeroth-order
ones (free responses) for several momentum transfers.  Here one
assesses the impact of the global two-body current contribution to the
responses.  First the overall effect of the two-body currents appears
sufficiently modest to justify our first-order treatment.  Next the
softening at large $q$ appears to be common to both L and T channels,
whereas at low $q$ the longitudinal response displays a hardening that
is absent in the transverse one.  Also evident is the already-noted
nearly vanishing of the two-body correlation contribution at the peak
of the free responses.  Finally the unrealistic dominance of the
self-energy contribution on the upper border is apparent. Summarizing,
the impact of the different first-order contributions --- MEC, vertex
correlations and self-energy --- to the total responses are all
comparable in size in the transverse channel (in the longitudinal one
the MEC are negligible), their relative contribution ranging from
$\sim$5 to $\sim$15\% depending upon the kinematics and the Fermi
momentum.


\section{Parity-violating electron scattering}
\label{sec:PV}


In this section we deal with the parity-violating (PV) effects arising
from the weak interaction between the electron and the nucleus. Such
effects, which are negligible in unpolarized electron processes, can
be brought to evidence by measuring the asymmetry associated with
longitudinally polarized electrons having opposite helicities, namely
\begin{equation}
{\cal A} = \frac{d\sigma^+-d\sigma^-}{d\sigma^++d\sigma^-}\ .
\label{Asym}
\end{equation}
In this case the purely electromagnetic cross sections cancel out and
one is left with the interference between the electromagnetic and
neutral weak currents, corresponding to the exchange of a photon and a
$Z^0$, respectively.

An important motivation of parity-violating experiments (see, for
example,~\cite{Mus94} for a general review and~\cite{Don92} for the
foundations of PV quasielastic scattering) is the measurement of the
single-nucleon form factors, in particular the strange and axial ones:
for this reason most experiments are presently being carried out on
light nuclei, where the uncertainties associated with the nuclear
model are minimized. Other motivations exist for such studies:
specifically, as discussed in the following, the PV response functions
display a different sensitivity to nuclear correlations compared with
the parity-conserving ones: hence they could not only shed light on
the part of the problem concerned with nucleon (and meson) structure,
but also are being used as a test of nuclear models. In the present
work we provide no details for the underlying formalism used in PV
electron scattering --- those discussions can be found
in~\cite{Mus94}. Our focus here is rather to place in context the
expectations for PV electron scattering of what role the modeling
discussed above plays.

\begin{figure}[ht]
\begin{center}
\leavevmode
\def\epsfsize#1#2{0.9#1}
\epsfbox[100 430 500 740]{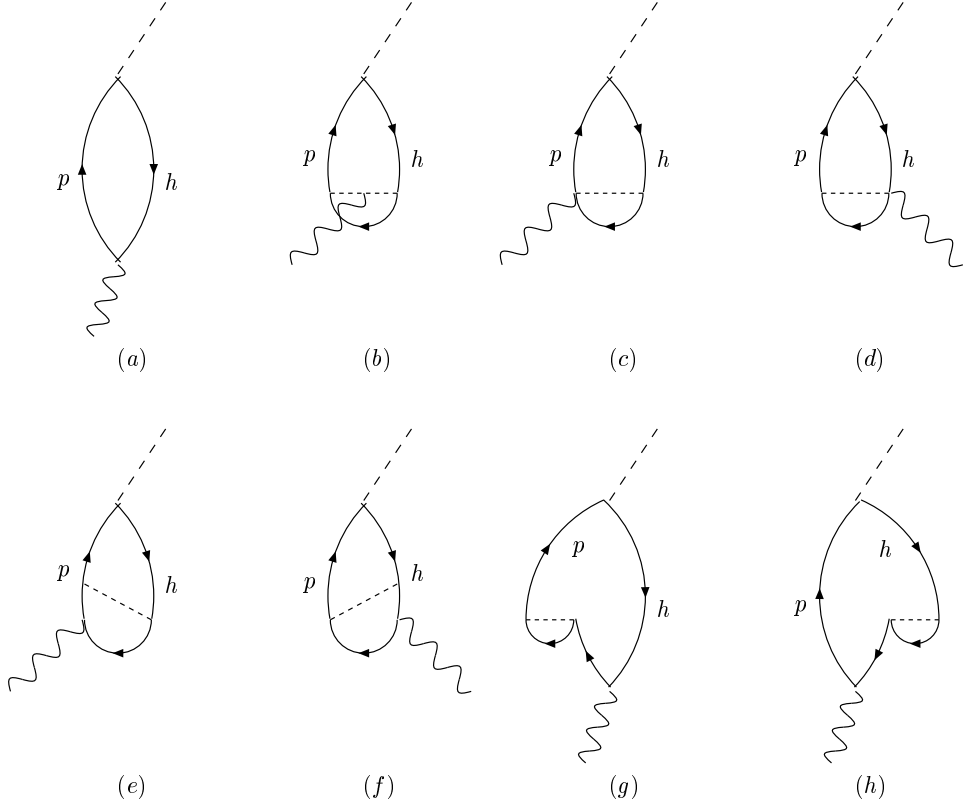}
\end{center}
\caption{
Feynman diagrams of the free (a) and first-order pion-in-flight (b), 
seagull (c and d), vertex correlation (e and f) and self-energy (g and h) 
PV polarization propagator. The external wavy and dashed lines represent a
photon and a
$Z^0$ boson, respectively.}
\end{figure}

\subsection{General formalism}
\label{sec:PVform}

The cross section for scattering of a polarized 
electron with helicity $h$ reads
\begin{eqnarray}
\frac{d\sigma^{(h)}}{d\Omega'_e d\omega}
=\frac{\varepsilon'}{\varepsilon}
\left(\frac{2\alpha^2}{Q^4} \eta_{\mu\nu} W^{\mu\nu}+
\frac{\alpha G}{2{\sqrt 2}\pi Q^2} {\tilde\eta_{\mu\nu}}{\tilde W^{\mu\nu}}
\right)\ .
\label{eq:sigh}
\end{eqnarray}
In eq.~(\ref{eq:sigh}) $G$ is the Fermi constant, $\eta_{\mu\nu}$,
$W^{\mu\nu}$ are the leptonic and hadronic electromagnetic tensors
defined in Section~\ref{sec:PCform} and ${\tilde\eta_{\mu\nu}}$,
${\tilde W^{\mu\nu}}$ are the tensors arising from the $\gamma-Z^0$
interference.  Here terms containing two weak currents have been
neglected. The interference tensors read
\begin{equation}
{\tilde\eta_{\mu\nu}}=(a_V-h a_A)\left(K_\mu K'_\nu+K'_\mu K_\nu-g_{\mu\nu}
K\cdot K'\right)+(a_A-h a_V) i \epsilon_{\mu\nu\rho\sigma} K^\rho K'^\sigma
\end{equation}
with $a_A=-1$ and $a_V=4\sin^2\theta_W-1$, $\theta_W$ being the
weak mixing angle, and
\begin{equation}
\tilde W^{\mu\nu}=
\overline{\sum_i}\sum_f\langle f|\hat{J}^\mu_{em}(Q) |i\rangle^\ast
\langle f|\hat{J}^\nu_{wn}(Q) |i\rangle \delta(E_i+\omega-E_f) \ ,
\label{eqPV3}
\end{equation}
$\hat{J}^\mu_{em}(Q)$ and $\hat{J}^\nu_{wn}(Q)$ being the nuclear 
electromagnetic and weak neutral currents, respectively.

When the difference of cross sections corresponding to opposite electron 
helicities is taken, the electromagnetic term in 
eq.~(\ref{eq:sigh}) cancels out and 
the resulting PV expression reads
\begin{eqnarray}
\left(\frac{d\sigma}{d\Omega'_e d\omega}\right)_{PV}
&\equiv& \frac{1}{2}
\left(\frac{d\sigma^{(+)}}{d\Omega'_e d\omega}-
\frac{d\sigma^{(-)}}{d\Omega'_e d\omega}
\right)
\nonumber\\
&=&
{\cal A}_0 \sigma_M\left[v_L R^L_{AV}(q,\omega) 
+ v_T R^T_{AV}(q,\omega)+v_T´ R^{T'}_{VA}(q,\omega)\right]
\ ,
\label{eqPV1}
\end{eqnarray}
where
\begin{equation}
{\cal A}_0 = \frac{G |Q^2|}{2\sqrt{2}\pi\alpha}\ ,
\end{equation}
$\sigma_M$ is the Mott cross section in eq.~(\ref{eq:Mott}), the leptonic
kinematical factors
$v_L$ and $v_T$ are given by eqs.~(\ref{eq:vl},\ref{eq:vt}) and
\be
v_{T'} = \tan\frac{\theta_e}{2}
\sqrt{-\left(\frac{Q^2}{q^2}\right)+\tan^2\frac{\theta_e}{2}} \, .
\ee
In terms of nuclear response functions the asymmetry in eq.~(\ref{Asym}) reads
\be
{\cal A} = {\cal A}_0 \frac{v_L R^L_{AV}+v_T R^T_{AV}+v_{T'}R^{T'}_{VA}}
{v_L R^L+v_T R^T}\ .
\label{Asym1}
\ee
 
\subsection{PV response functions}
\label{sec:PVresp}

The PV response functions appearing in eq.~(\ref{eqPV1}) are linked 
to the interference hadronic tensor in eq.~(\ref{eqPV3}) by the following 
relations:

\begin{eqnarray}
R^L_{AV}(q,\omega)
&=&
a_A \left(\frac{q^2}{Q^2}\right)^2\left[
\tilde W^{00}-\frac{\omega}{q}(\tilde W^{03}+\tilde W^{30})+
\frac{\omega^2}{q^2}\tilde W^{33}
\right]\label{eq2PVa} \\
R^T_{AV}(q,\omega)
&=&
a_A\left(\tilde W^{11}+\tilde W^{22}\right) \\
R^{T'}_{VA}(q,\omega)
&=&
-i a_V\left(\tilde W^{12}-\tilde W^{21}\right) \ .
\label{eq2PVb}
\end{eqnarray}
The subscript $AV$ in the PV responses denotes interferences of
axial-vector leptonic currents with vector hadronic currents, and the
reverse for the subscript $VA$.

Within the context of the RFG model the interference hadronic tensor is
\begin{equation}
\tilde W^{\mu\nu}
=\frac{3Z}{8\pi k_F^3q}
\int_{h_0}^{k_F} h dh (\omega+E_{\nh}) 
\int_0^{2\pi}d\phi_h \sum_{s_p,s_h}
\frac{m^2}{E_{\np}E_{\nh}}
2{\rm Re}\,
\left[j^\mu_{em}(\np,\nh)^* j^{\nu}_{wn}(\np,\nh)
\right] \ , 
\label{tilwrfg}
\end{equation}
where the electromagnetic current $j^\mu_{em}$ includes both the
single nucleon one-body and the two-body (MEC and correlation)
currents discussed in the previous section, {\em  i.e.}
$j^\mu_{em}=j^\mu_{OB}+j^\mu_{MEC}+j^\mu_{cor}$.  In this work we
include in the weak neutral current only the one-body contribution
(see Fig.~18), namely
\begin{equation}
j^\nu_{wn}(\np,\nh)=
\overline{u}({\bf p})
\left( \tilde F_1 \gamma^\nu
      +i\frac{\tilde F_2}{2m}\sigma^{\nu\rho}Q_\rho
      +\tilde G_A \gamma_5\gamma^\nu 
\right) u({\bf h})\ ,
\end{equation}
where the Pauli and Dirac form factors are
\ba
\tilde F_1 &=& \frac{\tilde G_E+\tau\tilde G_M}{1+\tau}
\\
\tilde F_2 &=& \frac{\tilde G_E-\tilde G_M}{1+\tau}\ .
\ea
Thus we neglect
the direct coupling of a $Z^0$ to the
pion 
(important clues for the understanding of the weak-neutral sector of
the MEC should be found in the study of pion electroproduction on the
nucleon, where a $Z^0$ is exchanged with the nucleon.  This topic has
recently been investigated in~\cite{Rek02}).

Within the standard model at tree level the weak neutral form factors 
are linked to the electromagnetic ones by the following relations
(possible contributions from the strange quark are neglected ---
these can be included in a straightforward way~\cite{Don92,Mus94} and
do not provide the primary focus of the present discussions):
\ba
\tilde G_{Ep(n)} &=& \beta_V^p G_{Ep(n)} + \beta_V^n G_{En(p)} 
\\
\tilde G_{Mp(n)} &=& \beta_V^p G_{Mp(n)} + \beta_V^n G_{Mn(p)} 
\\
\tilde G_{Ap(n)} &=& \beta_A^p G_{Ap(n)} + \beta_A^n G_{An(p)} \ ,
\ea
where
\ba
\beta_V^p = \frac{1}{2} \left(1-4\sin^2\theta_W\right)\ ,\ \ \ 
\beta_V^n =\beta_A^n =- \beta_A^p = \frac{1}{2}\ .
\ea
The one-body contribution to the three PV responses can be evaluated
analytically in RFG, yielding (see, for example~\cite{Don92})
\ba
R^{L,T}_{AV}(q,\omega) &=& a_A R_0(q,\omega) 
\left[\tilde U^{L,T}_p(q,\omega)+\tilde U^{L,T}_n(q,\omega)
\right]
\\
R^{T'}_{VA}(q,\omega) &=& a_V R_0(q,\omega) 
\left[\tilde U^{T'}_p(q,\omega)+\tilde U^{T'}_n(q,\omega)
\right] \, ,
\label{RLTPV}
\ea
where $R_0$ has been defined in eq.~(\ref{R0}) and
\ba
\tilde U^L_{p(n)}(q,\omega) &=& \frac{\kappa^2}{\tau} \left\{G_{Ep(n)}
\tilde G_{Ep(n)} +
\frac{\Delta}{1+\tau}\left[G_{Ep(n)}\tilde G_{Ep(n)}+
\tau G_{Mp(n)}\tilde G_{Mp(n)}\right]\right\} 
\\
\tilde U^T_{p(n)}(q,\omega) &=& 2\tau G_{Mp(n)}\tilde G_{Mp(n)} +
\frac{\Delta}{1+\tau}\left[G_{Ep(n)}\tilde G_{Ep(n)}+
\tau G_{Mp(n)} \tilde G_{Mp(n)}\right] 
\\
\tilde U^{T'}_{p(n)}(q,\omega) &=& 2\sqrt{\tau(1+\tau)} 
G_{Mp(n)}\tilde G_{Ap(n)} (1+\tilde\Delta)
\ ,
\ea
with $\Delta$ given by eq.~(\ref{Delta}) and
\be
\tilde\Delta \equiv \frac{1}{\kappa} \sqrt{\frac{\tau}{1+\tau}}
\left[\frac{1}{2}\left(\varepsilon_F+\varepsilon_0\right)
+\lambda\right]-1\ .
\ee
The two-body contributions  involve instead 
multidimensional integrals, to be numerically evaluated.

\subsection{Results}
\label{sec:PVresults}

In this section we analyze the PV response functions labeled L, T and
T$^\prime$ and the associated asymmetry for various values of the
momentum transfer. In~\cite{Ama96a} results for the PV responses in a
relativized continuum shell model where presented in the impulse
approximation for finite, closed shell nuclei. Two-body currents where
not included in that calculation. In~\cite{Bar93,Bar94} a
semi-relativistic analysis of the PV responses has been presented,
showing the dominance of pionic correlations in the longitudinal
channel. Here we perform a fully-relativistic calculation, which
confirms the above findings, and extend them to higher values of the
momentum transfer.

\begin{figure}[htp]
\begin{center}
\leavevmode
\def\epsfsize#1#2{0.9#1}
\epsfbox[100 265 500 745]{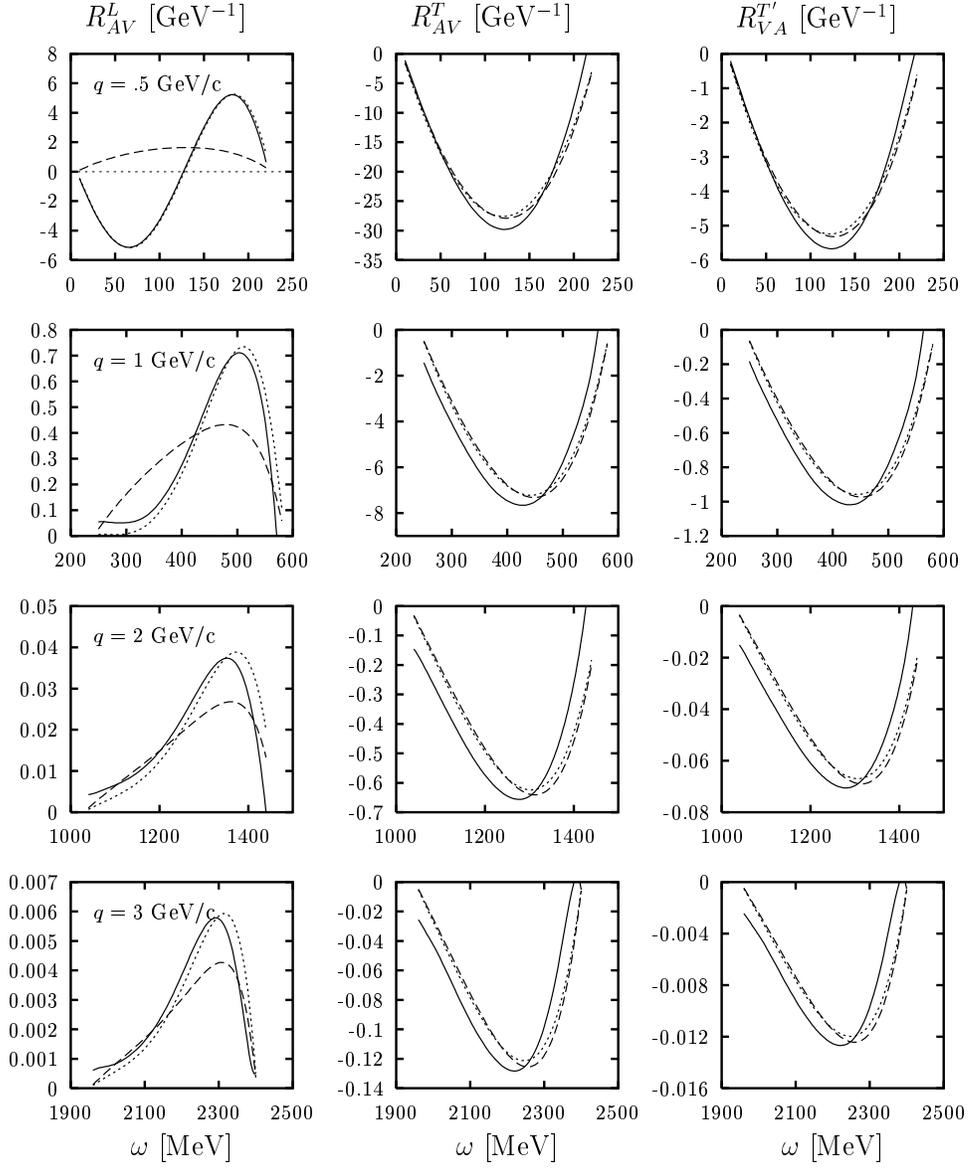}
\end{center}
\caption{The longitudinal (left panels), transverse (central panels)
and axial (right panels) PV responses plotted versus $\omega$.  Dashed
line: one-body contribution; dotted line: one-body+MEC+VC; solid
line: total (including SE).}
\end{figure}

\begin{figure}[hp]
\begin{center}
\leavevmode
\def\epsfsize#1#2{0.9#1}
\epsfbox[100 265 500 745]{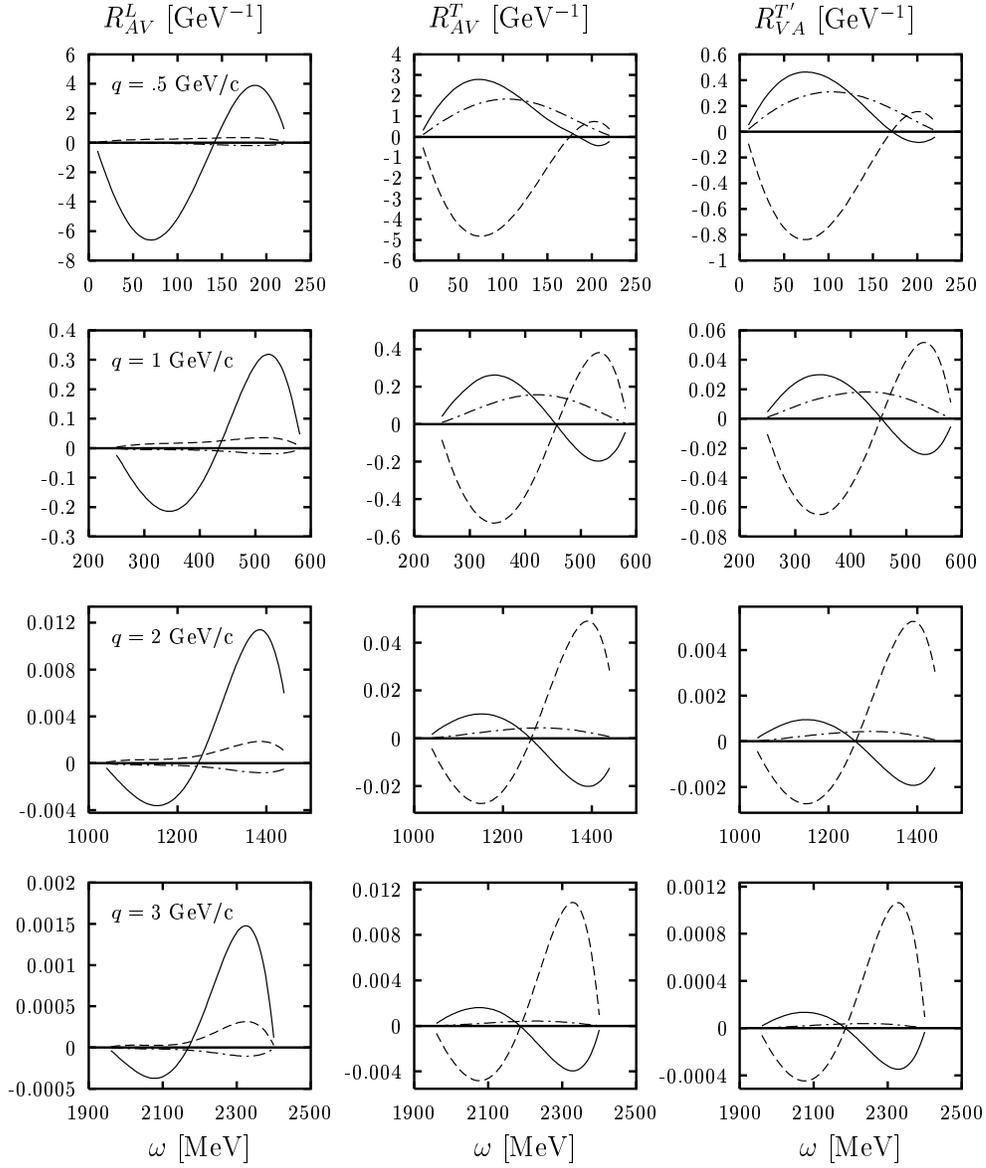}
\end{center}
\caption{Separate contributions to the PV longitudinal (left panels),
transverse (central panels) and axial (right panels) responses plotted
versus $\omega$.  Solid: VC; dashed: seagull; dot-dashed:
pion-in-flight.  }
\end{figure}

\begin{figure}[htp]
\begin{center}
\leavevmode
\def\epsfsize#1#2{0.9#1}
\epsfbox[100 265 500 745]{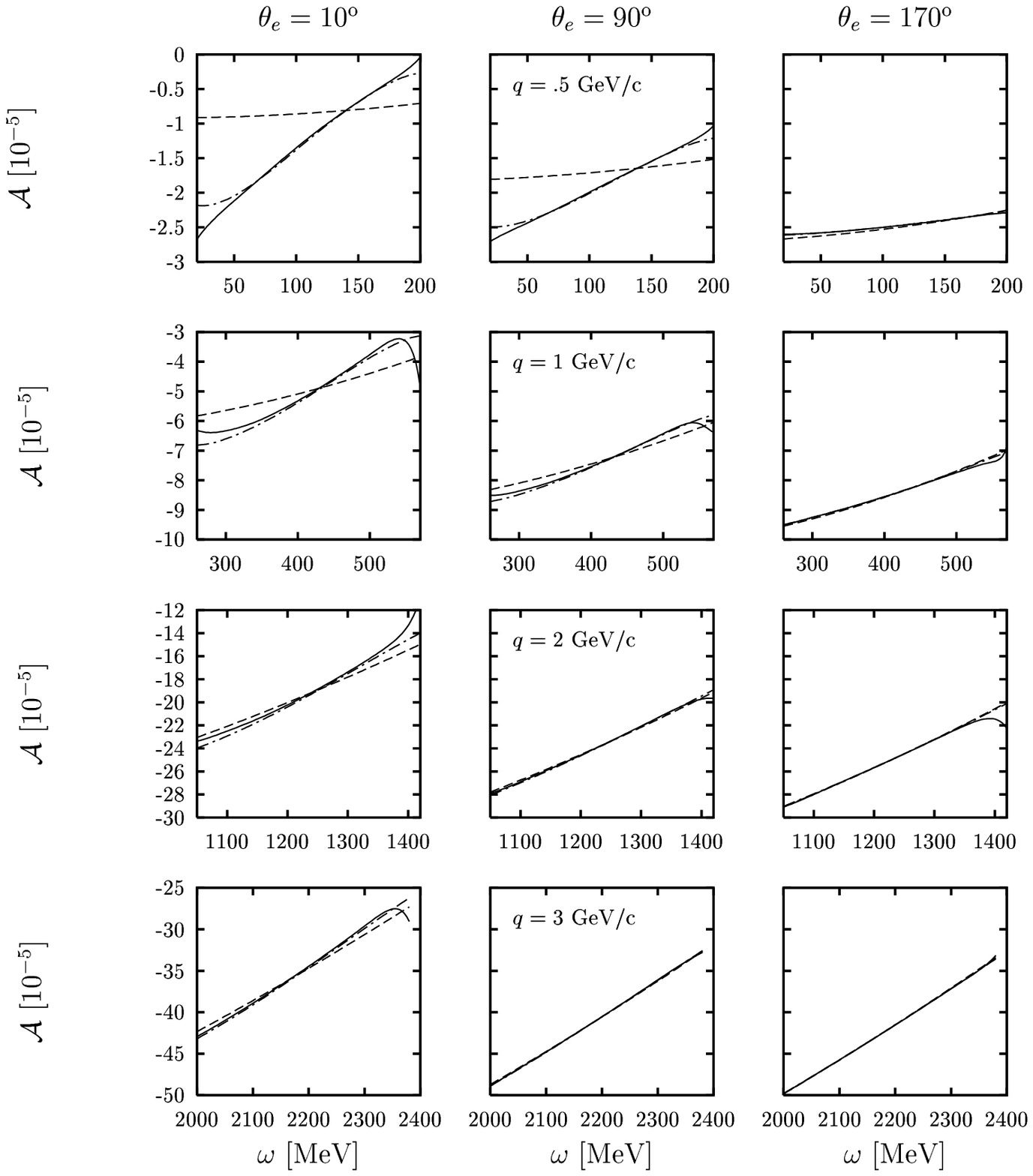}
\end{center}
\caption{The PV asymmetry displayed versus $\omega$ for various values of 
the momentum transfer $q$ and the scattering angle $\theta_e$.
Dashed: one-body; dot-dashed: one-body+MEC+VC; solid: total.
}
\end{figure}

In Fig.~19 we display the PV responses for four values of the momentum
transfer $q$.  The dashed line corresponds to the free RFG, the dotted
line includes MEC and vertex correlations, whereas the solid line also
includes the self-energy contribution.

One observes that
in the T and T$^\prime$ channels (central and right
columns) the main effect arises from the self-energy, which tends to
soften the response function, similarly to what happens in the
parity-conserving case, whereas the MEC and VC effect is very
tiny.  Note also that the axial response is proportional to the
transverse one, the factor between the two being roughly
$\sqrt{1+1/\tau} a_V G_A^{(1)}/G_M^{(1)}$: this agrees with the
conjecture of~\cite{Alv01,Bar94}, which is proven here to be
valid within a fully-relativistic context.  It also appears that the
self-energy contribution increases in going from $q=500$ MeV/c to $1$
GeV/c, then saturates for higher values of $q$ (thus scaling of first kind
is fulfilled). This is due to the same particle-hole cancellation
mechanism occurring in the electromagnetic case.

In the longitudinal channel the self-energy gives instead a very small
contribution compared with the MEC and vertex correlations. The effect
of the SE contribution is found to increase with $q$, but always
remains smaller than the one arising from the other correlations.
Indeed the one-body longitudinal response is suppressed due to a
delicate cancellation between the isoscalar and isovector
responses~\cite{Don92}. Physically this occurrence reflects the fact
that the electric form factor in one of the two vertices of diagram
18a is always very small, for both protons and neutrons.  When isospin
correlations are taken into account this balance can be disrupted, as
pointed out in~\cite{Bar93,Bar94,Bar96}, thus yielding the large
effects observed in Fig.~19. Indeed in the diagrams 18e-f a proton can
be converted into a neutron, leading to two large couplings, $G_{Ep}$
and $\tilde G_{En}$. Using different language, the MEC and VC are more
effective than the SE in $R^L_{AV}$, since they act differently in the
two isospin channels. Indeed the VC carries a factor -3 in the
isoscalar response and +1 in the isovector one and the MEC are purely
isovector, whereas the SE has almost the same impact in the two
channels.

This is clearly 
illustrated
in Fig.~20, where the separate seagull (dashed),
pion-in-flight (dot-dashed) and VC (solid) contributions are
displayed.  In the L channel the role of MEC is almost negligible, in
agreement with the findings for the electromagnetic $R^L$, whereas the
effect of the vertex correlations is dominant, especially at small
values of $q$.  In the T and T$^\prime$ channels the balance
between MEC and VC is similar to that occurring for the electromagnetic
$R^T$ (see Fig.~15): the pion-in-flight gives the smallest
contribution, particularly for large $q$, whereas the seagull
dominates for all $q$ and tends to cancel the VC contribution.
Note also that the seagull and VC vanish exactly at the same
value of $\omega$, which, for high $q$, coincides with the QEP.

Since the three PV responses are not at present experimentally
separable, we now explore the effect of 
the pionic physics
on the asymmetry in eq.~(\ref{Asym1}).  
In Fig.~21 we show ${\cal A}$ at various
values of the momentum transfer $q$ and of the electron scattering
angle $\theta_e$ for the free RFG (dashed), and including the MEC and
VC (dot-dashed) or 
the MEC, VC and SE (solid) contributions.
Clearly the pionic correlations are mostly felt at low values of
$\theta_e$ (left panel), where the longitudinal response is enhanced
by the kinematical factor $v_L$, and at low values of $q$, where the
vertex correlations dominate.  At high values of $\theta_e$ (right
panel) the asymmetry is totally insensitive to 
pions,
because the effect of the SE (which gives the main contribution)
cancels between the PV and PC responses appearing in the numerator and
denominator of eq.~(\ref{Asym1}).

We thus conclude that the extraction (at large electron angles) of the
axial nucleonic form factor $G_A$ is almost independent of the nuclear
model. On the contrary at small angles PV experiments can measure the
strange electric content of the nucleon only if a good control of the
nuclear dynamics is achieved, since the isospin correlations give very
large effects.  Conversely, interesting insight into the latter can in
principle be gained here.  Our results show that only at very large
momentum transfer does the forward-angle asymmetry become insensitive
to pionic correlations and hence suitable for assessing the
strangeness content of the nucleon.


\section{Non-relativistic reductions}
\label{sec:NR}


For years most of the effects introduced by the two-body pionic
currents in electron scattering reactions have been explored assuming
different types of non-relativistic 
reduction~\cite{Ama94a,Koh81,BoRa91,Van94,Van95,Ama99b}. Not only non-relativistic wave
functions have been used, but also non-relativistic current operators
derived from a direct Pauli reduction have been considered. Focusing
on the pionic effects on the hadronic $(e,e')$ response functions one
has to deal with the single-nucleon electromagnetic (electroweak in
general) current and the various two-body pionic currents discussed
previously.  Concerning the former, an improved version of the
single-nucleon electromagnetic current has been 
suggested in~\cite{Ama96a}, where the expression of the current is derived as a
non-relativistic expansion in terms of the dimensionless parameter
$\eta\equiv p/m$, $p$ being the three-momentum of the struck
nucleon. In Appendix F we review this approach --- which we call {\em
semi-relativistic} (SR)--- and compare it with the traditional
non-relativistic reduction, where the non-relativistic expansion is
performed with the additional assumption $\kappa\equiv q/2m<<1$ and
$\lambda\equiv \omega/2m <<1$.  As shown in~\cite{Ama98a,Ama96a} and
in Appendix F, the expansion of the current {\em  to first order} in
the variable $\eta$ yields quite simple expressions; moreover the
various pieces of the relativized current differ from the traditional
non-relativistic expressions only by multiplicative
$(q,\omega)$-dependent factors, and therefore are easy to implement in
already existing non-relativistic models.

The semi-relativistic form of the OB electromagnetic current operator
was first checked in~\cite{Ama96a}, where the inclusive longitudinal
and transverse responses of a non--relativistic Fermi gas were found
to agree with the exact relativistic result within a few percent if
one uses relativistic kinematics when computing the energy of the
ejected nucleon.  Recently the same expansion has been tested with
great success by comparing with the relativistic exclusive polarized
responses for the $^2$H$(e,e'p)$ reaction at high momentum
transfers~\cite{Jes98}.  This relativized current has also been
applied to the calculation of inclusive and exclusive responses that
arise in the scattering of polarized electrons from 
unpolarized~\cite{Maz02} and polarized nuclei~\cite{Ama96b,Ama98b,Ama99c,Ama02c}.
Finally, it also has been compared with a fully relativistic DWIA
calculation of $(e,e'p)$ observables for $|Q|^2=0.8\rm (GeV/c)^2$ in~\cite{Udi99,Cab01}. 
A systematic analysis of the semi-relativistic
approximation in the case of $(\vec{e},e'\vec{N})$ reactions has been
presented in~\cite{Cris1}.

\begin{figure}[htp]
\begin{center}
\leavevmode
\def\epsfsize#1#2{0.8#1}
\epsfbox[100 160 500 720]{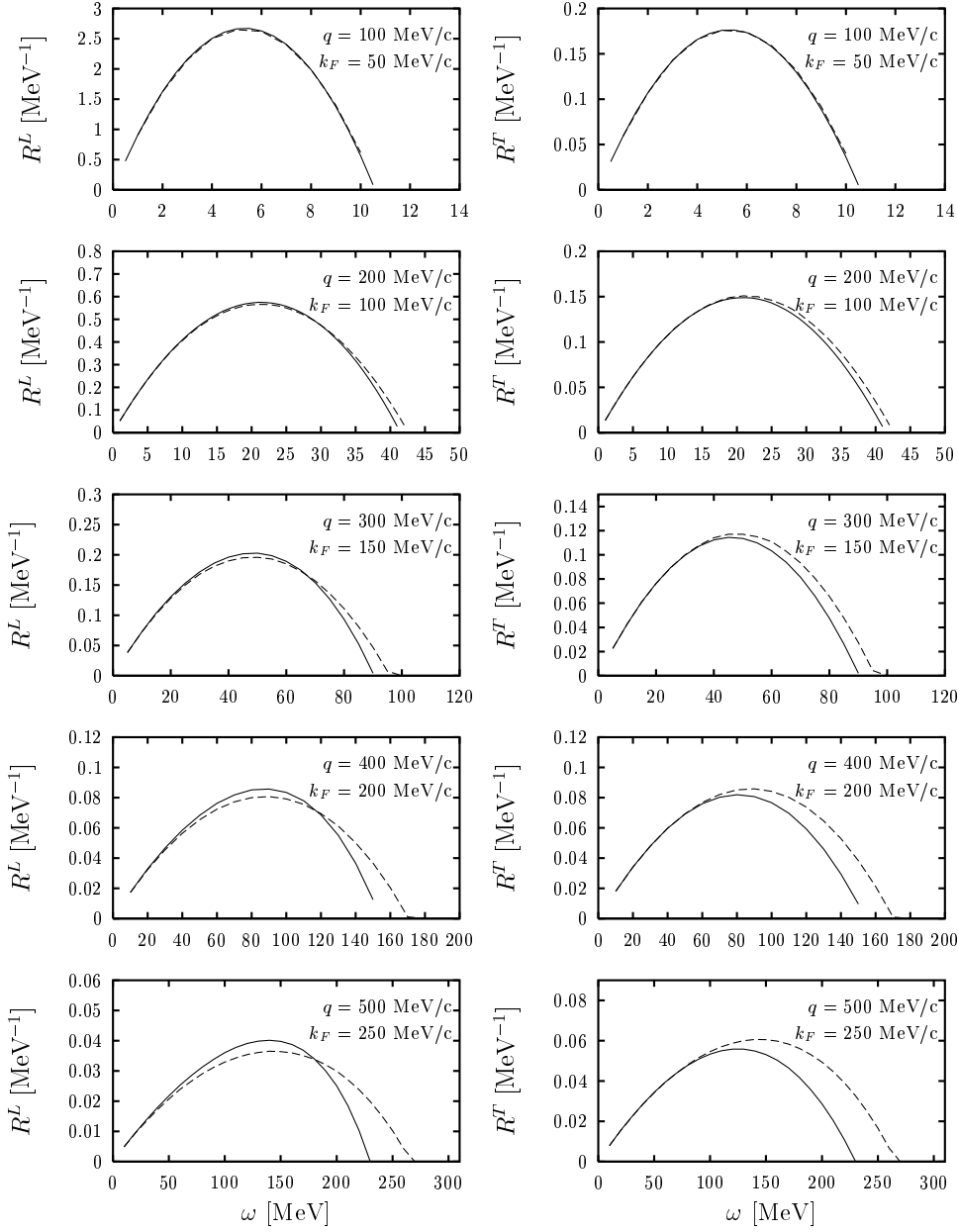}
\end{center}
\caption{The one-body longitudinal (left panels) and transverse (right
panels) responses displayed versus $\omega$ for various values of
the momentum transfer $q$ and of the Fermi momentum $k_F$.  Dashed: 
non-relativistic; solid: relativistic.  }
\end{figure}

Alternative expansions, in powers of the initial nucleon momentum, of
the structure functions of nuclei have recently been
proposed~\cite{Gur02} and ``recipes'' to obtain a relativistic
structure function from its non-relativistic analog by changing the
scaling variable and performing an energy shift have been suggested.
The so called three-dimensional reduction model, which includes
final-state interactions, has been tested in the case of a deuteron
target, but not for $A\geq 3$.

The necessity of a semi-relativistic form for the current even for
moderate momentum transfer is demonstrated in Fig.~22, where we
compare the traditional non-relativistic results for the electromagnetic responses
with the fully-relativistic calculation for a non-interacting system.
It clearly appears that for low densities and momentum transfers the
two approaches are equivalent, but that the two curves deviate from
each other as $q$ and $k_F$ increase. One of the effects of relativity
is the shrinking of the response region~\cite{Alb81} and is already
significant at $q=400$ MeV/c.  This effect, which arises from the
relativistic kinematics in the energy-conserving delta-function
appearing in the responses, can be accounted for approximately by the
replacement
\begin{equation} \label{pres}
\lambda \to \lambda(1+\lambda) \ .
\end{equation}
Another effect, stemming from the non-relativistic reduction of the
currents, relates to the enhancement of the longitudinal response and
to the reduction of the transverse one due to relativity.  Such an
effect can be mimicked by the kinematical factors $\kappa^2/\tau$ (in
the $L$ channel) and $\tau/\kappa^2$ (in the $T$ channel), which
naturally emerge from the $\eta$ expansion illustrated in Appendix F.
When included in the non-relativistic responses these factors,
together with the prescription of eq.~(\ref{pres}), allow one to
reproduce the fully-relativistic responses even for very high
$q$-values (see, for example,~\cite{Ama96a}).

In this  section we explore the impact of relativity on the
meson-exchange currents.

\subsection{Pion exchange currents}
\label{sec:NRMEC}

We first compare the fully-relativistic transverse MEC responses with
the traditional non-relativistic calculation developed in~\cite{Ama94b}, 
where the seagull p-h matrix element is evaluated
analytically, while the pion-in-flight contribution is reduced to a
one-dimensional integral. For this comparison the value $\Gamma_\pi=1$
for the $\pi NN$ form factor and the static pion propagator have been
used in the relativistic calculation.  The effect of static versus
dynamic pion propagator will be discussed later on.

    From Fig.~23 it emerges that the two calculations give the same results
for small density and momentum transfer.  As $q$ and $k_F$ increase we
see that, apart from the difference stemming from the relativistic
kinematics, which shrinks the response domain, the relativistic
responses are smaller than the non-relativistic ones: this reduction
amounts to about 30\% for $q$=500 MeV/c and $k_F$=250 MeV/c,
indicating that relativity plays an important role even for not so
high $q$-values.

The same curves are displayed for $k_F$=250 MeV/c and
$q$=500, 600 and 700 MeV/c in Fig.~24, where it is shown that 
the effect of relativity clearly grows with the momentum transfer.

In Fig.~25 the relativistic MEC-correlated transverse response
(dotted) is compared with the corresponding non-relativistic one
(dot-dashed) as well as with the relativistic (solid) and
non-relativistic (dashed) one-body response for three values of
$q$. The figure shows that for low values of $q$ (500 MeV/c) the
effects of MEC and relativity are roughly of the same size, the former
acting mainly to the left of the QEP, the latter to the right. As $q$
increases, the effect of relativity becomes dominant, pointing to the
necessity of a relativistic treatment for momentum transfers larger
than 500 MeV/c.

\begin{figure}[htp]
\begin{center}
\leavevmode
\def\epsfsize#1#2{0.8#1}
\epsfbox[100 160 500 740]{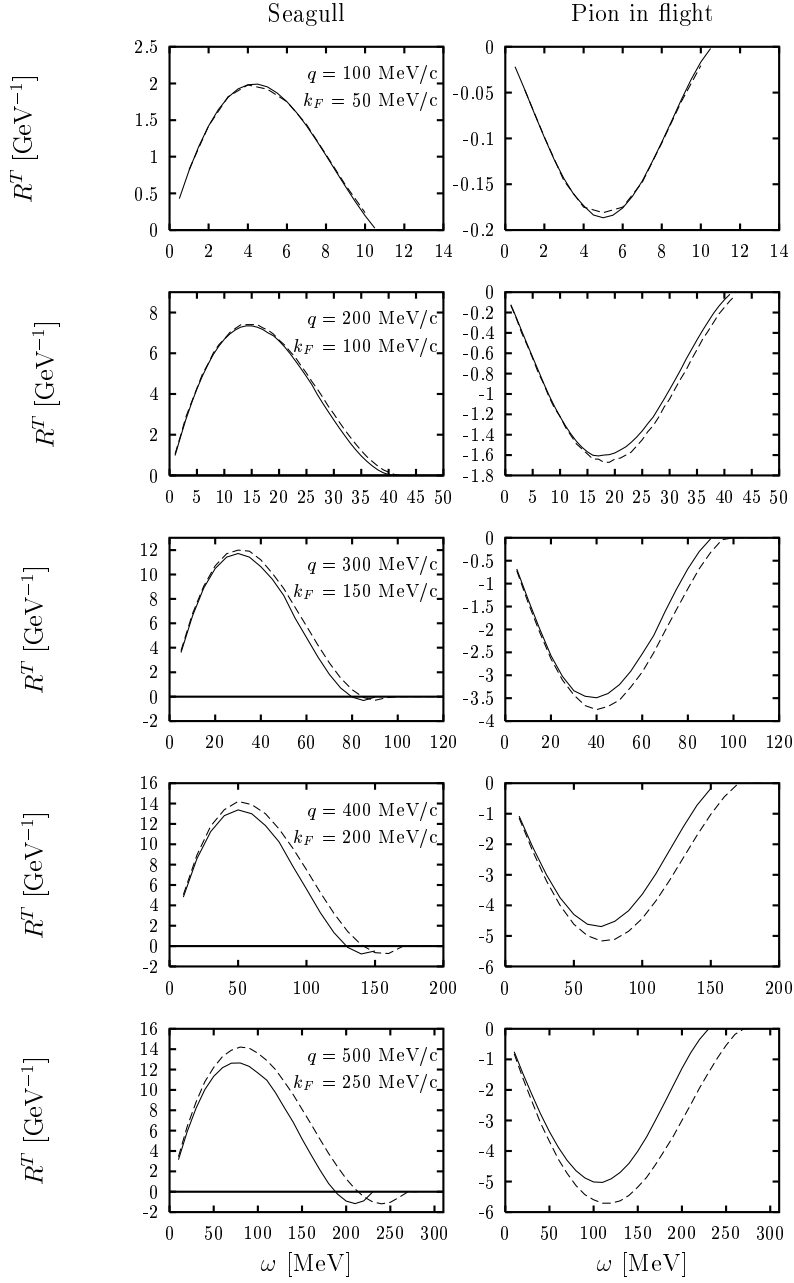}
\end{center}
\caption{The seagull (left panels) and pion-in-flight (right panels)
contributions to the transverse response displayed versus $\omega$ for
various values of the momentum transfer $q$ and of the Fermi momentum
$k_F$.  Dashed: non-relativistic; solid: relativistic.  }
\end{figure}

\begin{figure}[htp]
\begin{center}
\leavevmode
\def\epsfsize#1#2{0.9#1}
\epsfbox[100 385 500 740]{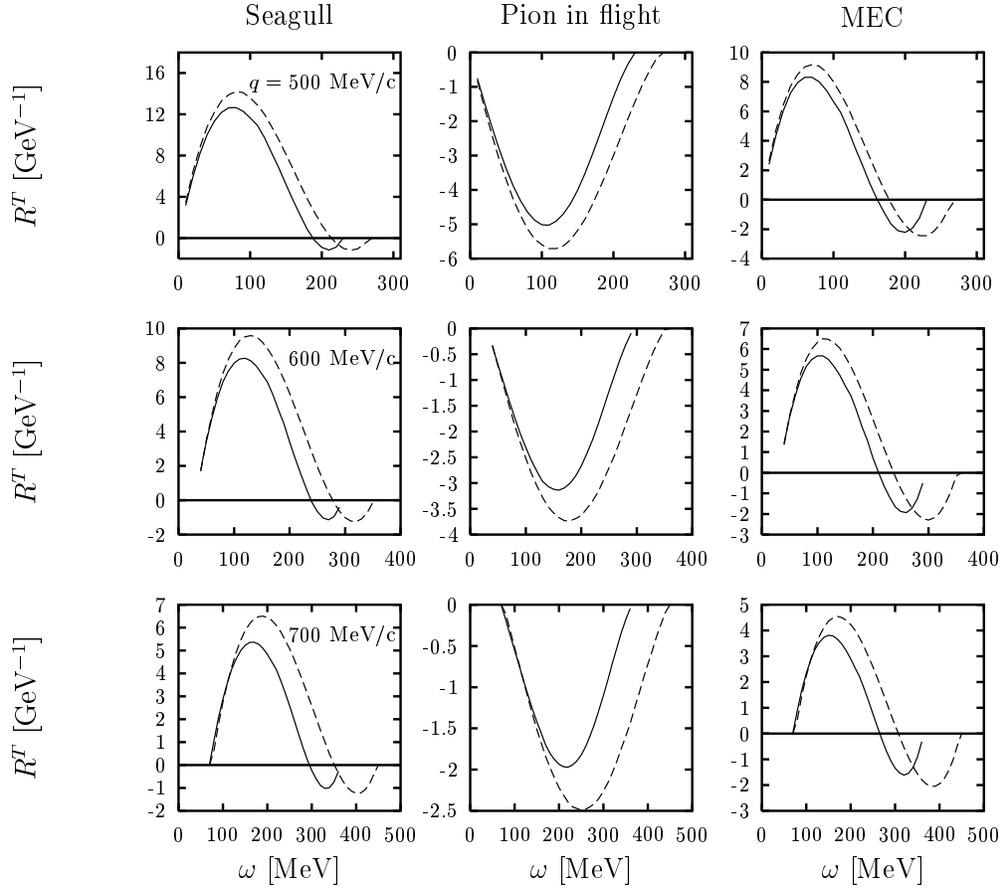}
\end{center}
\caption{The seagull (left panels), pion-in-flight (central panels)
and total MEC (right panels) contributions to the transverse response
displayed versus $\omega$ for $k_F$=250 MeV/c and various values of
the momentum transfer $q$.  Dashed: non-relativistic; solid:
relativistic.  }
\end{figure}

\begin{figure}[htp]
\begin{center}
\leavevmode
\def\epsfsize#1#2{0.9#1}
\epsfbox[240 325 430 725]{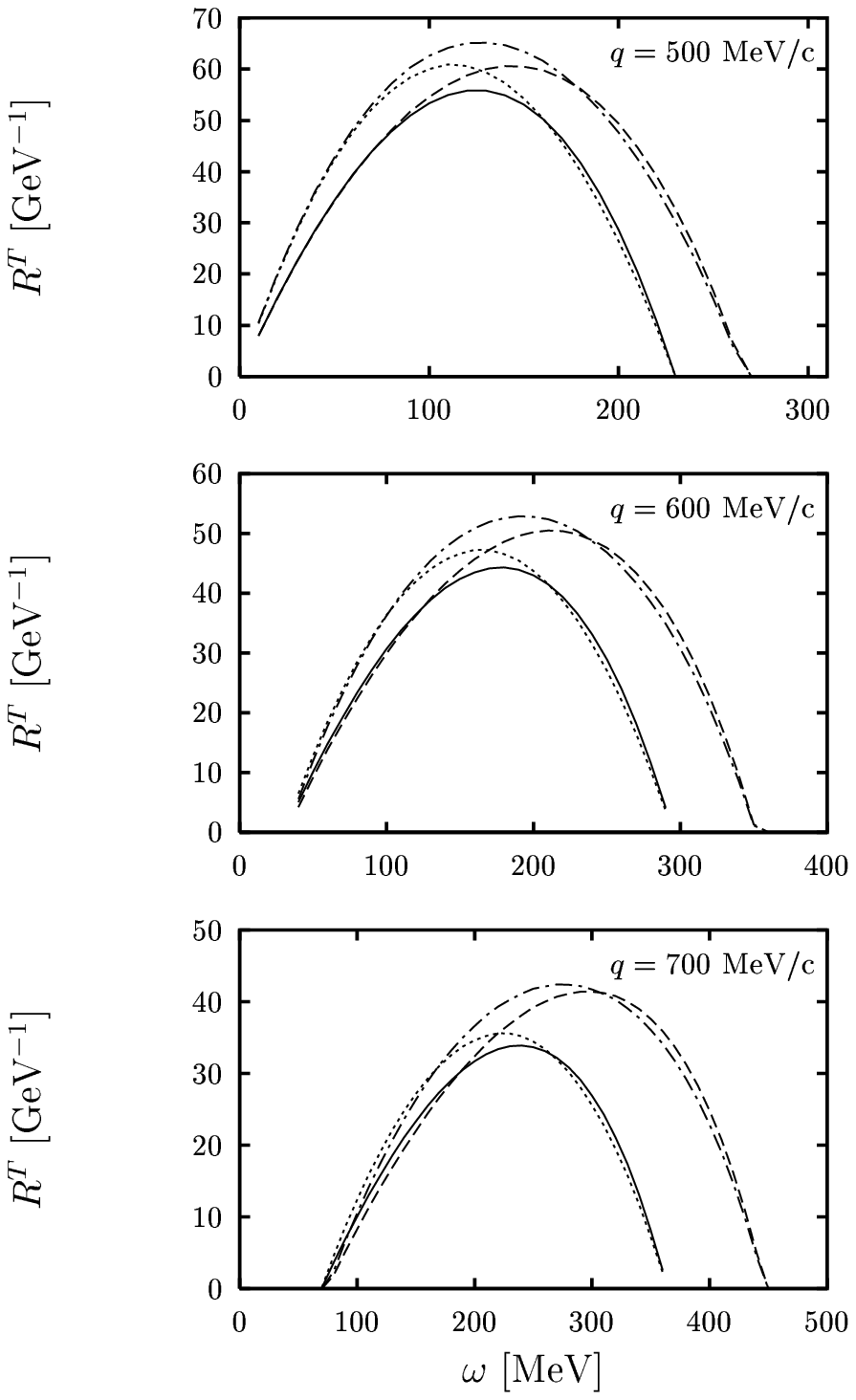}
\end{center}
\caption{The transverse response displayed versus $\omega$ for
$k_F$=250 MeV/c and various values of the momentum transfer $q$.
Dashed: one-body non-relativistic; solid: RFG; dot-dashed: one-body+MEC
non-relativistic; dotted: RFG+MEC relativistic.}
\end{figure}

Finally the impact on the responses of the relativistic propagator
$\Delta_\pi(K)=(K^2-m_\pi^2)^{-1}$ as compared with the static one
$\Delta_\pi^{(n.r.)}({\bf k})=-({\bf k}^2+m_\pi^2)^{-1}$, which is
commonly used in non-relativistic calculations, is explored.  In
Fig.~26 the pion-in-flight, seagull and total MEC contributions to
$R^T$ are evaluated for $q$=0.5 and 2 GeV/c using the two versions of
the propagator.  It appears that the dynamical propagator affects the
pion-in-flight contribution more than the seagull term (it increases
the latter by more than a factor 2 at $q$=2 GeV/c); however, the two
effects tend to cancel, so that their net effect is not very
significant.

\begin{figure}[htp]
\begin{center}
\leavevmode
\def\epsfsize#1#2{0.9#1}
\epsfbox[150 550 470 690]{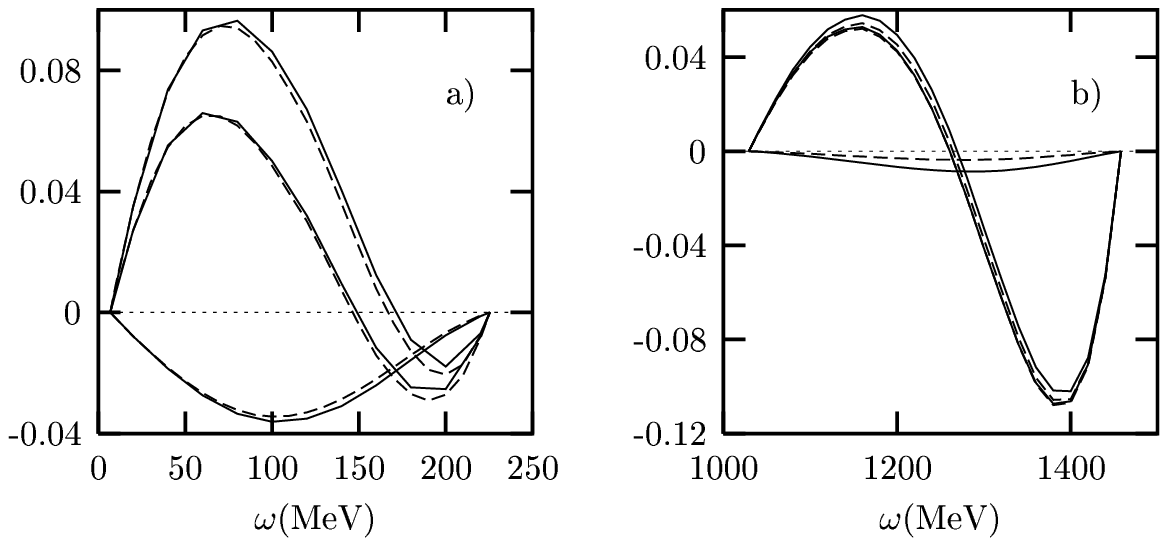}
\end{center}
\caption{
MEC contribution to $R^T$ (in $GeV^{-1}$) versus $\omega$ with 
dynamic (solid curves) 
and static (dashed curves) pion propagator at (a) $q$= 0.5 and (b) 2 GeV/c. 
The separate pion-in-flight and seagull contributions 
are displayed.
}
\end{figure}

\subsubsection{The $\eta_F$ expansion}
\label{sec:NRetaF}

In view of the relevance of relativistic effects illustrated above and
following the ideas and methods developed in the case of the
single-nucleon electromagnetic current operator and its
non-relativistic reduction~\cite{Ama96a}, a new semi-relativistic
reduction of the MEC has been developed in~\cite{Ama98a}, where the
transferred energy and momentum have been left unexpanded while
expanding only the initial nucleon momentum.  The expressions thus
obtained retain important aspects of relativity not included in the
traditional non--relativistic MEC used throughout the literature.
Here we summarize the basic results.

\begin{figure}[htp]
\begin{center}
\leavevmode
\def\epsfsize#1#2{0.9#1}
\epsfbox[100 170 550 730]{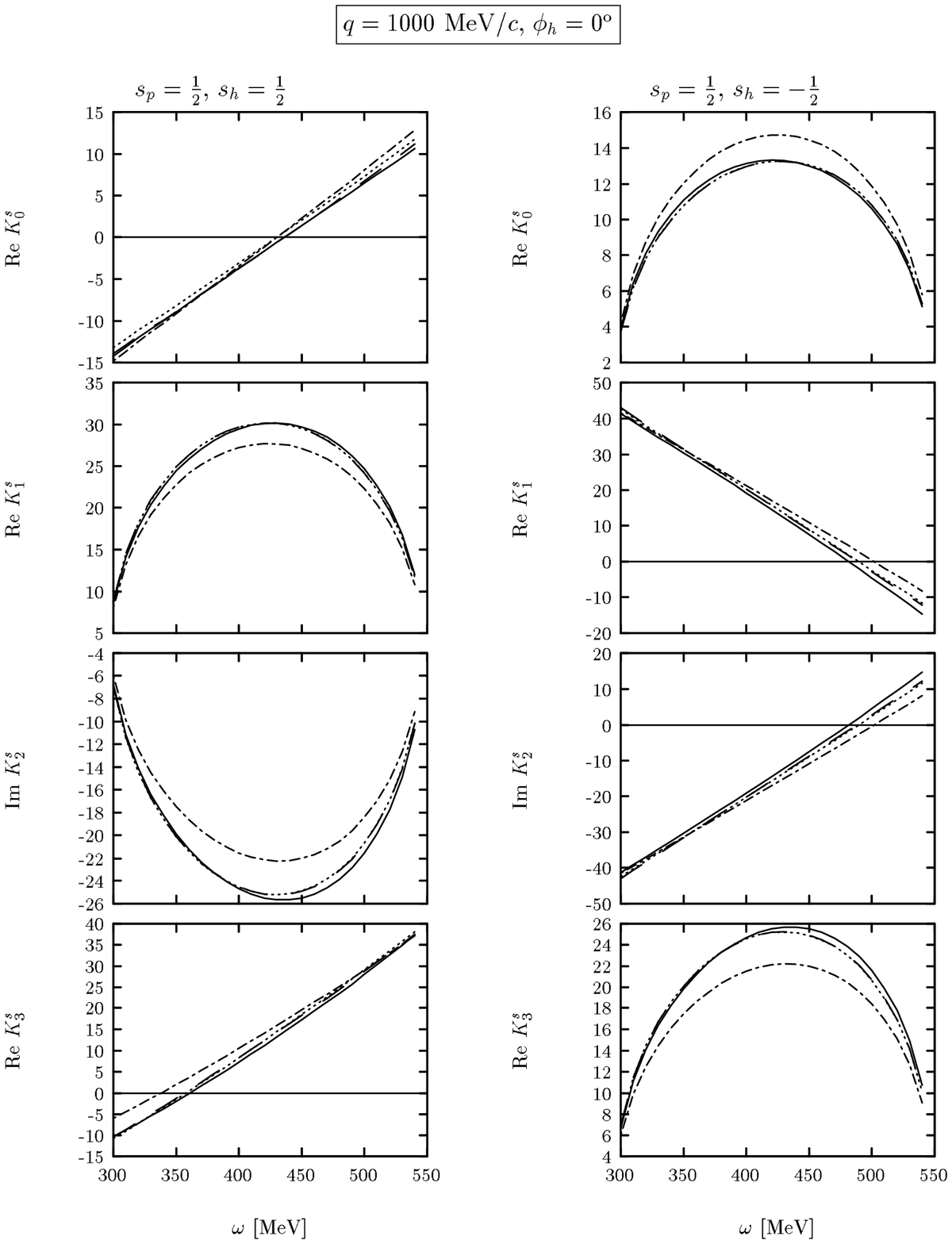}
\end{center}
\caption{ Seagull current matrix element $K_\mu^s$ -- 
see eq.~(\protect\ref{eq:Kmu}) --
for $q$=1 GeV/c and
$k_F$= 250 MeV/c.  The kinematics for the hole are $h$=175 MeV/c and
$\phi_h=0$. First column: spin $(1/2,1/2)$ component; second column:
spin $(1/2,-1/2)$ component. Solid: fully-relativistic; dashed: SR1
approximation; dot-dashed: traditional non-relativistic; dotted: SR2
approximation.  }
\end{figure}

\begin{figure}[htp]
\begin{center}
\leavevmode
\def\epsfsize#1#2{0.9#1}
\epsfbox[100 170 550 730]{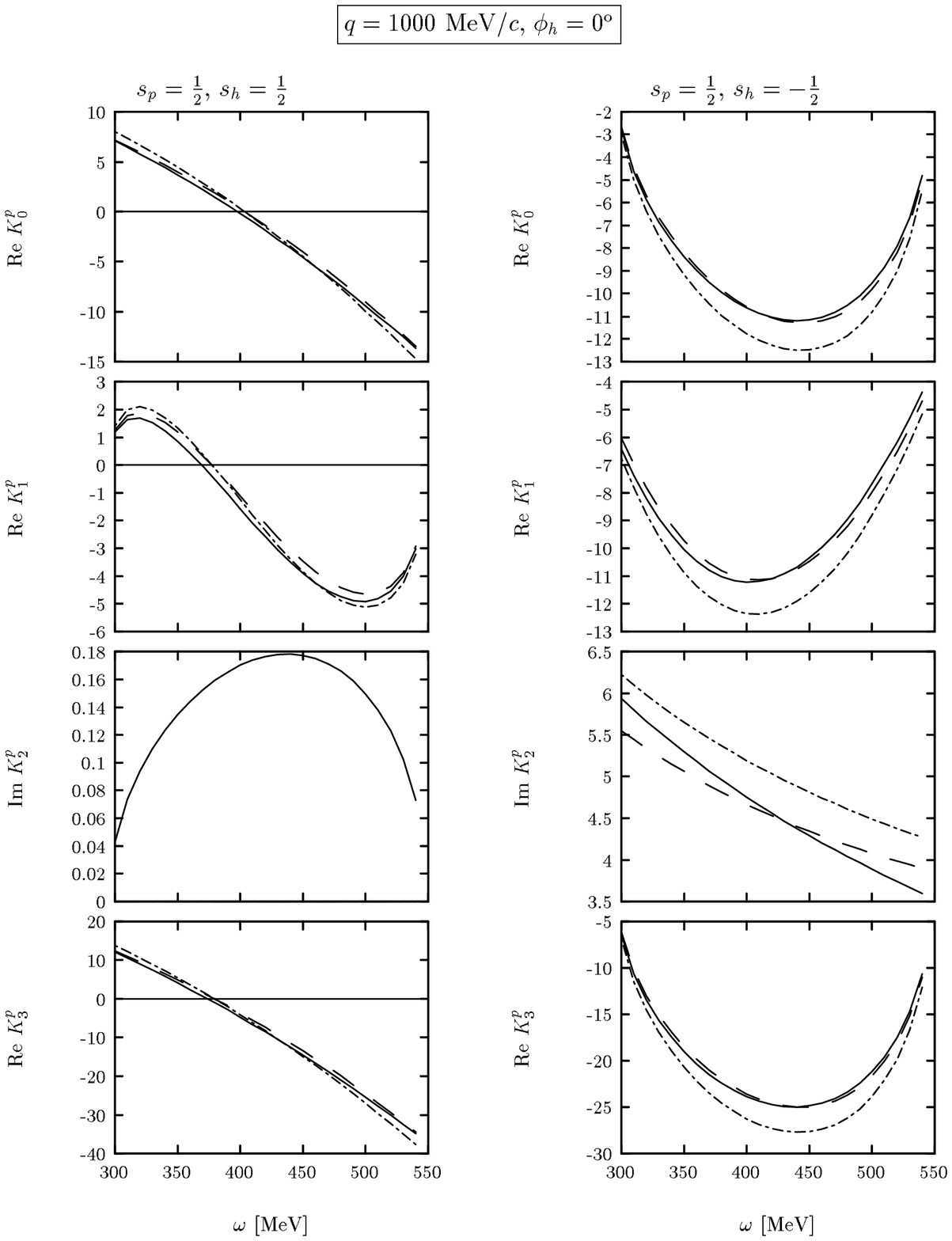}
\end{center}
\caption{ Pion-in flight current matrix element $K_\mu^p$ for $q$=1
GeV/c and $k_F$= 250 MeV/c. The kinematics for the hole are $h$=175
MeV/c and $\phi_h=0$. First column: spin $(1/2,1/2)$ component; second
column: spin $(1/2,-1/2)$ component. Solid: fully-relativistic;
dashed: semi-relativistic approximation; dot-dashed: traditional
non-relativistic.  }
\end{figure}

We are interested in the evaluation of the particle--hole matrix
elements $\langle pk|j_{MEC}^\mu|kh\rangle$ and their {\em new}
semi-relativistic expressions.  The resulting expansion for the MEC
should be used together with the single--nucleon current, developed to
first order in $\eta$ (see~\cite{Ama98a} and Appendix F), to
set up the various responses.  Therefore, in order to be
consistent, the expansion of the MEC should also be performed to first
order in the corresponding small quantities $\{\eta_k\equiv k/m,
\eta_h\equiv h/m\}$, whereas $\{\eta_p\equiv p/m,
\kappa\equiv q/2m\}$ are treated exactly.

After some algebra~\cite{Ama98a} the following semi-relativistic (SR)
expressions of the MEC currents (referred to as NR1 approximation
in~\cite{Ama98a}) are obtained:

\paragraph{Seagull Current Operator.}
\begin{eqnarray}
\lefteqn{j_s^0(\np,\nk,\nk,\np)_{SR1}}
\nonumber\\
&=& 
\frac{{\cal F}}{2\sqrt{1+\tau}} 
\chi_{s_p}^\dagger \left\{
\frac{\sigvec\cdot\left[2\kappavec+\etavec_h-(1+\tau)\etavec_k\right]
\chi_{s_k}\chi_{s_k}^\dagger
\sigvec\cdot(\etavec_k+\etavec_h)}{(P-K)^2-m_\pi^2} 
\right.
\nonumber\\
&-&
\left.
\frac{\sigvec\cdot\left[2\kappavec+\etavec_h+(1+\tau)\etavec_k\right]
\chi_{s_k}\chi_{s_k}^\dagger
\sigvec\cdot(\etavec_k-\etavec_h)}{(K-H)^2-m_\pi^2} 
\right\}\chi_{s_h}
\label{bai1}
\\ \nonumber 
\lefteqn{\jvec_s(\np,\nk,\nk,\np)_{SR1}}
\nonumber\\
&=&  
\frac{{\cal F}}{\sqrt{1+\tau}} \chi_{s_p}^\dagger\left\{ \left[
2\sigvec\cdot\kappavec
\left(1-\frac{\kappavec\cdot\etavec_h}{2(1+\tau)}\right) +
\sigvec\cdot(\etavec_h -\etavec_k) -\tau \sigvec\cdot\etavec_k \right]
\chi_{s_k} \right.\nonumber\\  &\times&\left.\chi_{s_k}^\dagger
\frac{\sigvec}{(P-K)^2-m_\pi^2} -(1+\tau)
\frac{\sigvec}{(K-H)^2-m_\pi^2} \chi_{s_k}\chi_{s_k}^\dagger
\sigvec\cdot(\etavec_k-\etavec_h) \right\} \chi_{s_h}\ ,
\label{bai2} 
\end{eqnarray}
where the factor 
\begin{equation}
{\cal F}=-\frac{f^2m}{m_\pi^2}i\varepsilon_{3ab} 
\langle t_p|\tau_a|t_k\rangle\langle t_k|\tau_b|t_h\rangle F_1^{V} 
\end{equation}
has been introduced.  Note that if the terms
$\etavec_h-(1+\tau)\etavec_k$ and $\etavec_h+(1+\tau)\etavec_k$ are
neglected (this approximation will be referred to as $SR2$) the
expression for the time component is similar to the one obtained in
the traditional non-relativistic reduction~\cite{Ama96a} except for
the factor $1/\sqrt{1+\tau}$, which accordingly incorporates important
aspects of relativity not considered in the traditional
non-relativistic reduction.  Analogously, in the space component, if
the terms $(\kappavec\cdot\etavec_h)/[2(1+\tau)]$ and
$\tau(\sigvec\cdot\etavec_k)$ are neglected ($SR2$ approximation), the
traditional non--relativistic expression~\cite{Alb90} is simply
recovered, except for the factors $1/\sqrt{1+\tau}$ and
$\sqrt{1+\tau}$ that multiply the contributions given by the two
diagrams involved.  Thus, as in the case of the time component, also
here important relativistic effects are simply accounted for by these
multiplicative factors.

To illustrate this point we plot in Fig.~27 the current matrix element
$K_\mu^s$ defined through 
\begin{equation}
\frac{1}{V}\sum_{k<k_F} 
\frac{m}{E_{\nk}}
j_\mu^s(\np,\nk,\nk,\nh) =
- \frac{m^2}{\sqrt{E_{\np}E_{\nh}E_{\nk}^2}}
\frac{k_F^2}{m} K_\mu^s(q,\omega,\nh) \sum_{t_k} {\cal F}
\label{eq:Kmu}
\end{equation}
for $q$=1 GeV/c, $h$=175 MeV/c and $\phi_h$=0.  The curves represent
the fully-relativistic result (solid), the traditional
non-relativistic approximation, including relativistic kinematics
through eq.~(\ref{pres}) (dot-dashed), the SR1 approximation of
eqs.~(\ref{bai1},\ref{bai2}) (dashed) and the SR2 approximation
(dotted). Only the relevant components are shown, the other vanishing
for symmetry reasons (see~\cite{Ama98a} for details).  It clearly
appears that, while the traditional non-relativistic reduction,
although corrected by the replacement $\lambda\to\lambda(1+\lambda)$,
fails to reproduce the exact results by roughly 10 to 20\% (this
deviation increasing with $q$, as shown in~\cite{Ama98a}), both the
SR1 and SR2 approaches yield excellent agreement with the
fully-relativistic current.

Finally, we examine the limit $\eta_F\rightarrow 0$, since this
provides some understanding of how the MEC effects are expected to
evolve in going from light ($\eta_F$ very small) to heavy nuclei
($\eta_F\cong 0.29$).  In this limit the seagull current simply
reduces to
\begin{equation}\label{eq49}
\lim_{\eta_F\rightarrow 0} j_s^0(\np,\nk,\nk,\nh)=0\ .
\end{equation}
This is because the time component of the seagull current is of first
order in the small momenta involved or, equivalently, it is
$O(\eta_F)$.  On the contrary, the vector component in the limit
$\eta_F\rightarrow 0$
becomes
\begin{equation}\label{eq54}
\lim_{\eta_F\rightarrow 0} \jvec_s(\np,\nk,\nk,\nh)
=\frac{2{\cal F}}{\sqrt{1+\tau}}
\chi_{s_p}^\dagger\frac{(\sigvec\cdot\kappavec)\chi_{s_k}
\chi_{s_k}^\dagger\sigvec}{Q^2-m_\pi^2}\chi_{s_h}\ ,
\end{equation}
which shows that the space components of the seagull current are $O(1)$ and 
contribute even for nucleons at rest, as happens for the charge and 
magnetization pieces of the one--body current. 

\paragraph{Pion-in-flight Current Operator.}\mbox{}

Keeping only linear terms in the small momenta, one obtains
for the semi-relativistic pion-in-flight current
\begin{eqnarray}
j_p^0(\np,\nk,\nk,\nh)_{SR}
&=& 
-\frac{{\cal F}}{\sqrt{1+\tau}} 4 m^2\tau \chi_{s_p}^\dagger
\frac{\sigvec\cdot\kappavec \chi_{s_k}\chi_{s_k}^\dagger
\sigvec\cdot(\etavec_k-\etavec_h)}
{\left[(P-K)^2-m_\pi^2\right]\left[(K-H)^2-m_\pi^2\right]}\chi_{s_h}
\label{bai3}\\
\jvec_p(\np,\nk,\nk,\nh)_{SR}
&=& -\frac{{\cal F}}{\sqrt{1+\tau}} 4 m^2 \chi_{s_p}^\dagger
\frac{\sigvec\cdot\kappavec \chi_{s_k}\chi_{s_k}^\dagger
\sigvec\cdot(\etavec_k-\etavec_h)}
{\left[(P-K)^2-m_\pi^2\right]\left[(K-H)^2-m_\pi^2\right]}\chi_{s_h}
\kappavec \, .
\label{bai4}
\end{eqnarray}
Again, these expressions are similar to the traditional
non--relativistic currents~\cite{Alb90} except for the common factor
$1/\sqrt{1+\tau}$, which includes important aspects of relativity not
taken into account in the traditional non--relativistic reduction.
Note that the space component of the pionic current is, in leading
order, purely longitudinal; its transverse components are in fact of
second order in $\eta_F$.

In Fig.~28 we display the current matrix element $K_\mu^p$, defined
analogously to eq.~(\ref{eq:Kmu}), for $q$=1 GeV/c, $h$=175 MeV/c and
$\phi_h$=0.  As for the seagull, the fully-relativistic result (solid)
is very well approximated by the SR prescription (dashed), whereas
the traditional non-relativistic approach with relativistic kinematics
(dot-dashed) deviates from the exact result by 10-20\%.

Finally the limit $\eta_F\rightarrow 0$ implies that
$j^\mu_p(\np,\nk,\nk,\nh)=0$, since all components of the pionic
current are $O(\eta_F)$ in the expansion.

Summarizing, the $\eta$ expansion shows that relativistic kinematics
can be very easily implemented in MEC semi-relativistic calculations
by applying the prescription $\lambda\to\lambda(1+\lambda)$ and by
multiplying the exchange currents by the kinematical factors indicated
in eqs.~(\ref{bai1}-\ref{bai4}).

\subsubsection{ Results for the responses using the relativized MEC model}

In this section we discuss the validity of the relativizing
prescriptions introduced above, when they are implemented in a
traditional non-relativistic model of the reaction.  We begin with the
non-relativistic Fermi gas of~\cite{Ama94b}, which includes also MEC
in the transverse response.  One of the advantages of this model is
that the integral over the Fermi sea appearing in the seagull matrix
elements can be performed analytically, while the pion-in-flight is
reduced to an one-dimensional integral.

Next we will relativize this model by implementing relativistic
kinematics through the substitution
$\lambda\rightarrow\lambda(1+\lambda)$ in all places except in the
nucleon and pion form factors $F(q,\omega)$, which should be evaluated
at the correct $\omega$-value.  Second, we use the new
semi-relativistic expansion of the electromagnetic OB+MEC operators in
powers of $\eta$.  For the OB operators we use the following expressions
(see Appendix F):
\begin{eqnarray}
\rho_{OB} 
&=& 
\frac{\kappa}{\sqrt{\tau}}G_E
+i\frac{G_M-G_E/2}{\sqrt{1+\tau}}(\nkappa\times\neta)\cdot\nsigma
\\
\nJ^T_{OB}
&=&
\frac{\sqrt{\tau}}{\kappa}
\left[ iG_M(\nsigma\times\nkappa) + G_E\neta_T \right] \, .
\end{eqnarray}
Note that near the QEP it makes little difference to use the factors
$1+\tau$ or $\kappa^2/\tau$.  In these factors lies the main
difference with the traditional non-relativistic charge and transverse
current operators. Note that in addition we include a first-order
spin-orbit term in the charge operator.  The contribution of this term
is small in the longitudinal unpolarized response, since its
interference with the leading order is exactly zero in PWIA and hence
it gives a negligible contribution of second order in $\eta$.  However
one should be careful in including this term in the more complex cases
in which there is an interference TL response (when the nucleus is
polarized or in the exclusive reactions $(e,e'p)$,
see~\cite{Ama96b,Ama98b,Cris1}), and where it gives a significant
contribution, since in this response the leading order is zero in
PWIA.

In the case of the MEC we use the following simplified prescription to
relativize transverse operators:
\begin{equation}
J^T_{MEC}= \frac{1}{\sqrt{1+\tau}}J^T_{MEC,\rm non\, rel}\ ,
\end{equation}
namely we introduce a factor $1/\sqrt{1+\tau}$ to take into account
relativistic corrections coming from the free Dirac spinors. Note that
in the case of the seagull we have neglected a further correction
factor $1+\tau$ in the hole part of the seagull current.  However here
we choose to present results with the above simplified version of the
transverse current, since it is the easier to implement in already
existing models of the reaction; otherwise one has to identify the
different pieces of the operator, which may be difficult.
Furthermore, this correction is not of much importance, its main
effect being to correct slightly the position of the zero in the
seagull response.  Be it as it may this {\em ad hoc} prescription for
the seagull current is supported by the quality of the results shown
below.
  
\begin{figure}[ht]
\begin{center}
\leavevmode
\def\epsfsize#1#2{0.9#1}
\epsfbox[60 460 500 730]{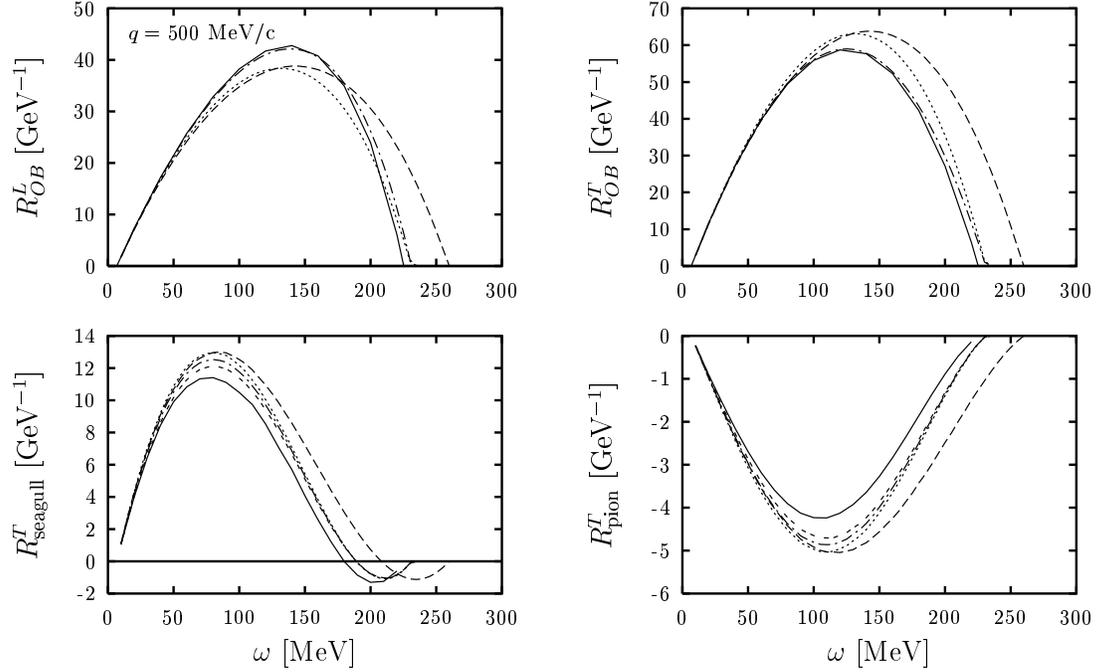}
\end{center}
\caption{\label{rep28} 
\small One-body longitudinal and transverse
response functions (top panels), and transverse responses of interference
between MEC and OB currents (bottom panels) for $q=500$ MeV/c and $k_F=237$
MeV/c.  Solid: exact relativistic results with static propagator and
without $\pi N$ form factor.  The rest of the curves have been
computed using the non-relativistic Fermi gas model, with or without
relativistic corrections.  Dashed: traditional non-relativistic
results.  Dotted: including relativistic kinematics in the 
non-relativistic calculations.  Dot-dashed: including in addition the new
expansion of the OB currents.  Double-dashed: including in addition a
correction to the MEC operators with a factor $1/\protect\sqrt{1+\tau}$.  }
\end{figure}

\begin{figure}
\begin{center}
\leavevmode
\def\epsfsize#1#2{0.9#1}
\epsfbox[60 460 500 730]{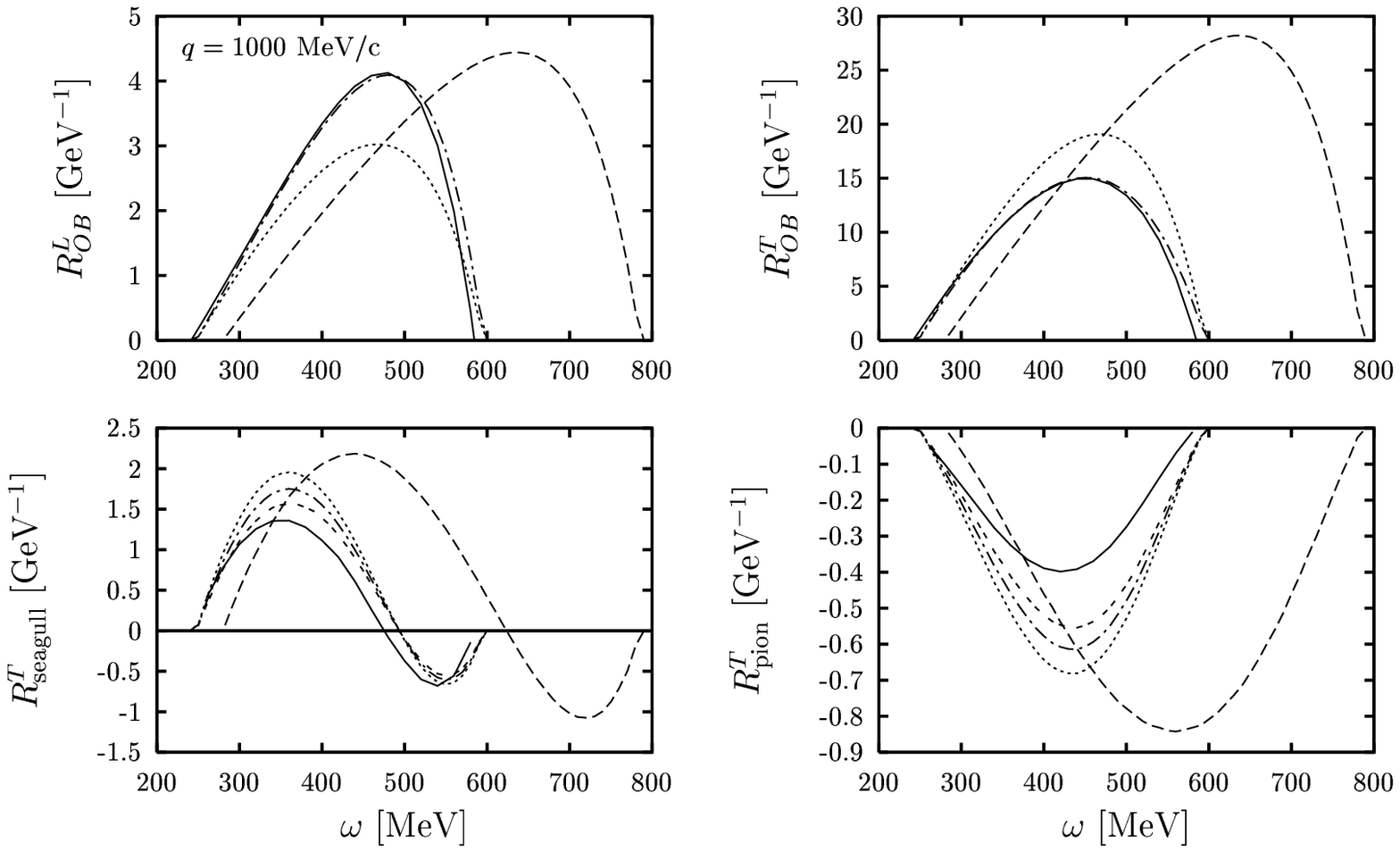}
\end{center}
\caption{\label{rep29}
\small The same as Fig. \ref{rep28}, now for $q=1000$ MeV/c.  }
\end{figure}

Results for the $^{40}$Ca nucleus for $q=500$ MeV/c and $q=1000$ MeV/c
are shown in Figs.~\ref{rep28} and \ref{rep29}, respectively.  In the
upper part of these figures we show the one-body (OB) separated
longitudinal and transverse responses.  The solid lines are the exact
relativistic results.  These are very different from the traditional
non-relativistic results shown with dashed lines.  Note that the same
nucleon form factors and the same $k_F=237$ MeV/c are used in both
calculations.  If we include relativistic kinematics, then we obtain
the dotted lines, which are still different from the exact result,
even if now the region where the response is nonzero is similar to the
relativistic case.  Finally, using in addition the new
semi-relativistic corrections (factors $\kappa/\sqrt{\tau}$ in the
charge and $\sqrt{\tau}/\kappa$ in the current) we obtain the
relativistic approximation shown with dot-dashed lines, which is very
similar to the exact result.  Hence we can safely say that the new
expansion of the OB current is very good, giving essentially the exact
answer.

The case of the MEC transverse responses is shown in the lower part of
Figs.~\ref{rep28} and~\ref{rep29}. There we show the separate
contribution of seagull and pion-in-flight currents to the transverse
response (interference with the OB current).  Again we show with solid
lines the exact relativistic results, and with dashed lines the
traditional non-relativistic results.  If again we include the
relativistic kinematics we obtain the dotted lines.  With dot-dashed
lines we display results which include in addition the relativistic
correction to the OB current, amounting to a factor
$1/\sqrt{1+\tau}\simeq \sqrt{\tau}/\kappa$.  This correction produces
a small reduction of the responses.  Finally, with double-dashed lines
we show the results computed using in addition the relativistic
corrections in the MEC, which amounts to another factor
$1/\sqrt{1+\tau}$.  This correction produces a further reduction of
the responses, giving a result which is closer to the exact one.

  From these results it appears that our expansion of MEC currents is
not as good (at least fractionally) as the OB expansion.  This is
likely related to the fact that the OB currents have been expanded in
powers of $\eta=h/M$, where $h$ is the momentum of the hole, and they
are exact, by construction, for $h=0$.  However in the case of the MEC
there is another variable in the expansion: the momentum of the second
hole $h'/M$, which is small, but is never zero and is being integrated
up to $k_F$.  Therefore our currents are not constructed to agree with
the exact ones for $h=0$.

\begin{figure}[hp]
\begin{center}
\leavevmode
\def\epsfsize#1#2{0.8#1}
\epsfbox[100 175 450 740]{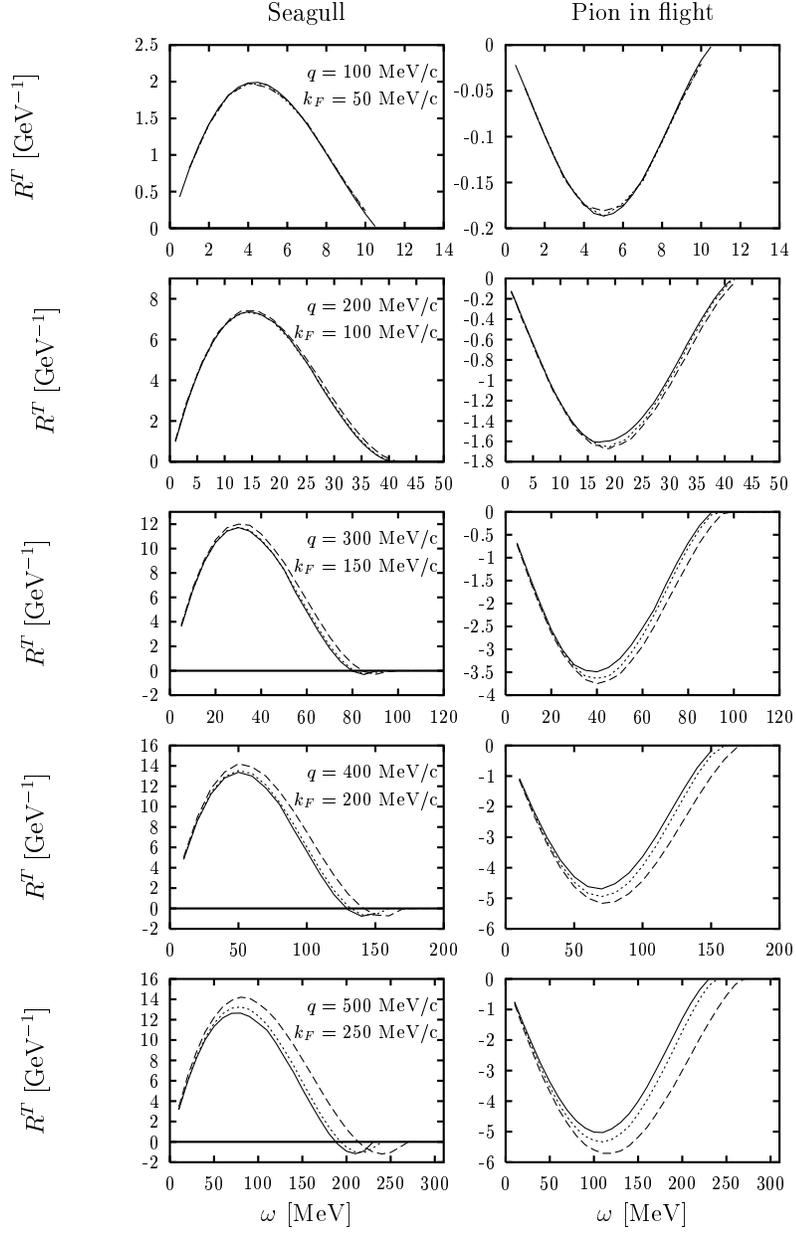}
\end{center}
\caption{\label{rep30}
\small Seagull and pionic responses computed for several
values of $q$ and $k_F$. Solid lines: exact relativistic results.
Dashed lines: non-relativistic results. Dotted lines: approximated
semi-relativistic results using relativistic kinematics 
and relativized currents.
Static propagators without a $\pi NN$ form factor have been used here.  }
\end{figure}

On the other hand, we have explicitly showed before that in the limit
$q\rightarrow0$ and $k_F\rightarrow0$, the relativistic and
non-relativistic results agree.  This is also the case for the present
results of the relativized currents, as it is illustrated in
Fig.~\ref{rep30}.  There we show the seagull and pionic responses for
several small values of $q=100,\ldots,500$ MeV/c and for $k_F=q/2$.
With solid lines we show the exact relativistic results, while with
dashed lines we show the traditional non-relativistic results. Finally
we also show with dotted lines the results using the present
semi-relativized approach.  It is seen that the last are always much
closer to the exact result than the non-relativistic ones, and that
they converge faster to the exact results.

\begin{figure}[ht]
\begin{center}
\leavevmode
\epsfbox[60 455 500 730]{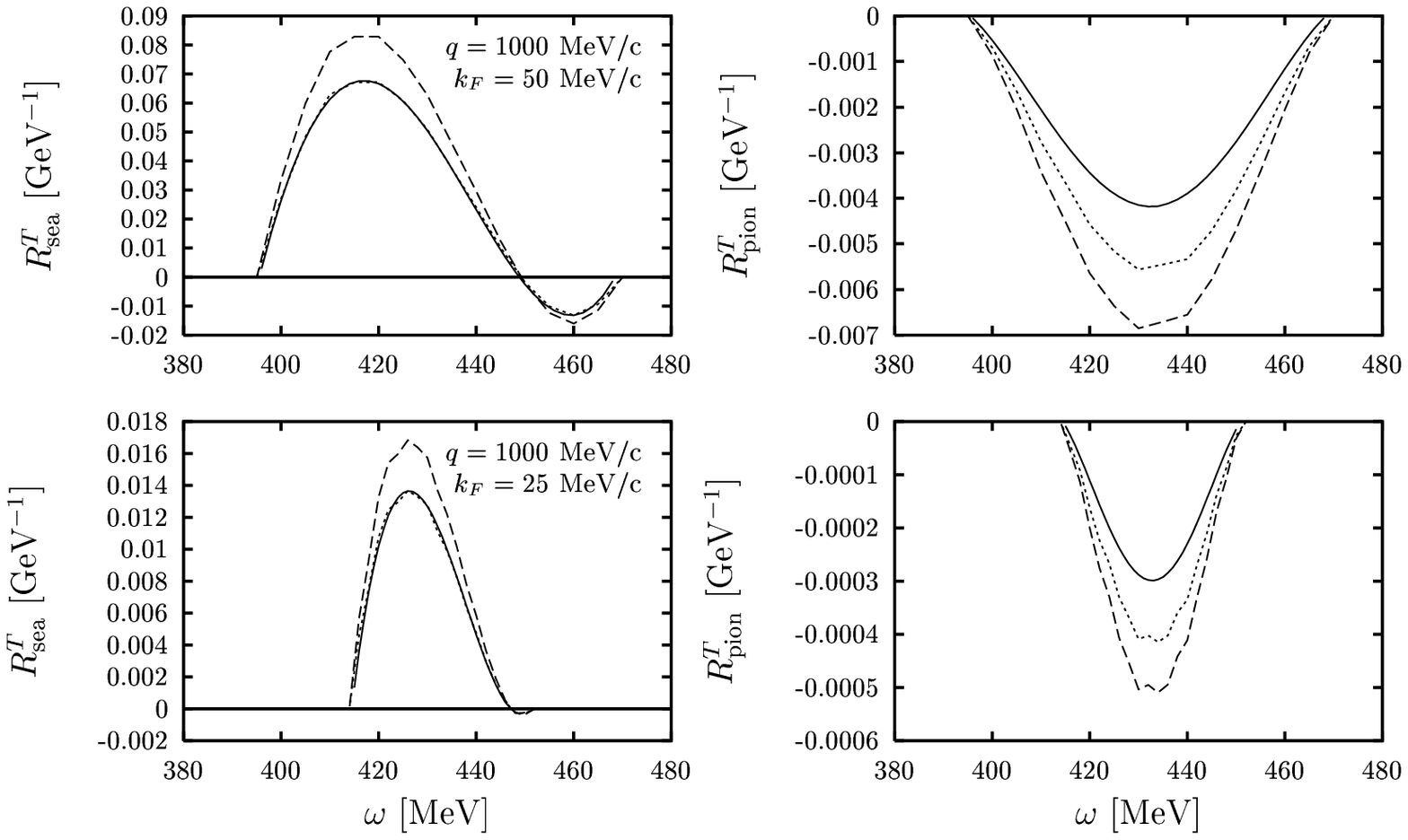}
\end{center}
\caption{\label{rep31}
\small MEC-OB transverse responses for for
$q=1000$ MeV/c and $k_F=50$ MeV/c.  Solid: exact relativistic results
with static propagator and without a $\pi N$ form factor.  Dashed: 
non-relativistic results, but including relativistic kinematics.  Dotted:
using in addition the relativized currents.  }
\end{figure}

Better agreement between the exact and the relativized models 
for the MEC responses
is also expected in the limit $\eta_F\rightarrow 0$
in the quasielastic peak, since in this case both momenta $h$ and $h'$
are forced to be small, which are the conditions assumed in our
expansion. Results for the transverse MEC responses in this limit are
shown in Fig.~\ref{rep31} for $q=1000$ MeV/c and for two values of 
$k_F=50$ and  25 MeV/c. 
With solid lines we show the exact relativistic results, while  
with  dashed lines we display the non-relativistic ones, but
including relativistic kinematics. Finally, the dotted lines correspond to
the 
semi-relativized results, which fully agree with the exact ones in the
case of the seagull current, while in the pionic case there is still 
a difference between the two calculations. However this is not very important
in this limit, since the transverse pionic contribution
which we are computing here is of second order in the hole momenta,
and so is negligible compared with the seagull one, as can be seen in the
figure.

In order to  improve  the present results for the MEC responses one should 
look for an expansion of the MEC in the form
\begin{equation}
J(p,h)=\frac{N(q,\omega,k_F)}{\sqrt{1+\tau}}J_{\rm non\, rel}(p,h) \, ,
\end{equation}
where $N(q,\omega,k_F)$ is an appropriate normalization factor defined
by requiring
\begin{equation}
\lim_{h\rightarrow 0}\frac{J(p,h)}{J_{\rm rel}(p,h)}=1 \, ,
\end{equation}
{\em  i.e.}, the coincidence between the relativistic and the
approximate results at the quasielastic peak.  Obviously the factor
$N(q,\omega,k_F)$ is a function of $k_F$ also, since an integral over
the Fermi sphere is implicit in the definition of the MEC in the 1p-1h
channel, and it can be written in the form
\begin{equation}
N(q,\omega,k_F)=\sqrt{1+\tau}\frac{J_{\rm rel}(q,0)}{J_{\rm non\ rel}(q,0)} 
\, .
\end{equation}
A simple approximation for this function is not easy to obtain, since
it requires the knowledge of the exact relativistic answer.

\begin{figure}[hp]
\begin{center}
\leavevmode
\def\epsfsize#1#2{0.9#1}
\epsfbox[160 185 400 710]{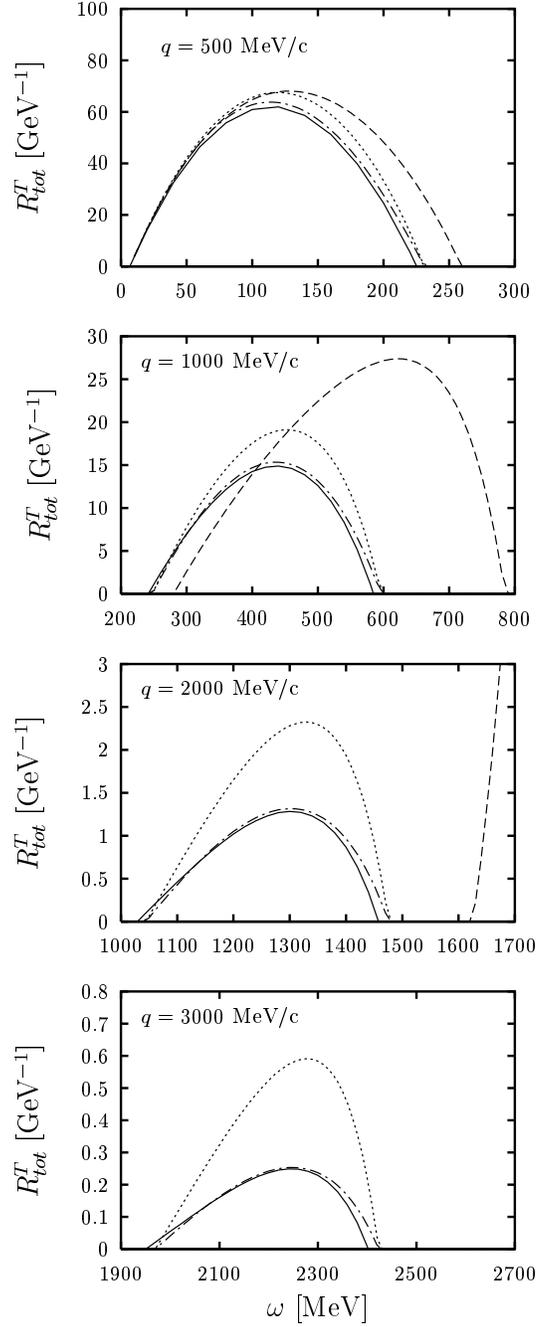}
\end{center}
\caption{\label{rep32} \small Total transverse response function of
$^{40}$Ca including MEC for several values of the momentum transfer,
and for $k_F=237$ MeV/c.  Solid: exact relativistic results.  The rest
of the curves have been computed using the non-relativistic Fermi gas
model, with or without relativistic corrections.  Dashed: traditional
non-relativistic results.  Dotted: including relativistic kinematics
in the non-relativistic calculations.  Dot-dashed: including in
addition the new expansion of the OB+MEC currents.  The relativistic
calculations include a dynamical propagator and $\pi N$ form factor,
while the non-relativistic calculations do not include these
corrections.  }
\end{figure}

Despite these difficulties, the quality of the OB expansion 
plus the approximated improvement of the MEC currents obtained in 
the present expansion are good enough to ensure a quite satisfactory 
description of the exact relativistic transverse response using the
relativized OB plus MEC operators altogether with relativistic
kinematics. This is shown in Fig.~\ref{rep32}, where we show the
total transverse response, including OB+MEC operators, for 
$q=500,$ 1000, 2000, and 3000 MeV/c. The solid lines are the exact
relativistic result. Again with dashed lines we display the traditional
non-relativistic results, which together with relativistic kinematics
give the dotted lines. Finally, with dot-dashed lines we show the 
results using the semi-relativized OB+MEC currents. 
The agreement between the two models is quite good even for very high 
$q$, since the major part of the relativistic effects is included in
the approximated model, and therefore these currents are very
appropriate and easy to implement in already existing non-relativistic
models of the reaction. 

\subsubsection{Comparison with the traditional relativistic
               corrections}
\label{Darwin-Foldy}
Here we discuss the reasons why the present expansion of
electromagnetic operators is preferable to other kinds of relativistic
corrections existing in the literature. The most common of these is
the Darwin-Foldy correction to the charge operator of 
the nucleus~\cite{Car98,deF66,Fri73,Cio80}. This correction is usually
derived
from a Foldy-Wouthuysen transformation~\cite{Bjo65}, but can
also be obtained from a Pauli reduction of the spin matrix element
(we do not write the spin indices for simplicity)
\begin{equation}
J_E^{\mu}(\np,\nh)
=\overline{u}_E(\np)\Gamma^{\mu}(Q) u_E(\nh)
=\overline{u}_E(\np)
\left( F_1\gamma^{\mu}+i\frac{F_2}{2m}\sigma^{\mu\nu}Q_{\nu}\right)
u_E(\nh) \, ,
\end{equation}
where we use the sub-index $E$ to denote the spinors normalized to 
$u_E^{\dagger}(\np) u_E(\np)=1$, {\em  i.e.}, 
namely
\begin{equation}
u_E(\np)=\left(\frac{E+m}{2E}\right)^{1/2}
\left( \begin{array}{c}  \chi   \\
                         \frac{\nsigma\cdot\np}{E+m}\chi
       \end{array}
\right) \, .
\end{equation}
This is in contrast to the Bjorken and Drell spinor normalization 
used in the present  work, where spinors are normalized to 
$u^{\dagger}(\np) u(\np)=E/m$. The  relation between the two sets of
spinors obviously is
\begin{equation}
u(\np)=\sqrt{\frac{E}{m}}u_E(\np) \, .
\end{equation}
Of course both formalisms based on the different spinor sets
$u_E(\np)$ or $u(\np)$ should give the same results for the observable
quantities. For instance, if we like to think in terms of wave
functions, for a nucleon in a box of volume $V$ this would be
\begin{equation}
\psi(x)=\sqrt{\frac{m}{EV}}u(\np) e^{-i p\cdot x}
=\frac{1}{\sqrt{V}}u_E(\np) e^{-i p\cdot x} \, .
\end{equation}
This means that observables
(expectation values, probabilities  or cross sections)
computed using the Bjorken and Drell normalization always 
contain additional phase-space factors $m/E$, while these factors do
not appear explicitly if one uses the $E$-scheme, since they
are already included inside the spinors $u_E(\np)$. 

As an example let us consider the  case of the 
longitudinal response function for protons
\begin{eqnarray}
R^L
&=&
\sum_{\np\nh}\sum_{s_p s_h}\delta(E_\np-E_\nh-\omega)
|\langle ph^{-1}|\rho(\nq)|F\rangle|^2 
\nonumber\\
&=&
\sum_{\np\nh}\delta(E_\np-E_\nh-\omega)
\frac{m^2}{E_\np E_\nh}\delta_{\np,\nh+\nq}
{\rm Tr}\, \left[\rho(\np,\nh)^{\dagger}\rho(\np,\nh)\right]
\nonumber\\
&=&
\left(\frac{3Z}{8\pi k_F^3}\right)
\int_{h<k_F} d^3 h \; \delta(E_\np-E_\nh-\omega)
\frac{m^2}{E_\np E_\nh}
{\rm Tr}\,  \left[\rho(\np,\nh)^{\dagger}\rho(\np,\nh)\right] \ ,
\end{eqnarray}
where in the last line $\np=\nh+\nq$, and we have used the replacement
$\sum_{\nh}\longrightarrow \frac{V}{(2\pi)^3}\int d^3h$, with
$V/(2\pi)^3=3Z/8\pi k_F^3$.  Note that we use the Bjorken and Drell
normalization and so the energy denominators appear explicitly. The
charge matrix element used here is the fully-relativistic one
$\rho(\np,\nh)=\overline{u}(\np)\Gamma^0(Q)u(\nh)$.

The interesting (and crucial) point is that
the energy denominator $E_\np$
cancels out when we perform the integral over $\cos\theta$ --- the
angle between $\nh$ and $\nq$ --- using the energy-conserving
delta function. In fact, from
\begin{equation} \label{rel-kinematics}
E_\np^2 = \np^2+m^2 = \nh^2+\nq^2+2hq\cos\theta+m^2
\end{equation}
we have $E_\np dE_\np= hqd\cos\theta$. Therefore the angle $\theta$ becomes
fixed by the energy conservation $E_\np=E_\nh+\omega$ and we obtain
\begin{eqnarray}
R^L(q,\omega)
&=&
\frac{3Z}{8\pi k_F^3}
\int_0^{k_F} hdh \int_0^{2\pi} d\phi \frac{E_\np}{q}\frac{m^2}{E_\np E_\nh}
{\rm Tr}\,  \left[\rho(\np,\nh)^{\dagger}\rho(\np,\nh)\right]
\nonumber\\
&=&
\frac{3Z}{8\pi k_F^3}\frac{m}{q}
\int_0^{k_F} hdh \int_0^{2\pi} d\phi \frac{m}{E_\nh}
{\rm Tr}\,  \left[\rho(\np,\nh)^{\dagger}\rho(\np,\nh)\right] \ .
\label{relativistic-response}
\end{eqnarray}

This expression has to be compared with the 
the non-relativistic response function, which can be computed 
by repeating the steps above using instead the 
non-relativistic energies $\epsilon_\np=\np^2/2m$, {\em  i.e.}, 
\begin{equation}
R^L(q,\omega)_{\rm non\, rel}=
\left(\frac{3Z}{8\pi k_F^3}\right)
\int_{h<k_F} d^3 h \; \delta(\epsilon_\np-\epsilon_\nh-\omega)
{\rm Tr}\,  \left[\rho(\np,\nh)^{\dagger}\rho(\np,\nh)\right]_{\rm n.r.} \, ,
\end{equation}
where again $\np=\nh+\nq$, no energy denominators 
appear and the non-relativistic charge operator is used.

Now the integral over $\cos\theta$ can again be performed using the
non-relativistic identity
\begin{equation} \label{non-rel-kinematics}
\epsilon_\np=\frac{\np^2}{2m}=\frac{\nh^2+\nq^2+2hq\cos\theta}{2m}\, .
\end{equation}
Hence $d\epsilon_\np= \frac{hq}{m}\cos\theta$
and the integral over the new variable $\epsilon_\np$ can be performed. 
The latter is $\epsilon_\np=\epsilon_\nh+\omega$. We obtain
\begin{equation}
R^L(q,\omega)_{\rm non\, rel}=
\frac{3Z}{8\pi k_F^3}\frac{m}{q}
\int_0^{k_F} hdh \int_0^{2\pi} d\phi
{\rm Tr}\,  \left[\rho(\np,\nh)^{\dagger}\rho(\np,\nh)\right]_{\rm n.r.} \, ,
\label{non-relativistic-response}
\end{equation}
which has formally the same structure as eq.~(\ref{relativistic-response})
with the exception of the factor $m/E_\nh \simeq 1$ included there.
Therefore, the relativistic response can be reproduced using a 
non-relativistic model if we introduce in the non-relativistic response 
in eq.~(\ref{non-relativistic-response}) a good
approximation for $\rho(\np,\nh)$, and  in addition  we use relativistic
kinematics, {\em  i.e.}, we use the relativistic relation between
$\cos\theta$
and $\omega$. This can be approximately accomplished 
starting from eq.~(\ref{rel-kinematics}). Indeed we have
\begin{eqnarray} 
h^2+q^2+2hq\cos\theta
&=&
(E_\nh+\omega)^2-m^2 = h^2+\omega^2+2E_\nh+\omega  
\nonumber\\
&\simeq&  h^2+\omega^2+2m\omega+2\epsilon_\nh\omega
\nonumber\\
&=&
2m\left(\omega+\epsilon_\nh+\frac{\omega^2+2\epsilon_\nh\omega}{2m}\right)
\nonumber\\
&\simeq& 2m\left(\omega+\epsilon_\nh+\frac{\omega^2}{2m}\right) \ ,
\end{eqnarray}
where we have neglected the term $\epsilon_\nh\omega/m=O(h^2/m^2)$. 
Comparing with the non-relativistic relation (\ref{non-rel-kinematics})
we see that the relativistic one can be approximately obtained with
the replacement $\omega\rightarrow\omega(1+\omega/2m)$.
The validity of this approximate method of relativization was demonstrated
numerically in the last sections. 

If instead we use the other spinor normalization  $u_E(\np)$
in the relativistic model, then the
matrix element is $\rho_E(\np,\nh)$ and  now there are no 
explicit energy denominators in the expression of the response, which hence
becomes, after integration over $\theta$,
\begin{equation}
R^L(q,\omega)=
\frac{3Z}{8\pi k_F^3}\frac{m}{q}
\int_0^{k_F} hdh \int_0^{2\pi} d\phi \frac{E_\np}{m}
{\rm Tr}\,  \left[\rho_E(\np,\nh)^{\dagger}\rho_E(\np,\nh)\right] \ .
\end{equation}
Comparing with eq.~(\ref{non-relativistic-response}) we see that if we
start with a non-relativistic model and use an approximate
non-relativistic form for $\rho_E(\np,\nh)$, as happens with the
Darwin-Foldy correction, an additional factor $E_\np/m$ is needed in
order to reproduce the relativistic response.  For this reason, a
careless introduction of relativistic corrections alone in
non-relativistic models can produce incorrect results.

\subsection{Pionic correlations}
\label{sec:NRcorr}

The analysis carried out in the previous section for the MEC could in
principle be extended to the correlation current.  However in this
case the calculation becomes extremely cumbersome and has not yet been
performed.  A semi-relativistic calculation of the vertex and
self-energy correlations has been carried out
in~\cite{Bar93,Bar94,Alb88}, where the relativistic energy-conserving
delta function has been accounted for via the replacement in
eq.~(\ref{pres}) and the form factors in the two-body current have
been modified to implement relativistic effects. The response
functions so obtained are in qualitative agreement with the
fully-relativistic ones for not too high $q$.  However, for high
values of $q$ a careful treatment of the relativistic effects is
needed.

In what follows we briefly examine the non-relativistic limit of the
vertex correlations and self-energy diagrams in order to bring to
light some differences with respect to the fully-relativistic case.

The non-relativistic leading order of the pionic correlation currents
in eqs.~(\ref{VC},\ref{self-energy-ph}) is obtained by using the
following prescriptions, valid in the static limit:
\begin{eqnarray}
E_\nk &\simeq& m 
\\
\gamma_5\Kbar &\simeq& \nsigma\cdot\nk 
\\
\frac{1}{K^2-m_\pi^2} &\simeq& -\frac{1}{\nk^2+m_\pi^2}
\\
S_F(P) &\simeq& S_{nr}(P) = \frac{1}{p_0-\frac{\np^2}{2m}} \ .
\label{nr-propagator}
\end{eqnarray}
The electromagnetic form factor $\Gamma^\mu(Q)$ is also replaced by
$\Gamma^{\mu}_{nr}(q)$, representing the usual non-relativistic
one-body current acting over bi-spinors~\cite{Ama98a,Alb90}.  Using
the above relations and performing the sums over spin and isospin
indices, the VC and SE current matrix elements read

\begin{eqnarray}
j^{\mu}_{VC}(\np,\nh)_{nr}
&=& 
\frac{f^2}{Vm_\pi^2} 
\chi_{s_p}^{\dagger}\sum_{\nk\leq k_F}
\left\{ \frac{\sigvec\cdot(\nk-\nh)}{(\nk-\nh)^2+m_\pi^2}
S_{nr}(K+Q) \tau_a \Gamma^{\mu}_{nr}(Q) \tau_a 
\sigvec\cdot(\nk-\nh)
\right.
\nonumber\\
&+& 
\left. \sigvec\cdot(\np-\nk)
\tau_a \Gamma^{\mu}_{nr}(Q) \tau_a S_{nr}(K-Q) 
\frac{\sigvec\cdot(\np-\nk)}{(\np-\nk)^2+m_\pi^2} \right\} 
\chi_{s_h}
\label{vcnr}
\end{eqnarray}
and

\begin{equation}
j^{\mu}_{SE}(\np,\nh)_{nr}
\simeq 
\chi_{s_p}^{\dagger}
\left[
       \Sigma_{nr}(\np)S_{nr}(P)\Gamma^{\mu}_{nr}(Q)
      +\Gamma^{\mu}_{nr}(Q)S_{nr}(H)\Sigma_{nr}(\nh)
\right]
\chi_{s_h} \ ,
\end{equation}
where $\chi_{s_p}$ and $\chi_{s_h}$ are two-components spinors.
The non-relativistic self-energy function is given by
\begin{equation} 
\label{nr-self-energy}
\Sigma_{nr}(\np)=3\frac{f^2}{Vm_\pi^2}\sum_{\nk\leq k_F}
\frac{(\np-\nk)^2}{(\np-\nk)^2+m_\pi^2}
=\Sigma_{nr}(|\np|) \ .
\label{Signr}
\end{equation}
The above expressions coincide with the traditional non-relativistic
currents used in the literature.

With the self-energy in eq.~(\ref{Signr})
one can then construct the non-relativistic Fock 
nucleon propagator
\begin{equation}
S^{HF}_{nr}(p_0,\np)=
\frac{1}{p_0-\frac{\np^2}{2m}}
+\frac{1}{p_0-\frac{\np^2}{2m}}
\Sigma_{nr}(\np)
\frac{1}{p_0-\frac{\np^2}{2m}}
+\cdots = 
\frac{1}{p_0-\frac{\np^2}{2m}-\Sigma_{nr}(\np)}\ .
\label{nr-full-propagator}
\end{equation}
As is well-known, this is a meromorphic function whose simple pole 
again defines the new energy of the nucleon in the medium, namely
\begin{equation}\label{nr-energy}
\epsilon_{nr}(\np)=\frac{\np^2}{2m}+\Sigma_{nr}(\np)\ ,
\end{equation}
since $\Sigma_{nr}(\np)$ is a function only of $\np$.

Since the non-relativistic self-energy function in
eq.~(\ref{nr-self-energy}) does not depend on spin, the nucleon wave
functions are not modified in the medium. In fact the corresponding
Schr\"odinger equation in momentum space, including the self-energy,
is simply given by
\begin{equation}
\left[\frac{\np^2}{2m}+\Sigma_{nr}(\np)
\right] \phi_{nr}(\np) = p_0\phi_{nr}(\np)\ ,
\end{equation}
with the bi-spinor $\phi_{nr}(\np)$ corresponding to the eigenvalue 
$p_0=\epsilon_{nr}(\np)$. 

The non-relativistic analysis of the nucleon self-energy
current~\cite{Fet71} is much simpler than its relativistic
counterpart. Indeed, in the former the self-consistency is immediately
achieved because the nucleon wave functions are not modified by the
self-energy interaction and thus the first iteration of the
``Hartree-Fock'' equations already provides the exact energy.  By
contrast, in the relativistic framework the spin dependence of the
self-energy~\cite{Ser86} modifies the Dirac-spinors, inducing an
enhancement of the lower components. Moreover, the field-strength
renormalization constant, namely the residue of the nucleon propagator
in eq.~(\ref{nr-full-propagator}) at the pole, in the non-relativistic
case is just unity.  Hence the enhancement of the lower components and
the spinors' field strength renormalization are genuine relativistic
effects absent in a non-relativistic analysis where only the
energy-momentum relation in the medium is altered by the self-energy
diagrams.  We have shown in Section~\ref{sec:PCSE} that the two
above-mentioned relativistic signatures can be incorporated as new
pieces in the electromagnetic current acting over free spinors.


\section{Conclusions}
\label{sec:concl}


Our goal in these studies has been to explore some of the ingredients
that enter at high energies where relativistic effects become
important in attempting to model the nuclear response functions for
inclusive quasielastic electron scattering. The full problem of
accounting for relativistic dynamics in nuclear physics is a daunting
one and far from being solved~\cite{Dek92,Dek94}.  While in many
papers it appears that a reasonable level of understanding has been
reached~\cite{Wil97,Orl91,Jou95,Jou96}, since the basic trends seen in
the data are reproduced, closer scrutiny reveals a different
situation. It is not only that contributions left out in various
analyses are far from being small, but, even more serious, fundamental
physics principles (Lorentz covariance, gauge invariance and
unitarity) turn out patently to be violated. Thus the successes in
reproducing the experiments often reflect more an adjusting of
parameters than a real understanding of the physics involved in the
quasielastic regime.

Our approach has been less to use a highly elaborated non-relativistic
model whose failings are expected at the outset than to employ a
simple model in which the important consequences of relativity are
hopefully present. For this we have begun with the relativistic Fermi
gas as our starting point~\cite{Cen97,Cen00}.  This approach is
motivated by several critical features of the model, namely, that it
is Lorentz covariant, that it allows the implementation of gauge
invariance and that is it simple enough to be tractable and yet not
obviously lacking at least for the quasielastic responses for which it
is designed. Clearly it is not an appropriate way to proceed if
near-Fermi-surface physics is the goal and this regime is not our
focus.

With these as basic motivations in a series of papers we have explored
the consequences of having a Lorentz covariant model.  In particular
in~\cite{Ama98a} we attempted to approximate the full theory by
identifying a dimensionless variable that is small enough to be
suitable in setting up a semi-relativistic expansion of the responses
(namely the momentum of a nucleon lying below the Fermi surface
compared with its mass). In contrast, in very recent 
work~\cite{Ama02} no expansion whatsoever is involved and the theory is
now fully-relativistic.

Our treatment proceeds in terms of nucleonic and mesonic degrees of
freedom (the latter viewed both as force and current carriers).  As
our aim is to study the quasielastic regime where the longest-range
hadronic ingredients may be expected to be dominant, we focus on
pions; studies using a larger set of hadrons can be undertaken and
some steps have already been taken by us in that direction. In our
model, the pions are dealt with to first order in a perturbative
framework, since their effects on the free responses of the RFG are
not expected to be too disruptive.

Gauge invariance is a fundamental property we have also addressed in very
recent work~\cite{Ama02}. We now understand how the continuity equation
is satisfied order by order in perturbation theory.
We have succeeded in showing that the continuity equation for the
one-body (single-nucleon) and the two-body (MEC and correlations) currents
is fulfilled, implying that our approach deals consistently with both forces
and currents. 

Given the point in our understanding of the quasielastic responses at
relatively high energies, we have been motivated to provide a
comprehensive set of discussions of progress made so far. In
particular, as a more in-depth presentation of the analysis carried
out in~\cite{Ama02}, where we first studied the fully-relativistic set
of one-pion-exchange operators that contribute to the electromagnetic
responses of nuclei in the 1p-1h channel, in the present work we have
gone further to answer the question of whether or not a {\em finite}
OPE self-energy current in nuclear matter exists.  Indeed we have
proven that the latter can be obtained through a renormalization of
the 1p-1h excitation vertex with a Fock self-energy insertion in the
particle or in the hole line. In~\cite{Ama02} these diagrams were
shown to diverge but, at the same time, to be crucial to preserve
gauge invariance. To overcome this impasse in that work we abandoned
the notion of current operators, using instead the polarization
propagator for computation of these diagrams. Indeed the double pole
appearing in the self-energy polarization propagator can be dealt with
employing the derivative of the nucleon propagator.

In assessing the role of the pions in the electromagnetic nuclear
responses, the MEC are not the only contributions that arise in
first-order perturbation theory. In fact the pionic correlations are
intimately linked to MEC through the continuity equation and, as we
have seen, only when the full set of Feynman diagrams with one
pion-exchange is considered can one expect gauge invariance to be
fulfilled.

Since all of these ingredients are required for a consistent theory,
a question we have addressed in this paper is whether or not a unified
treatment based on current operators at the level of the OPE can be
used even for the self-energy contribution. We succeeded in achieving
this goal introducing a new ingredient: the first-order correction to 
the wave function and energy of the nucleon in the medium, which is 
modified by the interaction with the other nucleons. Indeed the
iteration of the
self-energy diagrams generates a ``dressed'' propagator in the medium.
By the same token the self-energy generates ``dressed'' or
``renormalized'' wave functions in the medium, solutions of an
in-medium Dirac equation, where the self-energy plays the role of a
mean relativistic potential.  This equation also provides the
dispersion relation linking the energy and momentum of the nucleon in
the medium. Importantly, the new spinors should be multiplied by a
renormalization function $\sqrt{Z_2(\np)}$.

As the self-energy is generated by the interaction of a nucleon with
the other nucleons in the medium, the solutions of the new Dirac
equation should be used as input to re-compute the self-energy and so
on. The exact answer is obtained through a self-consistent procedure.
In this paper, however, we have just considered the first iteration:
we have thus computed the self-energy current confining ourselves to
first-order corrections to the energy and spinors -- or, equivalently,
to corrections linear in the self-energy -- which correspond to
diagrams with only one pionic line, in order to be consistent with the
MEC and vertex correlation currents.

Notably in the first-order expansion of the renormalized spinors two
new elements with respect to the non-relativistic approach emerge, one
arising from the negative-energy components in the wave function
produced by the interaction, the other from the renormalization
function $\sqrt{Z_2(\np)}$.  These two elements can be combined in a
new renormalized self-energy current, $j^{\mu}_{RSE}$, acting over
free spinors, and, together with renormalized self-energies, lead to
the same self-energy contribution of~\cite{Ama02}.  The introduction
of renormalized energies produces a shift of the response function.
Our results for the response functions for typical kinematics show
that the negative energy components constitute a correction to the
total self-energy contribution of roughly 10--20\%, whereas the
renormalization function for OPE is small, yet necessary if gauge
invariance is to be fulfilled exactly. Moreover, while at low momentum
transfers both particle and hole contributions play a role in the
response, at high $q$ only the hole contribution survives.  Finally,
the self-energy contribution to the response functions is comparable
in size to the one arising from the MEC and vertex correlations.

These formal developments have been discussed at length in the present
work; and not to interrupt the flow of the arguments unduly 
some details have been placed in a series of Appendices. 

In the remainder of the article we have presented some typical
results, both for parity-conserving and parity-violating quasielastic
electron scattering. Briefly we have found the following: we have
found that the MEC contributions are small enough to be well handled
in first order.  In particular both the pion-in-flight and seagull
contributions are very small in the L channel where the virtual photon
exchanged between the electron and the Fermi gas couples to the charge
of the pion, implying as expected that the MEC only marginally affect
the Coulomb sum rule. In contrast in the T channel the MEC are more
significant. There the seagull contribution dominates, and one sees
that the MEC contribution does not vanish when $q$
increases. In~\cite{Ama02} the scaling behaviors of the MEC were also
explored in detail: in summary it was seen that they break scaling of
the second kind everywhere, but, while breaking scaling of the first
kind at modest momentum transfers, tend to successful first-kind
scaling behavior at sufficiently high values of $q$.

The correlation contribution arising from the vertex corrections (VC)
display a different pattern: the L channel dominates over the T
channel by an amount of roughly 3:1. Thus the longitudinal response
effectively picks up only these correlation contributions, since the
MEC effects are so small there, and the former contribute to the total
at roughly the 10--15\% level. Indeed, were these to be the only
contributions needed in addition to the RFG response itself, then we
would expect the total to shift in $\omega$. Note that, since the
correlation contributions are roughly symmetrical about the
quasielastic peak, their impact on the Coulomb sum rule should be very
small, perhaps only at the few percent level. The correlation
contribution to $R^T$ is similar to the MEC contribution, but is
smaller, roughly 1/2 the size of the latter; since the two are of
opposite sign, they tend to cancel and thus the total is similar to
the MEC contribution but is cut down by a roughly factor of two.

In summary, the total contribution
(the sum of 1p-1h MEC + 1p-1h correlations) to be added to the RFG response
(1) is not insignificant, 
(2) is Lorentz covariant/gauge invariant and interestingly 
(3) does not go away as $q$ becomes very large.

\subsection*{Acknowledgments}
J.E.A. wants to thank J. Nieves for useful discussions. 
This work was partially supported by funds provided by DGICYT (Spain) 
under Contracts Nos. PB/98-1111, PB/98-0676 and PB/98-1367 and the Junta
de Andaluc\'{\i}a (Spain), by the Spanish-Italian Research Agreement
HI1998-0241, by the ``Bruno Rossi'' INFN-CTP Agreement, by the 
INFN-CICYT exchange and in part by the U.S. Department of Energy under
Cooperative Research Agreement No. DE-FC02-94ER40818.

\appendix


\section{Gauge invariance of two-body currents}
\label{app:A}


In this Appendix we prove that the total two-body current is gauge
invariant at the level of the two-body matrix elements in free space.
We start by evaluating the contraction of the four-momentum transfer
$Q_\mu$ with the correlation current
$j^\mu_{cor}(\np'_1,\np'_2,\np_1,\np_2)$.  It can be written as
\begin{equation}
Q_\mu j^\mu_{cor}(\np'_1,\np'_2,\np_1,\np_2)
=  \frac{f^2}{m_\pi^2}
              \overline{u}(\np'_1)\tau_a\gamma_5\Kbar_1u(\np_1)
              \frac{1}{K_1^2-m_\pi^2} {\cal M}_a 
             +(1 \leftrightarrow 2)
\end{equation}
with ${\cal M}_a$ given by
\begin{equation}
{\cal M}_a =\overline{u}(\np'_2)
    \left[    \tau_a\gamma_5\Kbar_1 
              S_F(P_2+Q)\Qbar F_1
            + F_1 \Qbar S_F(P'_2-Q)
              \tau_a\gamma_5\Kbar_1
    \right]u(\np_2) \ ,
\end{equation}
where we have used the relation $Q_\mu \Gamma^\mu(Q) = F_1(Q)\Qbar$. 
After some algebra, involving the nucleon propagator and the Dirac spinors, 
${\cal M}_a$ can be further simplified leading to 
\begin{equation}
{\cal M}_a=\overline{u}(\np'_2)
    \left[    \tau_a\gamma_5\Kbar_1 F_1 
            - F_1\tau_a\gamma_5\Kbar_1
    \right]u(\np_2) =\overline{u}(\np'_2)
    \left[\tau_a, F_1\right]
    \gamma_5\Kbar_1 
    u(\np_2) \ .
\end{equation}
To evaluate the commutator $[\tau_a,F_1]$
we now decompose the nucleon form factor into its isoscalar and isovector
pieces,
$F_1=\frac12\left( F_1^S+F_1^V\tau_3 \right)$. Then
\begin{equation}
\left[\tau_a,F_1\right] = -iF_1^V\epsilon_{3ab}\tau_b \ ,
\label{commutator}
\end{equation}
which entails the automatic conservation of the $\pi^0$ exchange current
($a$=3).
Using eq.~(\ref{commutator}) we can recast ${\cal M}_a$ as follows
\begin{equation}
{\cal M}_a = -i F_1^V \epsilon_{3ab}
\overline{u}(\np'_2)
    \tau_b\gamma_5\Kbar_1 
    u(\np_2) \ .
\end{equation}
Hence the divergence of the two-body correlation current matrix element
can finally be written as
\ba \label{div1}
Q_\mu j^\mu_{cor}(\np'_1,\np'_2,\np_1,\np_2)
&=& 2m \frac{f^2 }{m_\pi^2}
              i\epsilon_{3ab}
              \overline{u}(\np'_1)\tau_a\gamma_5 u(\np_1)
              \frac{F_1^V}{K_1^2-m_\pi^2}
\nonumber\\
&\times&              \overline{u}(\np'_2)
              \tau_b\gamma_5(\Qbar+2m) 
              u(\np_2)
             +(1 \leftrightarrow 2) \ .
\ea

The divergence of the seagull and pion-in-flight two-body current matrix 
elements can also be calculated in a straightforward way. The final result
reads
\begin{eqnarray}
Q_\mu j^\mu_s(\np'_1,\np'_2,\np_1,\np_2)
&=&  -2m\frac{f^2 }{m_\pi^2}
              i\epsilon_{3ab}
              \overline{u}(\np'_1)\tau_a\gamma_5 u(\np_1)
              \frac{F_1^V}{K_1^2-m_\pi^2}
\nonumber\\
&\times&
              \overline{u}(\np'_2)
              \tau_b\gamma_5\Qbar 
              u(\np_2)
             +(1 \leftrightarrow 2)
\label{div2}
\\
Q_\mu j^\mu_p(\np'_1,\np'_2,\np_1,\np_2)
&=& 4m^2 \frac{f^2}{m_\pi^2}
              i\epsilon_{3ab} F_\pi
              \frac{(K_1-K_2)\cdot Q}{(K_1^2-m_\pi^2)(K_2^2-m_\pi^2)}
\nonumber\\
&\times&      \overline{u}(\np'_1)\tau_a\gamma_5u(\np_1)
              \overline{u}(\np'_2)\tau_b\gamma_5u(\np_2) \ .
\label{div3}
\end{eqnarray}
Then, by summing up the contributions given by the correlation 
(eq.~(\ref{div1})) and seagull (eq.~(\ref{div2})) 
currents and writing the four-momentum transfer
as $Q_\mu=(K_1+K_2)_\mu$, we finally obtain
\begin{eqnarray} 
\lefteqn{Q_\mu 
\left[  j^\mu_{cor}(\np'_1,\np'_2,\np_1,\np_2)
      + j^\mu_s(\np'_1,\np'_2,\np_1,\np_2)
\right]}
\nonumber\\
&=&  
4m^2 \frac{f^2}{m_\pi^2}
F_1^V
              i\epsilon_{3ab}
  \frac{(K_2-K_1)\cdot Q}{(K_1^2-m_\pi^2)(K_2^2-m_\pi^2)}
            \overline{u}(\np'_1)\tau_a\gamma_5u(\np_1)
               \overline{u}(\np'_2)
              \tau_b\gamma_5 u(\np_2) \ ,
\end{eqnarray}
which cancels exactly the contribution of
pion-in-flight current in eq.~(\ref{div3}) provided the 
electromagnetic pion form factor is chosen to be $F_\pi=F_1^V$.
 

\section{Gauge invariance of the two-body current p-h matrix elements}
\label{app:B}


Following the study of gauge invariance at the level of the free-space
particle-particle matrix elements, here we extend the analysis to the
particle-hole channel, deriving the contribution to the continuity
equation of the isoscalar and isovector SE, VC and MEC particle-hole
matrix elements.  We start by evaluating the divergence of the
correlation particle-hole matrix element 
$j^\mu_{cor}(\np,\nh)$ for the SE and VC contributions; next we
address the MEC p-h matrix elements.

\paragraph{$\bullet$ Self energy (SE)}\mbox{}

\noindent From eqs.~(\ref{Sp},\ref{Sh}) we get 
\begin{eqnarray}
Q\cdot {\cal H}_p &=&
-\frac{3f^2}{2m Vm_\pi^2} 
\sum_{\nk\leq k_F}\frac{m}{E_\nk}
\overline{u}(\np) 
\frac{(\Pbar-\Kbar) (\Kbar - m) (\Pbar-\Kbar)}{(P-K)^2-m_\pi^2} 
              S_F(P) F_1 \Pbar u(\nh) \\
Q\cdot{\cal H}_h &=&
-\frac{3f^2}{2m Vm_\pi^2} 
\sum_{\nk\leq k_F}\frac{m}{E_\nk}
\overline{u}(\np) 
             F_1 \Qbar S_F(H)
              \frac{(\Kbar-\Hslash)(\Kbar - m)(\Kbar-\Hslash)}
{(K-H)^2-m_\pi^2} u(\nh) \ .
\end{eqnarray}
Note that $F_1$ cannot be taken out of the matrix element, since it acts 
on the isospinors. Now from the relations
\ba
S_F(P) \Qbar u(\nh) &=& u(\nh)
\\
\overline{u}(\np) \Qbar S_F(H) &=& - \overline{u}(\np) 
\\
\overline{u}(\np) (\Pbar-\Kbar)(\Kbar - m) &=& 2m \,\overline{u}(\np) 
(\Pbar-\Kbar) 
\\
(\Kbar - m) (\Kbar-\Hslash) u(\nh) &=& -2m (\Kbar - m)u(\nh) 
\ea
the following expressions are derived:
\ba
Q\cdot{\cal H}_p &=&
-\frac{3f^2}{Vm_\pi^2}
\sum_{\nk\leq k_F}\frac{m}{E_\nk}
\overline{u}(\np) 
\frac{(\Kbar - m) (\Pbar-\Kbar)}{(P-K)^2-m_\pi^2}  F_1 u(\nh)
\label{Spslash} \\
Q\cdot{\cal H}_h &=&
-\frac{3f^2}{Vm_\pi^2}
\sum_{\nk\leq k_F}\frac{m}{E_\nk}
\overline{u}(\np) F_1 
              \frac{(\Kbar-\Hslash) (\Kbar - m)}{(K-H)^2-m_\pi^2} u(\nh) \ .
\label{SHslash}
\ea

\paragraph{$\bullet$ Vertex correlations (VC)}\mbox{}

\noindent
From eqs.~(\ref{F},\ref{B}) the four-divergence of the VC matrix element 
is found to be
\ba
Q\cdot{\cal F} 
=  -\frac{f^2}{Vm_\pi^2}
\sum_{\nk\leq k_F}\frac{m}{E_\nk}
\overline{u}(\np) \gamma_5(\Kbar-\Hslash)
              S_F(K+Q)\tau_a F_1\tau_a \Qbar \gamma_5
              \frac{\Kbar-m}{(K-H)^2-m_\pi^2} u(\nh)
\\
Q\cdot{\cal B}
= -\frac{f^2}{Vm_\pi^2}
\sum_{\nk\leq k_F}\frac{m}{E_\nk}
\overline{u}(\np) 
              \frac{\Kbar-m}{(P-K)^2-m_\pi^2}  
            \tau_a F_1 \tau_a \gamma_5 \Qbar S_F(K-Q) \gamma_5
              (\Pbar-\Kbar)u(\nh) .
\ea
We now exploit the identities
\ba
S_F(K+Q) \Qbar (\Kbar + m) &=& +(\Kbar + m)
\\
(\Kbar + m) \Qbar S_F(K-Q) &=& -(\Kbar + m)
\label{ident}
\ea
to get finally 
\ba
Q\cdot{\cal F}
=            \frac{f^2}{Vm_\pi^2}
\sum_{\nk\leq k_F}\frac{m}{E_\nk}
\overline{u}(\np)  \tau_a F_1\tau_a 
\frac{(\Kbar-\Hslash)(\Kbar-m)}{(K-H)^2-m_\pi^2} u(\nh)
\label{Fslash}
\\
Q\cdot{\cal B}
=            \frac{f^2}{Vm_\pi^2}
\sum_{\nk\leq k_F}\frac{m}{E_\nk}
\overline{u}(\np) \tau_a F_1 \tau_a
              \frac{(\Kbar - m)(\Pbar-\Kbar)}{(P-K)^2-m_\pi^2} 
              u(\nh) .
\label{Bslash}
\ea
If the expressions 
(\ref{Spslash},\ref{SHslash},\ref{Fslash},\ref{Bslash})
are split into their isoscalar and isovector parts, as illustrated in 
Section~\ref{sec:PCph}, we get
\ba
Q\cdot{\cal H}^{(S)}_p &=&
-\frac{3f^2}{Vm_\pi^2} F_1^{S} 
\sum_{\nk\leq k_F}\frac{m}{E_\nk}
\overline{u}(\np) 
\frac{(\Kbar - m)(\Pbar-\Kbar)}{(P-K)^2-m_\pi^2} u(\nh)
\\
Q\cdot{\cal H}^{(V)}_p &=&
-\frac{3f^2}{Vm_\pi^2} F_1^{V} 
\sum_{\nk\leq k_F}\frac{m}{E_\nk}
\overline{u}(\np) 
\frac{(\Kbar - m) (\Pbar-\Kbar)\tau_3}{(P-K)^2-m_\pi^2} u(\nh)
\\
Q\cdot{\cal H}^{(S)}_h &=&
-\frac{3f^2}{Vm_\pi^2}  F_1^{S} 
\sum_{\nk\leq k_F}\frac{m}{E_\nk}
\overline{u}(\np) 
\frac{(\Kbar-\Hslash)(\Kbar - m)}{(K-H)^2-m_\pi^2} u(\nh) 
\\
Q\cdot{\cal H}^{(V)}_h &=&
-\frac{3f^2}{Vm_\pi^2}  F_1^{V} 
\sum_{\nk\leq k_F}\frac{m}{E_\nk}
\overline{u}(\np) 
\frac{(\Kbar-\Hslash)(\Kbar - m)\tau_3}{(K-H)^2-m_\pi^2} u(\nh) 
\\
Q\cdot{\cal F}^{(S)}
&=&            +\frac{3 f^2}{Vm_\pi^2} F_1^{S}
\sum_{\nk\leq k_F}\frac{m}{E_\nk}
\overline{u}(\np)
\frac{(\Kbar-\Hslash)(\Kbar - m)}{(K-H)^2-m_\pi^2} u(\nh)
\\
Q\cdot{\cal F}^{(V)}
&=&            +\frac{ f^2}{Vm_\pi^2} F_1^{V}
\sum_{\nk\leq k_F}\frac{m}{E_\nk}
\overline{u}(\np) 
\frac{(\Kbar-\Hslash)(\Kbar - m)}{(K-H)^2-m_\pi^2}
(\tau_3+i\varepsilon_{3ab}\tau_a\tau_b) u(\nh)
\\
Q\cdot{\cal B}^{(S)}
&=&            +\frac{3f^2}{Vm_\pi^2} F_1^{S} 
\sum_{\nk\leq k_F}\frac{m}{E_\nk}
\overline{u}(\np) 
\frac{(\Kbar - m)(\Pbar-\Kbar)}{(P-K)^2-m_\pi^2} u(\nh) 
\\
Q\cdot{\cal B}^{(V)}
&=&            +\frac{f^2}{Vm_\pi^2} F_1^{V} 
\sum_{\nk\leq k_F}\frac{m}{E_\nk}
\overline{u}(\np) 
\frac{(\Kbar - m) (\Pbar-\Kbar)}{(P-K)^2-m_\pi^2}
(\tau_3+i\varepsilon_{3ab}\tau_a\tau_b) u(\nh) \ .
\label{gauge}
\ea

   From these relations we learn that:

\begin{itemize}
\item In the isoscalar channel the self-energy and vertex contributions 
cancel
\be
Q\cdot{\cal H}^{(S)}_p + Q\cdot{\cal B}^{(S)}=
Q\cdot{\cal H}^{(S)}_h + Q\cdot{\cal F}^{(S)} = 0 .
\ee
This differs from the non-relativistic result~\cite{Alb90}, 
where the self-energy is by itself gauge invariant.
\item In the isovector channel we get
\ba
Q\cdot\left[{\cal H}^{(V)}_p + {\cal B}^{(V)} \right]&=& 
\frac{2f^2}{Vm_\pi^2} F_1^{V}i\varepsilon_{3ab}
\sum_{\nk\leq k_F}\frac{m}{E_\nk}
\overline{u}(\np) 
\frac{(\Kbar - m)(\Pbar-\Kbar)\tau_a\tau_b}{(P-K)^2-m_\pi^2}
u(\nh)
\\
Q\cdot\left[{\cal H}^{(V)}_h + {\cal F}^{(V)}\right] &=&  
\frac{2 f^2}{ Vm_\pi^2} F_1^{V}i\varepsilon_{3ab}
\sum_{\nk\leq k_F}\frac{m}{E_\nk}
\overline{u}(\np) 
\frac{(\Kbar-\Hslash)(\Kbar - m)\tau_a\tau_b}{(K-H)^2-m_\pi^2}
u(\nh).
\ea
\end{itemize}
These expressions, using the Dirac equations $\Hslash u(\nh)=m u(\nh)$ and
$\overline{u}(\np) \Pbar = m \overline{u}(\np)$, can be
further simplified to yield the following four-divergence of the correlation 
current
\ba
& & Q\cdot j_{cor}(\np,\nh) =
\frac{1}{2}Q\cdot\left[{\cal H}^{(V)}_p +{\cal B}^{(V)}+
{\cal H}^{(V)}_h + {\cal F}^{(V)}\right]
\nonumber\\
&=& 
\frac{2f^2}{Vm_\pi^2} F_1^{V}i\varepsilon_{3ab}
\sum_{\nk\leq k_F}\frac{m}{E_\nk}
\overline{u}(\np) \tau_a \left\{
\frac{K\cdot P-m^2}{(P-K)^2-m_\pi^2} - \frac{K\cdot H-m^2}{(K-H)^2-m_\pi^2}
\right\} \tau_b u(\nh).
\ea
This contribution is exactly canceled by that of the MEC (seagull and
pion-in-flight) as we illustrate in what follows.

\paragraph{$\bullet$ MEC}\mbox{}

\noindent
Using the expressions given in in eqs.~(\ref{Sph},\ref{Pph}) for the
p-h matrix elements corresponding to the seagull and pion-in-flight
currents the associated four-divergences are found to be 
\begin{eqnarray}
\lefteqn{ Q\cdot j_s(\np,\nh) }
\nonumber\\
&=&-\frac{f^2}{Vm_\pi^2} F_1^{V}
i\varepsilon_{3ab} \sum_{\nk\leq k_F}\frac{m}{E_\nk}
\overline{u}(\np)\tau_a\tau_b \left\{
\frac{(\Kbar-m)\Qbar}{(P-K)^2-m_\pi^2} + 
\frac{\Qbar(\Kbar-m)}{(K-H)^2-m_\pi^2} \right\}u(\nh) 
\\
\lefteqn{ Q\cdot j_p(\np,\nh)}
\nonumber\\
&=& \frac{2mf^2}{Vm_\pi^2} F_1^{V}
i\varepsilon_{3ab} \sum_{\nk\leq k_F}\frac{m}{E_\nk} \frac{(Q^2+2H\cdot
Q-2K\cdot Q)}{[(P-K)^2-m_\pi^2] [(K-H)^2-m_\pi^2]}
\overline{u}(\np)\tau_a(\Kbar-m) \tau_b u(\nh). 
\end{eqnarray}
Exploiting the Dirac equation and after some algebra the above can be
recast as follows 
\begin{eqnarray}
\lefteqn{Q\cdot j_{MEC}(\np,\nh) =Q\cdot j_s(\np,\nh)+ Q\cdot j_p(\np,\nh)}
\nonumber\\ 
&=& - \frac{2f^2}{Vm_\pi^2} F_1^{V} i\varepsilon_{3ab}
\sum_{\nk\leq k_F} \frac{m}{E_\nk} \overline{u}(\np)\tau_a\left\{ \frac{K\cdot
P-m^2}{(P-K)^2-m_\pi^2}-\frac{K\cdot H-m^2}{(K-H)^2-m_\pi^2} \right\}
\tau_b u(\nh)   
\end{eqnarray} 
We have thus proven that the correlation and MEC p-h matrix elements
satisfy current conservation, {\em  i.e.,} 
$ Q\cdot j_{cor}(\np,\nh)+ Q\cdot j_{MEC}(\np,\nh)=0$.


\section{Polarization propagator with nucleon self-energy}
\label{app:C}


Here we evaluate the Feynman diagrams for the polarization propagator
with self-energy ($\Sigma$) insertions in the particle and hole lines,
depicted in Fig.~6, subdiagrams (g) and (h). From the general Feynman rules
for the polarization propagator~\cite{Fet71} we have 
\begin{eqnarray}
\Pi^{\mu\nu}_{SE}(Q) 
&=& -i\int\frac{dh_0 d^3h}{(2\pi)^4} 
     {\rm Tr} \left\{
     \Gamma^{\mu}(Q) S_0(H)\Sigma(H) S_0(H) \Gamma^{\nu}(-Q) S_0(P) \right.
\nonumber\\ 
&& \left. \kern 2cm \mbox{} 
     +\Gamma^{\mu}(Q) S_0(H) \Gamma^{\nu}(-Q)S_0(P)\Sigma(P)S_0(P)
   \right\} \ ,
\label{se-pol-prop}
\end{eqnarray}
where $P=H+Q$ and $S_0$ is the free relativistic propagator for a nucleon 
in the nuclear medium in eq.~(\ref{S0}), which
can also be written in the equivalent ways:
\begin{eqnarray}
S_0(K)
&=& (\Kbar+m)
    \left[\frac{1}{K^2-m^2+i\epsilon}
          + 2\pi i \theta(k_F-k)
          \delta(K^2-m^2)\theta(k_0)
    \right]
\nonumber \\
&=& (\Kbar+m)
      \left[ \frac{\theta(k-k_F)}{K^2-m^2+i\epsilon}
            +\frac{\theta(k_F-k)}{K^2-m^2-i\epsilon k_0}
      \right] \ .
\label{eq45}
\end{eqnarray}
The self-energy function $\Sigma$ is given by
eq.~(\ref{SE1}).

In order to simplify the calculation of the above polarization
propagator, we will simultaneously compute the two diagrams
contributing to eq.~(\ref{se-pol-prop}). First we note that
eq.~(\ref{se-pol-prop}) can be rewritten as:
\begin{equation}\label{se-pol-prop-ph}
\Pi^{\mu\nu}_{SE}(Q)=\Pi^{\mu\nu}_{10}(Q)+\Pi^{\mu\nu}_{01}(Q) \, ,
\end{equation}
where we introduce $\Pi^{\mu\nu}_{nl}(Q)$ 
as the polarization propagator
shown in Fig.~\ref{fprop}, containing $n$ self-energy insertions $\Sigma(H)$ 
in the hole line and $l$ insertions $\Sigma(P)$ in the particle line, i.e., 
\begin{equation}
\Pi^{\mu\nu}_{nl}(Q)
\equiv
-i {\rm Tr}\int\frac{dh_0 d^3h}{(2\pi)^4}
\Gamma^{\mu}(Q)[S_0(H)\Sigma(H)]^n S_0(H)\Gamma^{\nu}(-Q)
[S_0(P)\Sigma(P)]^l S_0(P) \, ,
\label{C1}
\end{equation}
where again $P=H+Q$. From this expression one can derive, as
particular cases, the leading-order response ($n=l=0$, no interaction
lines) and the first-order self-energy response (with one interaction
line, given by eq.~(\ref{se-pol-prop-ph})).

\begin{figure}[tp]
\begin{center}
\leavevmode
\def\epsfsize#1#2{0.9#1}
\epsfbox[100 440 500 710]{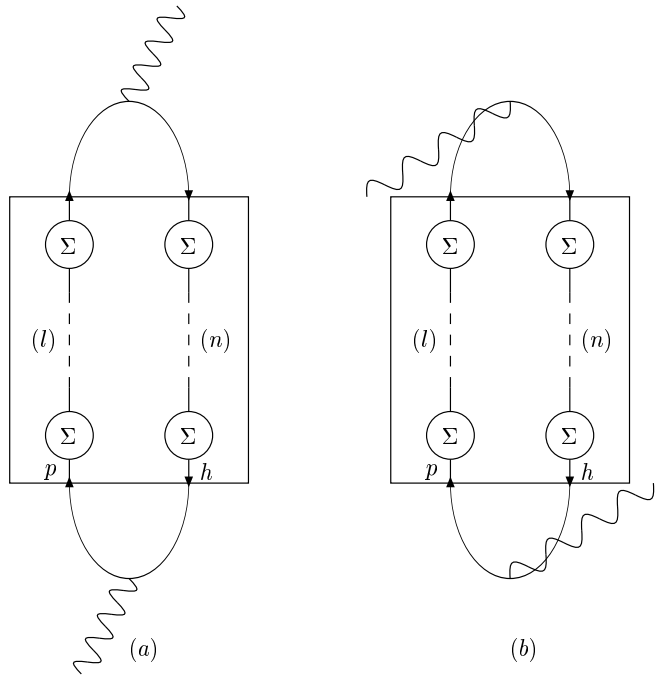}
\end{center}
\caption{\label{fprop} Diagrammatic definition of the polarization
propagator $\Pi^{\mu\nu}_{nl}$ for a $ph$ excitation with self-energy
insertions in the hole and particle lines.  Only the forward diagram
(a) contributes to the electromagnetic responses, while the backward
diagram (b) corresponds to a negative value of the energy transfer.}
\end{figure}

Using the nucleon propagator in the medium written in the form in
eq.~(\ref{eq45}),
the product of $n+1$ propagators appearing in
eq.~(\ref{C1}) 
can be expressed as a derivative of order $n$ according to
\begin{eqnarray}
[S_0(H)\Sigma(H)]^n S_0(H) 
&&
\nonumber\\
&&
\kern -8em
=
[(\Hslash+m)\Sigma(H)]^n
(\Hslash+m)
      \left[ \frac{\theta(h-k_F)}{(H^2-m^2+i\epsilon)^{n+1}}
            +\frac{\theta(k_F-h)}{(H^2-m^2-i\epsilon h_0)^{n+1}}
      \right]
\nonumber\\
&&
\kern -8em
=
[(\Hslash+m)\Sigma(H)]^n
(\Hslash+m)
\frac{1}{n!}
\left.\frac{d^n}{d\alpha^n}\right|_{\alpha=0}
      \left[ \frac{\theta(h-k_F)}{H^2-\alpha-m^2+i\epsilon}
            +\frac{\theta(k_F-h)}{H^2-\alpha-m^2-i\epsilon h_0}
      \right]
\nonumber\\
&&
\kern -8em
=
[(\Hslash+m)\Sigma(H)]^n
(\Hslash+m)
\nonumber\\
&&
\kern -8em
\times
\frac{1}{n!}
\left.\frac{d^n}{d\alpha^n}\right|_{\alpha=0}
\left[
       \frac{1}{H^2-\alpha-m^2+i\epsilon}
      + 2\pi i
        \theta(k_F-h)
        \delta(H^2-\alpha-m^2)\theta(h_0) 
\right] \, ,
\end{eqnarray}
where a parameter $\alpha$, which at the end is going to be zero,
has been introduced in the  propagator denominators.
A similar equation holds for the propagation of a particle introducing
a second parameter $\beta$.
The  polarization propagator $\Pi^{\mu\nu}_{nl}$ 
can then be written as
\begin{eqnarray}
\Pi^{\mu\nu}_{nl}(Q)
&=&
-i 
\left.\frac{d^n}{d\alpha^n}\right|_{\alpha=0}
\left.\frac{d^l}{d\beta^l}\right|_{\beta=0}
\int\frac{dh_0 d^3h}{(2\pi)^4}
I^{\mu\nu}_{nl}(H,P,Q)
\nonumber\\
&&\mbox{}\times
\left[
       \frac{1}{H^2-\alpha-m^2+i\epsilon}
      + 2\pi i
        \theta(k_F-h)
        \delta(H^2-\alpha-m^2)\theta(h_0) 
\right]
\nonumber\\
&&\mbox{}\times
\left[
       \frac{1}{P^2-\beta-m^2+i\epsilon}
      + 2\pi i
        \theta(k_F-p)
        \delta(P^2-\beta-m^2)\theta(p_0) 
\right]
\label{propagator-product}
\end{eqnarray}
with $P=H+Q$, and where we have introduced the functions
\begin{eqnarray}
\lefteqn{I^{\mu\nu}_{nl}(H,P,Q)=
I^{\mu\nu}_{nl}(h_0,\nh;p_0,\np;q_0,\nq)}
\nonumber\\
&\equiv&
\frac{1}{n!l!}
{\rm Tr}
\left\{
\Gamma^{\mu}(Q)[(\Hslash+m)\Sigma(H)]^n(\Hslash+m)\Gamma^{\nu}(-Q)
[(\Pbar+m)\Sigma(P)]^l (\Pbar+m)
\right\}.
\label{Inl-function}
\end{eqnarray}

The product of the two brackets inside the integral in 
eq.~(\ref{propagator-product}) 
gives rise to four terms. The first of these contains the product of the
two free propagators, namely
\begin{equation}
       \frac{1}{H^2-\alpha-m^2+i\epsilon}\times
       \frac{1}{P^2-\beta-m^2+i\epsilon} \ ,
\end{equation}
and yields a genuine vacuum contribution, $\Pi^{(0)\mu\nu}_{nl}(Q)$,
which diverges after integration. Therefore we subtract out its
contribution, since it pertains to a domain beyond nuclear physics.
Performing this subtraction of the vacuum propagator
we obtain
\begin{eqnarray}
\lefteqn{\Pi^{\mu\nu}_{nl}(Q)-\Pi^{(0)\mu\nu}_{nl}(Q)
=
2\pi
\left.\frac{d^n}{d\alpha^n}\right|_{\alpha=0}
\left.\frac{d^l}{d\beta^l}\right|_{\beta=0}
\int\frac{dh_0 d^3h}{(2\pi)^4}
I^{\mu\nu}_{nl}(H,P,Q)}
\nonumber\\
&&
\mbox{}\times
\left[
\rule[-7mm]{0mm}{14mm}
       \frac{\theta(k_F-p)
             \delta(P^2-\beta-m^2)\theta(p_0) 
            }{H^2-\alpha-m^2+i\epsilon}
+
        \frac{\theta(k_F-h)
             \delta(H^2-\alpha-m^2)\theta(h_0) 
            }{H^2-\beta-m^2+i\epsilon}
\right.
\nonumber\\
&&
\mbox{}+
\left.
\rule[-7mm]{0mm}{14mm}
        2\pi i 
        \theta(k_F-p)
        \theta(k_F-h)
        \delta(P^2-\beta-m^2)
        \delta(H^2-\alpha-m^2)
        \theta(P_0) \theta(H_0) 
\right] \, . 
\end{eqnarray}
Taking the imaginary part according to eq.~(\ref{eq4}) 
we obtain the corresponding hadronic tensor%
\footnote{The extra  factor $V$ appears since we are computing the
response function of an extended system, 
see eq.~(17.17) of ref.~\cite{Fet71}}
\begin{eqnarray}
\lefteqn{-\frac{V}{\pi} {\rm Im}\,
\left[\Pi^{\mu\nu}_{nl}(Q)-\Pi^{(0)\mu\nu}_{nl}(Q)\right]}
\nonumber \\
&=& 2\pi
\left.\frac{d^n}{d\alpha^n}\right|_{\alpha=0}
\left.\frac{d^l}{d\beta^l}\right|_{\beta=0}
\int\frac{dh_0 d^3h}{(2\pi)^4}
I^{\mu\nu}_{nl}(H,P,Q)
        \delta(P^2-\beta-m^2)
        \delta(H^2-\alpha-m^2)
\nonumber\\
&&\mbox{}\times
\left[ \theta(k_F-p)\theta(p_0)
      +\theta(k_F-h)\theta(h_0) 
      -2\theta(k_F-p)\theta(k_F-h)\theta(p_0)\theta(h_0) 
\right].
\end{eqnarray}
Now the factor containing the step functions can be expressed in the form
\begin{eqnarray}
\left[ \theta(k_F-p)\theta(p_0)
     + \theta(k_F-h)\theta(h_0) 
     -2\theta(k_F-p)\theta(k_F-h)\theta(p_0)\theta(h_0) 
\right]
\nonumber\\
\mbox{}=\theta(k_F-h)\theta(h_0) [1-\theta(k_F-p)\theta(p_0)]
       +\theta(k_F-p)\theta(p_0)[1-\theta(k_F-h)\theta(h_0)] 
\end{eqnarray}
so that the hadronic tensor can be written as a sum of two pieces
\begin{equation}
-\frac{V}{\pi} {\rm Im}\, [\Pi^{\mu\nu}_{nl} -\Pi^{(0)\mu\nu}_{nl}]
= W^{(+)\mu\nu}_{nl} + W^{(-)\mu\nu}_{nl} \ ,
\end{equation}
where 
\begin{eqnarray}
\lefteqn{W^{(+)\mu\nu}_{nl}(Q)}
\nonumber\\
&=&
2\pi V
\left.\frac{d^n}{d\alpha^n}\right|_{\alpha=0}
\left.\frac{d^l}{d\beta^l}\right|_{\beta=0}
\int\frac{dh_0 d^3h}{(2\pi)^4}
I^{\mu\nu}_{nl}(H,P,Q)
\delta(P^2-\beta-m^2)\delta(H^2-\alpha-m^2)
\nonumber\\
&&
\kern 3cm\mbox{}\times
\theta(k_F-h)\theta(h_0)[1-\theta(k_F-p)\theta(p_0)]
\label{wmunuplus}
\end{eqnarray}
corresponds to the hadronic tensor sought for
electron scattering (Fig.~\ref{fprop}(a)), whereas the second term
\begin{eqnarray}
\lefteqn{W^{(-)\mu\nu}_{nl}(Q)}
\nonumber\\
&=&
2\pi V
\left.\frac{d^n}{d\alpha^n}\right|_{\alpha=0}
\left.\frac{d^l}{d\beta^l}\right|_{\beta=0}
\int\frac{dh_0 d^3h}{(2\pi)^4}
I^{\mu\nu}_{nl}(H,P,Q)
\delta(P^2-\beta-m^2)\delta(H^2-\alpha-m^2)
\nonumber\\
&&
\kern 4cm\mbox{}\times
\theta(k_F-p)\theta(p_0)[1-\theta(k_F-h)\theta(h_0)]
\label{wmunuminus}
\end{eqnarray}
corresponds to a process with negative energy transfer
(Fig.~\ref{fprop}(b)); hence it does not contribute to the electron
scattering response and should be disregarded.

Finally, the integration with respect to $h_0$ 
in eq.~(\ref{wmunuplus}) can be explicitly performed
by using the $\delta$-functions.
One then gets the following expression for the $nl$-th SE contribution to the 
hadronic tensor 
\begin{eqnarray}
W^{\mu\nu}_{nl} 
\equiv
W^{(+)\mu\nu}_{nl}
&=&
V\left.\frac{d^n}{d\alpha^n}\right|_{0}
\left.\frac{d^l}{d\beta^l}\right|_{0}
\int\frac{d^3h}{(2\pi)^3}
\frac{I^{\mu\nu}_{nl}(E'_\nh(\alpha),\nh;E'_\np(\beta),\np;q)}
{4E'_\nh(\alpha)E'_\np(\beta)} 
\nonumber \\
&&\mbox{}\times
\delta(E'_\nh(\alpha)+q_0-E'_\np(\beta))
\theta(k_F-h)
        \theta(p-k_F) \, ,
\label{Wnl}
\end{eqnarray}
where $\np=\nh+\nq$ and we have defined the following energy 
functions of the parameters $\alpha$, $\beta$
\begin{eqnarray}
E'_\nh(\alpha) &=& \sqrt{\nh^2+\alpha+m^2} \label{energy-h}\\
E'_\np(\beta) &=& \sqrt{\np^2+\beta+m^2}.  \label{energy-p}
\end{eqnarray}
Expression (\ref{Wnl}) is the general equation for which we are searching.
It is one of the $(n+l)$-th order  contributions
to the full Hartree-Fock hadronic tensor, which is
an infinite sum of all perturbative orders. 
In the particular case  
$n=l=0$ it gives the well known free (OB) hadronic tensor
\begin{equation}
W^{\mu\nu}_{OB}= W^{\mu\nu}_{00}=
V \int\frac{d^3h}{(2\pi)^3}
\frac{I^{\mu\nu}_{00}(E_\nh,\nh;E_\np,\np;q)}
{4E_\nh E_\np}
\delta(E_\nh+q_0-E_\np)
        \theta(k_F-h)
        \theta(p-k_F) \ .
\end{equation}
Finally, the hadronic tensor corresponding to one self-energy insertion
in the particle or hole lines, corresponding to diagrams (g) and (h) in
Fig.~6 is
given by 
\begin{equation}
W^{\mu\nu}_{SE}= W^{\mu\nu}_{10}+W^{\mu\nu}_{01} \, ,
\end{equation}
where the $n=1$, $l=0$ term correspond to the the first-order hole
self energy diagram (Fig.~6(h)) 
\begin{eqnarray}\label{eq106}
W^{\mu\nu}_{10}&=&
V\left.\frac{d}{d\alpha}\right|_{\alpha=0}
\int\frac{d^3h}{(2\pi)^3}
\frac{I^{\mu\nu}_{10}(E'_\nh(\alpha),\nh;E_\np,\np;q)}
{4E'_\nh(\alpha)E_\np}
\delta(E'_\nh(\alpha)+q_0-E_\np)
        \theta(k_F-h)
        \theta(p-k_F) \nonumber \\
& &
\end{eqnarray}
and for $n=0$, $l=1$ the first-order particle self-energy 
diagram (Fig.~6(g))
\begin{eqnarray}\label{eq107}
W^{\mu\nu}_{01}&=&
V\left.\frac{d}{d\beta}\right|_{\beta=0}
\int\frac{d^3h}{(2\pi)^3}
\frac{I^{\mu\nu}_{01}(E_\nh,\nh;E'_\np(\beta),\np;q)}
{4E_\nh E'_\np(\beta)}
\delta(E_\nh+q_0-E'_\np(\beta))
        \theta(k_F-h)
        \theta(p-k_F) . \nonumber \\
& &
\end{eqnarray}

In the above expressions, after the derivatives with respect to the
parameters $\alpha$ and $\beta$ are taken, the integral over the hole
polar angle $\cos\theta_h$ can be performed analytically by exploiting
the $\delta$-function. Hence the SE contribution to the hadronic
tensor can finally be expressed as a double integral.  Since the
self-energy $\Sigma$ involves a triple integral, the contribution to
hadronic tensor turns out to be a 5-dimensional integral, to be
carried out numerically.


\section{Renormalized self-energy  
response using the polarization propagator}
\label{app:D}


In Appendix C we computed the first-order self-energy contribution to
the polarization propagator corresponding to the two diagrams of
Fig.~6.  The corresponding hadronic tensor splits into the sum of the
two terms given in eqs.~(\ref{eq106},\ref{eq107}) with Fock
self-energy insertions in the hole and particle lines respectively,
and reads
\begin{eqnarray}
\lefteqn{W^{\mu\nu}}
\nonumber\\
&&
=\left. V \frac{d}{d\alpha}\right|_{\alpha=0}
\int\frac{d^3h}{(2\pi)^3}
\frac{I^{\mu\nu}_{10}(E'_\nh(\alpha),\nh;E_\np,\np;q)}
{4E'_\nh (\alpha)E_\np}
\delta(E'_\nh (\alpha)+q_0-E_\np)
        \theta(k_F-h)
        \theta(p-k_F)
\nonumber\\
&&
+
\left. V \frac{d}{d\beta}\right|_{\beta=0}
\int\frac{d^3h}{(2\pi)^3}
\frac{I^{\mu\nu}_{01}(E_\nh,\nh;E'_\np(\beta),\np;q)}
{4E_\nh E'_\np(\beta)}
\delta(E_\nh+q_0-E'_\np(\beta))
        \theta(k_F-h)
        \theta(p-k_F) \, ,
\nonumber\\
\label{W-SE}
\end{eqnarray}
where $\np=\nh+\nq$, and the modified energies for holes and
particles have been introduced in
eqs.~(\ref{energy-h},\ref{energy-p}), with $\alpha$ and $\beta$ being
real parameters.  Finally the functions $I^{\mu\nu}_{nl}$ are 
defined in eq.~(\ref{Inl-function}).

In order to prove the equivalence
between the responses computed using the polarization propagator in
eq.~(\ref{W-SE}) and the result in eq.~(\ref{W-RSE}), obtained using
the renormalized current and energies, we proceed to perform the
derivative with respect to $\alpha$ and $\beta$. For a general
function $F(h_0)$ we have
\begin{equation}
\left.\frac{dF(E'_\nh(\alpha))}{d\alpha}\right|_{\alpha=0}= 
\frac{1}{2E_\nh}\left[\frac{dF(h_0)}{dh_0}\right]_{h_0=E_\nh} \ .
\end{equation}
Hence, interchanging the derivatives and the integral,
we get for the hadronic tensor the expression
\begin{eqnarray}
\lefteqn{W^{\mu\nu}}
\nonumber\\
&=&
V\int\frac{d^3h}{(2\pi)^3}
\frac{1}{4E_\nh E_\np}
\frac{d}{d h_0}
\left[\frac{I^{\mu\nu}_{10}(h_0,\nh;E_\np,\np;q)}{2h_0}\right]_{h_0=E_\nh}
\delta(E_\nh+q_0-E_\np)
        \theta(k_F-h)
        \theta(p-k_F)
\nonumber\\
&+&
V\int\frac{d^3h}{(2\pi)^3}
\frac{1}{4E_\nh E_\np}
I^{\mu\nu}_{10}(E_\nh,\nh;E_\np,\np;q)
\frac{1}{2E_\nh}
\frac{d}{d q_0}
\delta(E_\nh+q_0-E_\np)
        \theta(k_F-h)
        \theta(p-k_F)
\nonumber\\
&+&
V\int\frac{d^3h}{(2\pi)^3}
\frac{1}{4E_\nh E_\np}
\frac{d}{d p_0}
\left[\frac{I^{\mu\nu}_{01}(E_\nh,\nh;p_0,\np;q)}{2p_0}\right]_{p_0=E_\np}
\delta(E_\nh+q_0-E_\np)
        \theta(k_F-h)
        \theta(p-k_F)
\nonumber\\
&-&
V\int\frac{d^3h}{(2\pi)^3}
\frac{1}{4E_\nh E_\np}
I^{\mu\nu}_{01}(E_\nh,\nh;E_\np,\np;q)
\frac{1}{2E_\np}
\frac{d}{d q_0}
\delta(E_\nh+q_0-E_\np)
        \theta(k_F-h)
        \theta(p-k_F) \ . \nonumber \\
\end{eqnarray}
In differentiating the function $I^{\mu\nu}_{01}$ defined in 
eq.~(\ref{Inl-function}), we first consider the term:
\begin{eqnarray}
\lefteqn{\frac{d}{dp_0}
\left[\frac{1}{2p_0}(\Pbar+m)\Sigma(P)(\Pbar+m)\right]_{p_0=E_\np}} \nonumber
\\
&=&
\frac{\Sigma_0(\np)}{2E_\np}
\left[
-\frac{2m}{E_\np}(\Pbar+m)
+\gamma_0(\Pbar+m)
+(\Pbar+m)\gamma_0
\right]_{p_0=E_\np}
\nonumber\\
&&
+
\left[
\frac{1}{2E_\nh}(\Hslash+m)
\frac{\partial\Sigma(H)}{\partial p_0}
(\Hslash+m)
\right]_{p_0=E_\np} \ ,
\label{derivative}
\end{eqnarray}
where use has been made of the results
\begin{eqnarray}
\Sigma(P)(\Pbar+m) &=&\Sigma_0(\np)(\Pbar+m)
\label{spm}
\\
(\Pbar+m)(\Pbar+m) &=&  2m(\Pbar+m) \ ,
\label{pm2}
\end{eqnarray}
which hold for $P^\mu$ on-shell and
where $\Sigma_0(\np)$ is the eigenvalue of the self-energy 
for on-shell spinors.

Next we should compute the derivative of the self-energy
$\Sigma(P)$. This function has the general structure given in
eq.~(\ref{spin}),
and its derivative implies derivatives of the coefficients $A$, $B$,
and $C$, namely
\begin{equation}
\frac{\partial\Sigma(P)}{\partial p_0}
=
m\frac{\partial A(P)}{\partial p_0}
+\frac{\partial B(P)}{\partial p_0} \gamma_0 p_0 
-\frac{\partial C(P)}{\partial p_0} \ngamma\cdot\np
+B(P)\gamma_0 \ ,
\end{equation}
which must be evaluated for $P^{\mu}$ on-shell. 
Using again eq.~(\ref{pm2}) together with the identity
\begin{equation}
(\Pbar+m)\gamma^{\mu}(\Pbar+m) =  2P^{\mu}(\Pbar+m)
\end{equation}
we obtain, for $P$ on-shell, 
\begin{eqnarray}
\lefteqn{\frac{1}{2E_\np}(\Pbar+m)
\frac{\partial\Sigma(P)}{\partial p_0}
(\Pbar+m)}
\nonumber\\
&=&
\frac{1}{E_\np}
\left[ m^2  \frac{\partial A(P)}{\partial p_0}
       E_\np^2\frac{\partial B(P)}{\partial p_0} 
      -\np^2\frac{\partial C(P)}{\partial p_0}
      +E_\np B_0(\np)
\right]_{p_0=E_\np} (\Pbar+m)
\nonumber\\
&=& \alpha(\np)(\Pbar+m) \ ,
\end{eqnarray}
where the definition of the function $\alpha(\np)$
in eq.~(\ref{alpha}) has been used.

Finally, collecting the above results, the derivative in 
eq.~(\ref{derivative}) is found to read
\begin{eqnarray}
\lefteqn{
\frac{d}{dp_0}
\left[\frac{1}{2p_0}(\Pbar+m)\Sigma(P)(\Pbar+m)\right]_{p_0=E_\np}
}
\nonumber\\
&=&
\frac{\Sigma_0(\np)}{E_\np}
\left[ \frac{\gamma_0 E_\np-m}{2E_\np}(\Pbar+m)
      +(\Pbar+m)\frac{\gamma_0E_\np-m}{2E_\np}
\right]
+
\alpha(\np)(\Pbar+m) \ .
\end{eqnarray}
Hence the following expression 
\begin{eqnarray}
\lefteqn{
\frac{d}{dp_0}
\left[\frac{I^{\mu\nu}_{01}(H,P,Q)}{2p_0}\right]_{p_0=E_\np}
}
\nonumber\\
&=&
{\rm Tr}
\left\{\Gamma^{\mu}(Q)(\Hslash+m)\Gamma^{\nu}(-Q)
      \left[\frac{\Sigma_0(\np)}{E_\np}\frac{\gamma_0E_\np-m}{2E_\np}
            +\frac{\alpha(\np)}{2}
      \right]
(\Pbar+m)
\right\}
\nonumber\\
&+&
{\rm Tr}
\left\{\Gamma^{\mu}(Q)(\Hslash+m)\Gamma^{\nu}(-Q)(\Pbar+m)
       \left[\frac{\Sigma_0(\np)}{E_\np}\frac{\gamma_0E_\np-m}{2E_\np}
             +\frac{\alpha(\np)}{2}
       \right]
\right\}
\nonumber\\
\end{eqnarray}
yields the derivative of $I^{\mu\nu}_{01}(H,P,Q)$ with respect to $p_0$
and the similar result 
\begin{eqnarray}
\lefteqn{
\frac{d}{dh_0}
\left[\frac{I^{\mu\nu}_{10}(H,P,Q)}{2h_0}\right]_{h_0=E_\nh}
}
\nonumber\\
&=&
{\rm Tr}
\left\{\Gamma^{\mu}(Q)
       \left[\frac{\Sigma_0(\nh)}{E_\nh}\frac{\gamma_0E_\nh-m}{2E_\nh}
             +\frac{\alpha(\nh)}{2}
       \right]
       (\Hslash+m)\Gamma^{\nu}(-Q)(\Pbar+m)
\right\}
\nonumber\\
&+&
{\rm Tr}
\left\{ \Gamma^{\mu}(Q) (\Hslash+m)
        \left[\frac{\Sigma_0(\nh)}{E_\nh}\frac{\gamma_0E_\nh-m}{2E_\nh}
               +\frac{\alpha(\nh)}{2}
        \right]
        \Gamma^{\nu}(-Q)(\Pbar+m)
\right\} 
\nonumber\\
\end{eqnarray}
holds for the derivative of $I^{\mu\nu}_{10}(H,P,Q)$ with respect to $h_0$.
In addition, with the help of eq.~(\ref{spm}), we can write for 
on-shell momenta
\begin{eqnarray}
I^{\mu\nu}_{10}(H,P,Q)
&=&
{\rm Tr}
\left\{
  \Gamma^{\mu}(Q)(\Hslash+m)\Sigma(H)(\Hslash+m)\Gamma^{\nu}(-Q)(\Pbar+m)
\right\}
\nonumber\\
&=&
2m\Sigma_0(\nh){\rm Tr}
\left\{ \Gamma^{\mu}(Q)(\Hslash+m)\Gamma^{\nu}(-Q)(\Pbar+m)
\right\}
\end{eqnarray}
and, as well, 
\begin{equation}
I^{\mu\nu}_{01}(H,P,Q)
=
2m\Sigma_0(\np){\rm Tr}
\left\{ \Gamma^{\mu}(Q)(\Hslash+m)\Gamma^{\nu}(-Q)(\Pbar+m)
\right\} \ .
\end{equation}
Finally the response functions are found as
linear combinations of the diagonal components of the hadronic tensor,
{\em  i.e.} $W^{\mu\mu}$.
Using the above equations the latter reads
\begin{eqnarray}
W^{\mu\mu}
&=&
V\int\frac{d^3h}{(2\pi)^3}
\frac{1}{4E_\nh E_\np}
\delta(E_\nh+q_0-E_\np)\theta(k_F-h)\theta(p-k_F)
\nonumber\\
&&\mbox{}\times
{\rm Tr}
\left\{\Gamma^{\mu}(Q)
       \left[\frac{\Sigma_0(\nh)}{E_\nh}\frac{\gamma_0E_\nh-m}{2E_\nh}
             +\frac{\alpha(\nh)}{2}
       \right]
       (\Hslash+m)\Gamma^{\mu}(-Q)(\Pbar+m)
\right.
\nonumber\\
&&\kern 1cm
+\Gamma^{\mu}(Q)(\Hslash+m)
\left[ \frac{\Sigma_0(\nh)}{E_\nh}\frac{\gamma_0E_\nh-m}{2E_\nh}
       +\frac{\alpha(\nh)}{2}
\right]
\Gamma^{\mu}(-Q)(\Pbar+m)
\nonumber\\
&&\kern 1cm
+\Gamma^{\mu}(Q)(\Hslash+m)\Gamma^{\mu}(-Q)
\left[\frac{\Sigma_0(\np)}{E_\np}\frac{\gamma_0E_\np-m}{2E_\np}
      +\frac{\alpha(\np)}{2}
\right]
(\Pbar+m)
\nonumber\\
&&\kern 1cm
+\left.
     \Gamma^{\mu}(Q)(\Hslash+m)\Gamma^{\mu}(-Q)(\Pbar+m)
     \left[\frac{\Sigma_0(\np)}{E_\np}\frac{\gamma_0E_\np-m}{2E_\np}
           +\frac{\alpha(\np)}{2}
     \right]
\right\}
\nonumber\\
&+&
V\int\frac{d^3h}{(2\pi)^3}
\frac{1}{4E_\nh E_\np}{\rm Tr}
\left\{ 
      \Gamma^{\mu}(Q)(\Hslash+m)\Gamma^{\mu}(-Q)(\Pbar+m)
\right\}
\nonumber\\
&&\mbox{}\times
\left( 
    \Sigma_0(\nh)\frac{m}{E_\nh}-\Sigma_0(\np)\frac{m}{E_\np}
\right)
\frac{d}{d q_0}
\delta(E_\nh+q_0-E_\np) \theta(k_F-h) \theta(p-k_F) \ ,
\end{eqnarray}
which coincides with the result in eq.~(\ref{W-RSE}), obtained by
computing the response functions using the renormalized 
current and energy.


\section{On-shell self-energy and 
         field strength renormalization function}
\label{app:E}

In this Appendix we show in detail how to evaluate the on-shell
self-energy in eq.~(\ref{Sigma_0}) and the field strength
renormalization function in eq.~(\ref{alpha}). They can be expressed
in terms of the integrals $I(P)$ and $L^\mu (P)$ in
eqs.~(\ref{I},\ref{Lmu}) as follows: \be \Sigma_0(\np)=2m B(E_\np,\np)
=-12m\frac{f^2}{m_\pi^2} \left[p_0 L_0(p_0,\np)-p L_3(p_0,\np) -m^2
I(p_0,\np)\right]_{p_0=E_\np}
\label{Sigma0_app}
\ee
and
\begin{eqnarray}
\lefteqn{\alpha(\np)=}
\nonumber\\
&=&
B_0(\np) + 
\frac{1}{E_\np}
\left[ 
       m^2\frac{\partial A(p_0,\np)}{\partial p_0}
     + E_\np^2\frac{\partial B(p_0,\np)}{\partial p_0}
     -\np^2\frac{\partial C(p_0,\np)}{\partial p_0}
\right]_{p_0=E_\np}
\nonumber\\
&=&
-12 m^2\frac{f^2}{m_\pi^2} \left[
\frac{L_0(p_0,\np)}{p_0}-I(p_0,\np)+\frac{\partial L_0(p_0,\np)}{\partial p_0}
-\frac{p}{p_0}\frac{\partial L_3(p_0,\np)}{\partial p_0}
-\frac{m^2}{p_0} \frac{\partial I(p_0,\np)}{\partial p_0}\right]_{p_0=E_\np} 
\nonumber\\
\label{alpha_app}
\end{eqnarray}
where we have used eqs.~(\ref{A-definition}-\ref{C-definition}) and 
the derivatives
\begin{eqnarray}
\left(
\frac{\partial A(p_0,\np)}{\partial p_0}
\right)_{p_0=E_\np}
&=&
-6\frac{f^2}{m_\pi^2}
\left[ 
          p_0 \frac{\partial L_0(p_0,\np)}{\partial p_0}
          -p \frac{\partial L_3(p_0,\np)}{\partial p_0}
          -m^2   \frac{\partial I(p_0,\np)}{\partial p_0} \right.
\nonumber \\
        & & \left.  + L_0 (p_0,\np)-p_0 I(p_0,\np)
\right]_{p_0=E_\np}
\\
\left(
\frac{\partial B(p_0,\np)}{\partial p_0}
\right)_{p_0=E_\np}
&=&
-6\frac{f^2}{m_\pi^2}
\left[ 
          p_0 \frac{\partial L_0(p_0,\np)}{\partial p_0}
          -p \frac{\partial L_3(p_0,\np)}{\partial p_0}
          -m^2   \frac{\partial I(p_0,\np)}{\partial p_0}
\right]_{p_0=E_\np}
\\
\left(
\frac{\partial C(p_0,\np)}{\partial p_0}
\right)_{p_0=E_\np}
&=&
-6\frac{f^2}{m_\pi^2}
\left[ 
          p_0 \frac{\partial L_0(p_0,\np)}{\partial p_0}
          -p \frac{\partial L_3(p_0,\np)}{\partial p_0}
          -m^2   \frac{\partial I(p_0,\np)}{\partial p_0} \right.
\nonumber \\
        & & \left.
          + L_0(p_0,\np)-\frac{p_0}{p}L_3(p_0,\np)
\right]_{p_0=E_\np} \, .
\end{eqnarray}

By choosing the z-axis in the
direction of $\np$ the angular integrals can be
performed analytically, yielding
\ba
I(E_\np,\np) &=& \frac{1}{(2\pi)^2}\int_0^{k_F} dk 
\frac{k}{4 p E_\nk}
    \ln\frac{\gamma(p,k)+2pk}{\gamma(p,k)-2pk}
\\
L_0(E_\np,\np)  &=& \frac{1}{(2\pi)^2}\int_0^{k_F} dk 
\frac{k}{4 p}
    \ln\frac{\gamma(p,k)+2pk}{\gamma(p,k)-2pk}
\\
L_3(E_\np,\np)  &=& \frac{1}{(2\pi)^2}\int_0^{k_F} dk 
\left\{
\frac{k^2}{2 p E_\nk}-\frac{k\gamma(p,k)}{8p^2E_\nk}
    \ln\frac{\gamma(p,k)+2pk}{\gamma(p,k)-2pk}
\right\}
\\
\left.\frac{\partial I(p_0,\np)}{\partial p_0}\right|_{p_0=E_\np}
&=& \frac{1}{(2\pi)^2}\int_0^{k_F} dk
\frac{k}{2 p E_\nk} (E_\np-E_\nk) 
\left[\frac{1}{\gamma(p,k)+2pk}-\frac{1}{\gamma(p,k)-2pk}\right]
\\
\left.\frac{\partial L_0(p_0,\np)}{\partial p_0}\right|_{p_0=E_\np}
&=& \frac{1}{(2\pi)^2}\int_0^{k_F} dk
\frac{k}{2 p} (E_\np-E_\nk) 
\left[\frac{1}{\gamma(p,k)+2pk}-\frac{1}{\gamma(p,k)-2pk}\right]
\\
\left.\frac{\partial L_3(p_0,\np)}{\partial p_0}\right|_{p_0=E_\np}
&=& -\frac{1}{(2\pi)^2}\int_0^{k_F} dk
\frac{k}{4 p^2 E_\nk} (E_\np-E_\nk) 
\left\{
\ln\frac{\gamma(p,k)+2pk}{\gamma(p,k)-2pk} \right.
\nonumber \\
& & \left.
+\gamma(p,k)\left[\frac{1}{\gamma(p,k)+2pk}-\frac{1}{\gamma(p,k)-2pk}\right]
\right\}\ ,
\ea
where we have defined the function
\be
\gamma(p,k) \equiv (E_\np-E_\nk)^2-
np^2-k^2-m_\pi^2=2m^2-m_\pi^2-2E_\np E_\nk\ .
\ee

By replacing the above integrals in 
eqs.~(\ref{Sigma0_app},\ref{alpha_app}) we obtain
\be
\Sigma_0(p) = \frac{3 m f^2}{2\pi^2 m_\pi^2} \int_0^{k_F} dk \frac{k^2}{E_\nk}
\left[
1+\frac{m_\pi^2}{4pk} \ln\frac{\gamma(p,k)+2pk}{\gamma(p,k)-2pk}
\right]
\ee
and
\be
\alpha(p) = \frac{3 m^2 f^2}{\pi^2 E_\np} \int_0^{k_F} dk \frac{k^2}{E_\nk}
\cdot\frac{E_\nk-E_\np}{\gamma^2(p,k)-4 p^2 k^2}\, .
\ee

It is interesting to note that for large $p$-values the following limits
hold
\ba
\lim_{p\to \infty}\alpha(p)&=& 0 \\
\lim_{p\to \infty}\Sigma_0(p) &=&
\frac{3 m f^2}{2\pi^2 m_\pi^2} \int_0^{k_F} dk \frac{k^2}{E_\nk}
\frac{3 m f^2}{4\pi^2 m_\pi^2}\left(E_Fk_F
-m^2\ln\frac{E_F+k_F}{m}\right) \, ,
\ea
where $E_F=\sqrt{k_F^2+m^2}$ is the Fermi energy. For $k_F=237$ MeV/c
the on-shell self-energy limit is $\sim 34$ MeV.


\section{The electromagnetic current operator}
\label{app:F}


In this Appendix we provide a simple derivation of the non--relativistic 
reduction of the single--nucleon on--shell electromagnetic
current operator (see~\cite{Ama96a,Ama96b,Jes98}).
The single--nucleon electromagnetic current reads
\be
J^{\mu}(P's';Ps) = 
\ubar \left[ F_1(Q^2)\gamma^\mu + \frac{i}{2m}F_2(Q^2)\sigma^{\mu\nu}Q_\nu 
\right] \uu\ ,
\label{eq1}
\ee
where $P^\mu = (E,\np)$ is the four--momentum of the incident nucleon,
$P'^\mu = (E',\np')$ the four--momentum of the outgoing nucleon and
$Q^\mu = P'^\mu-P^\mu =(\omega,{\bf q})$ 
the transferred four--momentum. The spin projections
for incoming and outgoing nucleons are labeled $s$ and $s'$, respectively.
We follow the conventions of Bjorken and Drell~\cite{Bjo65} for the 
$u$--spinors.
For convenience in the discussions that follow of the scales in the problem 
we introduce the dimensionless variables:
$\etavec = {\bf p}/m$, 
$\varepsilon = E/m=\sqrt{1+\eta^2}$,
$\lambda = \frac{\omega}{2m}$,
$\kappa = \frac{q}{2m}$ and
$\tau = -\frac{Q^2}{4m^2} = \kappa^2-\lambda^2$.
For the outgoing nucleon, $\etavec'$ and $\varepsilon'$ are defined
correspondingly. 

For any general operator whose $\Gamma$-matrix form is given by
\be
\Gamma=
\left(
\begin{array}{cc}
\Gamma_{11} & \Gamma_{12}\\
\Gamma_{21} & \Gamma_{22}
\end{array}
\right)
\label{eq11}
\ee
one has $\ubar\Gamma\uu=\chi^\dagger_{s'}\overline{\Gamma}\chi_s$, with the
current operator $\overline{\Gamma}$ given by 
\be
\overline{\Gamma}= \frac{1}{2}\sqrt{(1+\varepsilon)(1+\varepsilon')}
\left(\Gamma_{11}+
\Gamma_{12}\frac{\sigvec\cdot\etavec}{1+\varepsilon}-
\frac{\sigvec\cdot\etavec'}{1+\varepsilon'}\Gamma_{21}-
\frac{\sigvec\cdot\etavec'}{1+\varepsilon'}\Gamma_{22}
\frac{\sigvec\cdot\etavec}{1+\varepsilon}\right)\ .
\label{eq12}
\ee

An important point in our approach is that we
expand only in powers of the bound nucleon momentum 
$\eta$, not in the transferred momentum $\kappa$ or
the transferred energy $\lambda$. This is a very reasonable approximation
as the momentum of the initial nucleon is relatively low in most
cases, since the typical values of $\eta$ lie below $\eta_F\equiv k_F/m$,
where $k_F$ is the Fermi momentum ($\eta_F$ is typically about 1/4). 
However, for those cases corresponding to short--range properties
of the nuclear wave functions it will be necessary to be very careful with
the approximations made. Indeed, for large values of $\eta$ a 
fully--relativistic
approach will likely prove necessary. Expanding up to first order in powers of
$\eta$ we get
$\varepsilon \simeq 1$ and
$\varepsilon' \simeq 1+2\lambda$.

Thus, the non--relativistic
reductions of the time and space components of the single--nucleon
electromagnetic current operator can be evaluated in a rather simple form.

Let us consider first the case of the time component. We have
\be
J^0(P's';Ps)=
\ubar J^0 \uu = \chi^\dagger_{s'}\overline{J^0}\chi_s\ ,
\label{eq17}
\ee
with the current operator $J^0=F_1\gamma^0+iF_2\sigma^{0\nu}Q_\nu/2m$.
Using the general result given by eq.~(\ref{eq12}) and
expanding up to first order in $\eta$, it is straightforward
to get the relation
\be
\overline{J^0}\simeq
        \frac{\kappa}{\sqrt{\tau}}G_E
        +\frac{i}{\sqrt{1+\tau}}\left(G_M-\frac{G_E}{2}\right)
        (\kappavec\times\etavec)
        \cdot\sigvec\ ,
\label{eq22}
\ee
where we have introduced 
the Sachs form factors $G_E=F_1-\tau F_2$ and
$G_M=F_1+F_2$, and have used the relations
\ba
\lambda &\simeq& \tau+\kappavec\cdot\etavec \label{eq15}\\
\kappa^2 &\simeq & \tau(1+\tau+2\kappavec\cdot\etavec) \label{eq16}\ .
\ea

The expression~(\ref{eq22}) coincides with the leading-order expressions 
obtained in previous work~\cite{Ama96a,Jes98}; in those studies a different 
approach was taken which, while more cumbersome, does yield terms of higher 
order than the ones considered in the present work. 
It is important to remark again that
no expansions have been made in terms of the transferred energy and 
transferred momentum; indeed, $\kappa$, $\lambda$ and $\tau$ may be
arbitrarily large in our approach.

Let us consider now the case of space components. Thus,
we have 
\begin{equation}
{\bd J}(P's';Ps)=\ubar{\bd J}\uu = 
\chi^\dagger_{s'}\overline{\bd J}\chi_s.
\end{equation}
Introducing the matrix form of the vector component for the
single--nucleon electromagnetic current operator in
the general relation 
(\ref{eq12}), one can finally write
\ba
\overline{\bd J} &\simeq & 
        \frac{1}{\sqrt{1+\tau}}\left\{
        iG_M(\sigvec\times\kappavec)    +
        \left(G_E+\frac{\tau}{2}G_M\right)\etavec + G_E\kappavec \right.
\nonumber \\
& -&    \left. \frac{G_M}{2(1+\tau)}(\kappavec\cdot\etavec)\kappavec-
\frac{iG_E}{2(1+\tau)}(\sigvec\times\kappavec)\kappavec\cdot\etavec
\right. \nonumber \\
        &-& \left. i\tau(G_M-G_E/2)(\sigvec\times\etavec)+
        \frac{i(G_M-G_E)}{2(1+\tau)}(\kappavec\times\etavec)
        \sigvec\cdot\kappavec \right\}\ ,
\label{eq24}
\ea
where we have used the relations given by eqs.~(\ref{eq15},\ref{eq16}).

In order to compare with~\cite{Jes98},
we write the expression for the transverse component of the current, 
{\em i.e.},
$\overline{\bd J}^\perp =\overline{\bd J}-
\frac{\overline{\bd J}\cdot\kappavec}{\kappa^2}
\kappavec$. After some algebra we get the final result
\ba
\overline{\bd J}^\perp &\simeq&
\frac{1}{\sqrt{1+\tau}}\left\{
        iG_M(\sigvec\times\kappavec)+
        \left(G_E+\frac{\tau}{2}G_M\right)
        \left(\etavec
        -\frac{\kappavec\cdot\etavec}{\kappa^2}\kappavec\right) 
\right. \nonumber \\
&-& \left.
        \frac{iG_M}{1+\tau}
        (\sigvec\times\kappavec)\kappavec\cdot\etavec 
+       \frac{iG_M}{2(1+\tau)}(\etavec\times\kappavec)
        \sigvec\cdot\kappavec \right\}\ .
\label{eq28}
\ea
It is straightforward to prove that this expression coincides with the
result given by eq.~(25) in~\cite{Jes98} for an expansion in powers
of $\eta$ up to first order.

Therefore, as can be seen from eqs.~(\ref{eq22},\ref{eq28}), at linear
order in $\eta$ we retain the spin--orbit part of the charge and one
of the relativistic corrections to the transverse current, the
first--order convective spin--orbit term.  It is also important to
remark here that the current operators given by
eqs.~(\ref{eq22},\ref{eq24}) satisfy the property of current
conservation $\lambda J_0 = \kappavec\cdot{\bd J}$.  Finally, it is
also interesting to quote the results obtained in the traditional
non--relativistic reduction~\cite{Jes98,Bof80,Giu80,Ada97,Rit97},
where it is assumed that $\kappa <<1$ and $\lambda << 1$: 
\ba
\overline{J^0}_{nonrel} 
&=& 
G_E 
\\ 
\overline{\bd J}_{nonrel}^\perp 
&=& -
iG_M[\kappavec \times \sigvec]+
G_E\left[\etavec
         -\left(\frac{\kappavec\cdot\etavec}{\kappa^2}
          \right)\kappavec
    \right]\ .
\label{eq30}
\ea Note that this traditional non--relativistic reduction contains
both terms of zeroth and first order in $\eta$, {\em i.e.}, the
convection current, and is therefore not actually of lowest order in
$\eta$.

We see that the expansion of the current {\em to first order} in the
variable $\eta=p/m$ yields quite simple expressions; moreover the
various surviving pieces of the relativized current ({\em  i.e.}, charge and
spin--orbit in the longitudinal and magnetization and convection in
the transverse) differ from the traditional non--relativistic
expressions only by multiplicative $(q,\omega)$-dependent factors such
as $\kappa/\sqrt{\tau}$ or $1/\sqrt{1+\tau}$, and therefore are easy
to implement in already existing non--relativistic models.



\begin{thebibliography}{Expo92}

\bibitem{Wil97} C.F. Williamson {\em et al.,}
                Phys. Rev. {C56} (1997) 3152. **

\bibitem{Vol99} V.D. Burkert,
                Prog. Part. Nucl. Phys. {44} (2000) 273-291.

\bibitem{Arr99} J. Arrigton {\em et al.},
                Phys. Rev. Lett. {82} (1999) 2056.

\bibitem{Gil97a} A. Gil, J. Nieves, E. Oset,
                 Nucl. Phys. {A627} (1997) 543. *

\bibitem{Gil97b} A. Gil, J. Nieves, E. Oset,
                 Nucl. Phys. {A627} (1997) 599. 

\bibitem{Car02} J. Carlson, J. Jourdan, R. Schiavilla, I. Sick,
                Phys.Rev. { C65} (2002) 024002.

\bibitem{Car98} J. Carlson, R. Schiavilla,
                Rev. Mod. Phys. {70} (1998) 743. *

\bibitem{Fab97} A. Fabrocini, Phys.Rev. C55 (1997) 338. 

\bibitem{Kim01} K.S. Kim, L.E. Wright and D.A. Resler,
                Phys. Rev. { C64} (2001) 044607.

\bibitem{Hor81} C.J. Horowitz and B.D. Serot, *
                Nucl. Phys. { A368} (1981) 503.

\bibitem{Yan92} Yanhe Jin, D.S. Onley and L.E. Wright,
                Phys. Rev. { C45} (1992) 1311. 

\bibitem{Kur02} H. Kurasawa and T. Suzuki,
                preprint nucl-th/0201035.

\bibitem{Amo96} P. Amore, M.B. Barbaro, A. De Pace,
                Phys. Rev. { C53} (1996) 2801.

\bibitem{Ben99} O. Benhar,
                Phys. Rev. Lett. {83} (1999) 3130.

\bibitem{Gur02} S.A. Gurvitz and S. Rinat,
                Phys. Rev. {C65} (2002) 024310.

\bibitem{Sim00} S. Simula,
                preprint nucl-th/0002030.

\bibitem{Par01} M.W. Paris and V.R. Pandharipande,
                preprint nucl-th/0110048.

\bibitem{Isg01} N. Isgur, S. Jeschonnek, W. Melnitchouk, J.W. Van Orden,
                 Phys.Rev. {D64} (2001) 054005.

\bibitem{Ama02} J.E. Amaro, M.B. Barbaro, J.A. Caballero, 
                T.W. Donnelly, A. Molinari,
                Nucl. Phys. {A697} (2002) 388. ***

\bibitem{Ama98a} J.E. Amaro, M.B. Barbaro, J.A. Caballero, 
                T.W. Donnelly, A. Molinari,
                Nucl. Phys. { A643} (1998) 349. **

\bibitem{Ama99a} J.E. Amaro, M.B. Barbaro, J.A. Caballero, 
                T.W. Donnelly, A. Molinari,
                Nucl. Phys. { A657} (1999) 161.

\bibitem{Che71} M. Chemtob and M Rho, 
                Nucl. Phys. A163 (1971) 1.

\bibitem{Ama98c} J.E. Amaro, A. M. Lallena, G. Co' and  A. Fabrocini,
                 Phys. Rev. C57 (1998) 3473.

\bibitem{Alb90} W.M. Alberico, T.W. Donnelly, A. Molinari,
                Nucl. Phys. { A512} (1990) 541. ***

\bibitem{Alb81} W.M. Alberico, A. Molinari,
                J.Phys. {G7} (1981) L93.

\bibitem{Ama92a} J.E. Amaro, G. Co', E.M.V. Fasanelli, A.M. Lallena, 
                 Phys. Lett. {B277} (1992) 249. 

\bibitem{Ama92b} J.E. Amaro, G. Co', A.M. Lallena, 
                 Ann. Phys. (N.Y.) {221} (1993) 306.

\bibitem{Ama94a}  J.E. Amaro, G. Co', A.M. Lallena, 
                 Nucl. Phys. {A578} (1994) 365. *

\bibitem{Blu89}  P.G. Blunden, M.N. Butler,
                 Phys. Lett. B219 (1989) 151.

\bibitem{Hor90}  C.J. Horowitz, J. Piekarewicz,
                 Nucl. Phys. {A511} (1990) 461. 

\bibitem{Bjo65}  J.D. Bjorken, S.D. Drell,
                 {Relativistic quantum mechanics}
                 (McGraw-Hill, 1965).

\bibitem{Pes95}  M.E. Peskin, D. V. Schroeder,
                 {An introduction to quantum field theory}
                 (Perseus, 1995).

\bibitem{Sch00}  E. Schiller and H. Mutter,
                 Eur. Phys. J. {A11} (2001) 15.

\bibitem{Don86} T.W.Donnelly, A.S. Raskin, 
                Ann. Phys. (N.Y.) { 169} (1986) 247. *

\bibitem{Don92} T.W. Donnelly, M.J. Musolf, W.M. Alberico, 
                M.B. Barbaro, A. De Pace, A. Molinari, 
                Nucl. Phys. {A541} (1992) 525.  *

\bibitem{Bof93} S. Boffi, C. Giusti, F.D. Pacati, 
                Phys. Rep. { 226} (1993) 1. **

\bibitem{Sar93} A. M. Saruis,
                Phys. Rep. 235 (1993) 57.

\bibitem{Bof96} S. Boffi, C. Giusti, F. Pacati, M. Radici,      
                {Electromagnetic Response of Atomic Nuclei} 
                (Oxford-Clarendon Press, 1996).

\bibitem{Bar01} M.B. Barbaro, A. De Pace, T.W. Donnelly and A. Molinari, 
                in: Electromagnetic Response Functions of Nuclei, 
                R. Cenni Ed. 
                (Nova Science Publishers, Huntington, NY, 2001).

\bibitem{Ama01} J.E. Amaro, G. Co' and A.M. Lallena,
                in: Electromagnetic Response Functions of Nuclei, 
                R. Cenni Ed. 
                (Nova Science Publishers, Huntington, NY, 2001).

\bibitem{Alv01} L. Alvarez-Ruso, M.B. Barbaro, T.W. Donnelly, A. Molinari,
                Phys. Lett. {B497} (2001) 214.

\bibitem{Chi89} C.R. Chinn, A. Picklesimer, J.W. Van Orden, 
                Phys. Rev. { C40} (1989) 790.

\bibitem{Fet71} A.L. Fetter and J.D. Walecka, 
                Quantum theory of many-particle systems
                (McGraw-Hill, 1971).

\bibitem{Gal71} S. Galster {\em et al.}, 
                Nucl. Phys. {B32} (1971) 221.

\bibitem{Van81} J.W. Van Orden, T.W. Donnelly, 
                Ann. Phys. {131} (1981) 451. ***

\bibitem{Gro87} F. Gross, D.O. Riska,
                Phys. Rev. { C36} (1987) 1928.

\bibitem{War50} J.C. Ward, 
                Phys. Rev. {78} (1950) 182.

\bibitem{Tak57} Y. Takahashi,
                Nuovo Cimento {6} (1957) 371.

\bibitem{Cel86} L.S. Celenza, C. M. Shakin,
                Relativistic Nuclear Physics
                (World Scientific, Singapore, 1986). *

\bibitem{Ama02b} J.E. Amaro, M.B. Barbaro, J.A. Caballero, 
                 T.W. Donnelly, A. Molinari,
                 in preparation.

\bibitem{Bar93} W.M. Alberico, M.B. Barbaro, A. De Pace, 
                T.W. Donnelly, A. Molinari,
                Nucl. Phys. {A563} (1993) 605.

\bibitem{Dub76} J. Dubach, J.H. Koch, T.W. Donnelly, 
                Nucl. Phys. {A271} (1976) 279.

\bibitem{Ris89} D.O. Riska, 
                Phys. Rep. {181} (1989) 208.

\bibitem{Mat89} J.F. Mathiot, 
                Phys. Rep. {173} (1989) 64.

\bibitem{Bar94} M.B. Barbaro, A. De Pace, T.W. Donnelly and A. Molinari,
                Nucl. Phys. {A569} (1994) 701.

\bibitem{Ama96c} J.E. Amaro, B. Ameziane  and A.M. Lallena,
                Phys. Rev. C 53 (1996) 1430.

\bibitem{Bar99} P.J. Barneo, J.E. Amaro and A.M. Lallena,
                Phys. Rev. C60 (1999) 044615.

\bibitem{Ama93} J.E. Amaro, A.M. Lallena, G. Co',
                Ann. Phys. (N.Y.) {221} (1993) 306.

\bibitem{Ama93b} J.E. Amaro, 
                 Ph.D. Thesis, Granada 1993 (unpublished). 

\bibitem{Don99a} T.W. Donnelly and  I. Sick, 
                Phys. Rev. Lett. {82} (1999) 3212.

\bibitem{Don99b} T.W. Donnelly and  I. Sick, 
                Phys. Rev. {C60} (1999) 065502. 

\bibitem{Mai02} C. Maieron, T.W. Donnelly and I. Sick, 
                Phys. Rev. {C65} (2002) 025502. 

\bibitem{Ana81}  M.R. Anastasio, L.S. Celenza and  C.M. Shakin,
                 Phys. Rev. { 23} (1981) 569.

\bibitem{Mus94} M.J. Musolf, T.W. Donnelly, J. Dubach, S.J. Pollock,
                S. Kowalski and  E.J. Beise,
                Phys. Rep. {239} (1994) 1. ***

\bibitem{Rek02} M. Rekalo, J. Arvieux, E. Tomasi-Gustafsson
                Phys.Rev. C65 (2002) 035501.

\bibitem{Ama96a} J.E. Amaro, J.A. Caballero, E. Moya de Guerra, 
                 T.W. Donnelly, A.M. Lallena and  J.M. Udias,
                 Nucl. Phys. {A602} (1996) 263. *

\bibitem{Bar96} M.B. Barbaro, A. De Pace, T.W. Donnelly and  A. Molinari,
                Nucl. Phys. {A598} (1996) 503.

\bibitem{Koh81} M. Kohno and N. Ohtsuka, 
                Phys. Lett. { B98} (1981) 335.

\bibitem{BoRa91} S. Boffi and M. Radici,
                Nucl. Phys. A526 (1991) 602.

\bibitem{Van94} V. Van der Sluys, J. Ryckebusch and M. Waroquier,
                 Phys. Rev. {C49} (1994) 2695.

\bibitem{Van95} V. Van der Sluys, J. Ryckebusch and M. Waroquier, 
                Phys. Rev. {C51} (1995) 2664.

\bibitem{Ama99b} J.E. Amaro, A.M. Lallena, J.A. Caballero,
                Phys. Rev C60 (1999) 014602.

\bibitem{Jes98} S. Jeschonnek and  T.W. Donnelly, 
                Phys. Rev. {C57} (1998) 2438.

\bibitem{Maz02} M. Mazziotta, J.E. Amaro and F. Arias de Saavedra,
                Phys. Rev. C65 (2002) 034602.

\bibitem{Ama96b} J.E. Amaro, J.A. Caballero, E. Moya de Guerra
                 and  T.W. Donnelly,
                 Nucl. Phys. A611 (1996) 163.

\bibitem{Ama98b} J.E. Amaro and T.W. Donnelly, 
                 Ann. Phys. (N.Y.) {263} (1998) 56.

\bibitem{Ama99c} J.E. Amaro and T.W. Donnelly, 
                 Nucl. Phys. A646 (1999) 187.

\bibitem{Ama02c} J.E. Amaro and T.W. Donnelly, 
                 Nucl. Phys. A (2002) in press.

\bibitem{Udi99} J.M. Udias, J.A. Caballero, E. Moya de Guerra,
                 J.E. Amaro and T.W. Donnelly,
                Phys. Rev. Lett. 83 (1999) 5451

\bibitem{Cab01} J.A. Caballero, M.C. Martinez, E. Moya de Guerra,
                J.M. Udias,  J.E. Amaro and T.W. Donnelly,
                Nucl. Phys. A689 (2001) 449c.

\bibitem{Cris1} M.C. Mart\'{\i}nez, J.A. Caballero and T.W. Donnelly, 
                nucl-th/0201004,
                submitted to Nucl. Phys. A.

\bibitem{Ama94b} J.E. Amaro, G. Co' and  A.M. Lallena, 
                 Int. J. Mod. Phys. {E3} (1994) 735.

\bibitem{deF66} T. deForest and J.D. Walecka, 
                Adv. Phys. {15} (1966) 1.

\bibitem{Fri73} J.L. Friar,
                Ann. Phys. (N.Y.) {81} (1973) 332.

\bibitem{Cio80} C. Ciofi degli Atti,
                Prog. Part. Nucl. Phys. {3} (1980) 163.

\bibitem{Alb88} W.M. Alberico, A. Molinari, T.W. Donnelly, 
                E.L. Kronenberg and J.W. Van Orden, 
                Phys. Rev. {C38} (1988) 1801. *

\bibitem{Ser86} B.D. Serot and  J.D. Walecka, 
                Adv. Nucl. Phys. {16} (1986) 1.

\bibitem{Dek92} M.J. Dekker, P.J. Brussaard, J.A. Tjon, 
                Phys. Lett. {B289} (1992) 255.

\bibitem{Dek94} M.J. Dekker, P.J. Brussaard, J.A. Tjon, 
                Phys. Rev. {C49} (1994) 2650;

\bibitem{Orl91} G. Orlandini, M. Traini, 
                Rep. Prog. Phys. 54 (1991) 257.

\bibitem{Jou95} J. Jourdan, 
                Phys. Lett. { B353} (1995) 189; 

\bibitem{Jou96} J. Jourdan, 
                Nucl. Phys. { A603} (1996) 117.

\bibitem{Cen97} R. Cenni, T.W. Donnelly, A. Molinari,
                Phys. Rev. C56 (1997) 276.

\bibitem{Cen00} R. Cenni,
                Nucl.Phys. A696 (2001) 605.

\bibitem{Bof80} S.Boffi, C. Giusti and F.D. Pacati,
                 Nucl. Phys. {A336} (1980) 416.

\bibitem{Giu80} C. Giusti and  F.D. Pacati,
                 Nucl. Phys. {A336} (1980) 427.

\bibitem{Ada97} J. Adam Jr. and  H. Arenh\"ovel,
                Nucl. Phys. {A614} (1997) 289.

\bibitem{Rit97} F. Ritz, H. G\"oller, T. Wilbois and  H. Arenh\"ovel,
                Phys. Rev. {C55} (1997) 2214.

\end{thebibliography}
\end{document}